\documentclass{aa}

\usepackage{graphicx}
\usepackage{txfonts}
\usepackage{natbib}

\usepackage{stfloats} %double column image where it should be without [H]
\usepackage{hhline}
\usepackage{pdflscape}
\usepackage{lscape}
\usepackage{floatpag}
\usepackage{subcaption}
\usepackage{float} % here for H placement parameter for tale in section
\usepackage{newtxtext,newtxmath}

\usepackage[explicit]{titlesec}
\usepackage{multirow}
\usepackage{tabulary}
\usepackage[para]{threeparttable}
\usepackage{array,booktabs,longtable,tabularx}
\newcolumntype{L}{>{\raggedright\arraybackslash}X}
\usepackage{ltablex}

\newcommand{\MM}{$\rm{M}_\odot$\xspace}
\newcommand{\MS}{$\rm{M}_\star$\xspace}
\newcommand{\SFR}{$\rm{M}_{\odot}/\rm{yr}$\xspace}

\newcommand{\simba}{\mbox{{\sc Simba}}\xspace}
\newcommand{\gizmo}{\mbox{{\sc Gizmo}}\xspace}
\newcommand{\ramses}{\mbox{{\sc Ramses}}\xspace}
\newcommand{\gad}{\mbox{{\sc \small Gadget-3}}\xspace}
\newcommand{\grac}{\mbox{{\sc \small Grackle-3.1}}\xspace}
\newcommand{\hagn}{\mbox{{\sc Horizon-AGN}}\xspace}
\newcommand{\eagle}{\mbox{{\sc EAGLE}}\xspace}
\newcommand{\tng}{\mbox{{\sc IllustrisTNG}}\xspace}
\newcommand{\illustris}{\mbox{{\sc Illustris}}\xspace}
\newcommand{\tngone}{\mbox{{\sc TNG100}}\xspace}
\newcommand{\arepo}{\mbox{{\sc AREPO}}\xspace}

\newcommand\sfig[3]{
\begin{figure}[!htbp]
  \centering
  \includegraphics[width=#1\linewidth]{./im/#2}
  \caption{#3}
  \label{#2}
\end{figure}
}
\newcommand\sfigs[4]{
\begin{figure}[!htbp]
  \centering
  \includegraphics[width=#1\linewidth]{./im/#2}
  \caption{#3}
  \label{#2}
\end{figure}
}
\newcommand\bfig[3]{
\begin{figure*}[!htbp]
  \centering
  \includegraphics[width=#1\textwidth]{./im/#2}
  \caption{#3}
  \label{#2}
\end{figure*}
}
\newcommand\bfigs[4]{
\begin{figure*}[!htbp]
  \centering
  \includegraphics[width=#1\textwidth]{./im/#2}
  \caption{#3}
  \label{#2}
\end{figure*}
}

\usepackage{color}
\definecolor{Blue}{rgb}{0,0.08,0.65}
\definecolor{Red}{rgb}{0.65,0.08,0.05}
\definecolor{Green}{rgb}{0.2,0.55,0.25}
\definecolor{Orange}{rgb}{1.0,0.5,0.15}
\definecolor{Purple}{rgb}{0.5,0.0,0.5}

\begin{document}

\def\SFRD{\,{\rm M_\odot\,year^{-1}\,Mpc^{-3}}}

\title{HSC-CLAUDS survey: The star formation rate functions since $z\sim 2$ and comparison with hydrodynamical simulations}
\titlerunning{HSC-CLAUDS survey: The SFR functions up to $z\sim 2$, comparison with hydro-dynamical simulations}

    \author{V.~Picouet
          \inst{1,2}\fnmsep\thanks{\email{vincent@picouet.fr}}
          \and
          S.~Arnouts\inst{1}
          \and
          E.~Le~Floc'h\inst{3}
          \and
          T.~Moutard\inst{1}
          \and
          K. Kraljic\inst{1,4}
          \and
          O.~Ilbert\inst{1}
          \and
          M.~Sawicki\inst{5}
          \and
          G.~Desprez\inst{5}
          \and
          C.~Laigle\inst{6}
          \and
          D.~Schiminovich\inst{2}
          \and
          S.~de~la~Torre\inst{1}
          \and
          S.~Gwyn\inst{7}
          \and
          H.J.~McCracken\inst{6}
          \and
          Y.~Dubois\inst{6}
          \and
          R. Dav\'e\inst{8,9,10}
          \and
          S. Toft \inst{11,12}
          \and
          J.R.~Weaver\inst{13}
          \and
          M. Shuntov\inst{11}
          \and
          O. B. Kauffmann\inst{1}
          }

   \institute{
             Aix Marseille Universit\'e, CNRS, CNES, LAM, Marseille, France
             \and
             Department of Astronomy, Columbia University, 550 W. 120$^{th}$ Street, New York, NY 10027, USA
             \and
             Laboratoire AIM, CEA/DSM-CNRS-Université Paris Diderot, IRFU/Service d’Astrophysique, Bat. 709, CEA-Saclay, 91191 Gif- sur-Yvette Cedex, France
             \and
             Universit\'e de Strasbourg, CNRS UMR 7550, Observatoire astronomique de Strasbourg, 
             F-67000 Strasbourg, France
             \and
             Department of Astronomy \& Physics and Institute for Computational Astrophysics, Saint Mary's University, 923 Robie Street, Halifax, Nova Scotia, B3H 3C3, Canada
           \and
            Institut d'Astrophysique de Paris, UMR 7095, CNRS, UPMC Univ. Paris VI, 98 bis boulevard Arago, 75014 Paris, France
            \and 
            Herzberg Astronomy and Astrophysics, National Research Council of Canada, 5071 West Saanich Rd., Victoria, BC V9E 2E7, Canada
            \and
            Institute for Astronomy, University of Edinburgh, Royal Observatory, Blackford Hill, Edinburgh, EH9 3HJ, United Kingdom
            \and
            University of the Western Cape, Bellville, Cape Town 7535, South Africa
            \and
            South African Astronomical Observatories, Observatory, Cape Town 7925, South Africa
            \and
            Cosmic Dawn Center (DAWN), Denmark
            \and
            Niels Bohr Institute, University of Copenhagen, Jagtvej 128, DK-2200 Copenhagen, Denmark
            \and
            Department of Astronomy, University of Massachusetts, Amherst, MA 01003, USA
             }

   \date{}

  \abstract
{Star formation rate functions (SFRFs) give an instantaneous view of the distribution of star formation rates (SFRs) in galaxies at different epochs. They are a complementary and more stringent test for models than the galaxy stellar mass function, which gives an integrated view of the past star formation activity. However, the exploration of SFRFs has been limited thus far due to difficulties in assessing the SFR from observed quantities and probing the SFRF over a wide range of SFRs.}
{We overcome these limitations thanks to an original method that predicts the infrared luminosity from the rest-frame UV/optical color of a galaxy and then its SFR  over a wide range of stellar masses and redshifts. We applied this technique to the  deep imaging survey HSC-CLAUDS combined with near-infrared and UV photometry.
We provide the first SFR functions with reliable measurements in the high- and low-SFR regimes up to $z=2$ and compare our results with previous observations and four state-of-the-art hydrodynamical simulations.}
{ The SFR estimates are based on the calibration of the infrared excess ($IRX=L_{\rm IR}/L_{\rm UV}$) in the NUVrK color-color diagram. We improved upon the original calibration in the COSMOS field by incorporating \textit{Herschel} photometry, which allowed us to extend the analysis to higher redshifts and to galaxies with lower stellar masses using stacking techniques. Our $NrK$ method leads to an accuracy of individual SFR estimates of $\sigma=$0.2/0.3dex at low/high stellar masses. We show that it reproduces the evolution of the main sequence up to $z=2$ and 
the behavior of the attenuation (or $\langle IRX \rangle$) with stellar mass. In addition to the known lack of evolution of this relation up to $z=2$ for galaxies with \MS$\le10^{10.3}$\MM, we observe a plateau in $\langle IRX \rangle$ at higher stellar masses that depends on redshift.      }  
{ We measure the SFR functions and cosmic SFR density up to $z=2$ for a mass-selected star-forming galaxy sample (with a mass limit of \MS$\ge 2.10^{9}$\MM at $z=2$). The SFR functions cover a wide range of SFRs ($0.01\le SFR\le 1000\ $\SFR), providing good constraints on their shapes. They are well fitted by a Schechter function after accounting for the Eddington bias. The high-SFR tails match the far-infrared observations well, and show a strong redshift evolution of the Schechter parameter, $SFR^{\star}$, as $\log_{10}(SFR^{\star})=0.58 z + 0.76$. The slope of the SFR functions, $\alpha$, shows almost no evolution up to $z=1.5-2$ with $\alpha=-1.3\pm 0.1$. We compare the SFR functions with predictions from four state-of-the-art hydrodynamical simulations. Significant differences are observed between them, and none of the simulations are able to reproduce the observed SFRFs over the whole redshift and SFR range. We find that only one simulation is able to predict the fraction of highly star-forming galaxies at high z, $1\leq z\leq 2$. This highlights the benefits of using SFRFs as a constraint that can be reproduced by simulations; however, despite efforts to incorporate more physically motivated prescriptions for star-formation and feedback processes, its use remains challenging.} 
{}

 \titlerunning{HSC-CLAUDS SFR functions since $z\sim 2$ and comparison with simulations}
\authorrunning{V. Picouet}
 \maketitle

\clearpage
\newpage
\newpage

% \tableofcontents

\section{Introduction}\label{intro}
Spectroscopic and multiwavelength imaging surveys have provided insights into galaxy properties and their evolution across cosmic time. A major result was the determination of the history of the cosmic star formation rate density (SFRD) thanks to analyses of galaxy luminosity functions in different wavelengths (from the UV/optical to the far-infrared and radio). After an increase from early time up to $z\sim3-4$ \citep[]{Smit2012,Mashian2016}, the SFRD reaches a maximum at cosmic noon, $z\sim 1.5-2.5$ \citep[][]{Gruppioni2013}, followed by a decline of an order of magnitude until today \citep[]{Schiminovich2005, Karim2011, Madau2014} despite a considerable amount of neutral and atomic gas available \citep[]{Peroux2020}. This is corroborated by analyses of the integrated quantity and the galaxy stellar mass function (GSMF), which shows that half of the stellar mass density has already been assembled since $z\sim2$ \citep[]{Arnouts2007}, with remarkably little evolution of the GSMF of the star-forming population since then \citep[]{Moutard2016b}. Star-forming galaxies (SFGs) gradually transition toward quiescent systems and have contributed to the buildup and evolution of the passive GSMF up to now \citep[]{Ilbert2013,Davidzon2017}.   

 Characterizing the main mechanisms involved in the evolution of the SFG population and understanding what triggers their star formation and what ultimately causes their migrations into passives, is thus a major challenge. The SFGs appear to lie on a tight sequence that links their stellar mass to their star formation rate (SFR), the so-called main sequence \citep[MS; ][]{Noeske2008, Salim2007, Speagle2014,Whitaker2012}.
 The tightness of the MS suggests that the SFGs grow through a secular evolution in an equilibrium between gas accretion, star formation, and outflows \citep[]{Bouche2010, Dave2012, Lilly2013}. The slope of the MS also suggests that lower-mass SFGs tend to be more efficient at forming stars with a higher specific star formation rate (sSFR; the SFR per stellar mass) than more massive ones. The depletion timescale based on the molecular gas content reveals that galaxies rapidly consume their gas, $t_{\rm dep}=M_{\rm mol}/SFR\sim1-2$ Gyr \citep[]{Bigiel2008,Tacconi2020},
 which must then be replenished for galaxies to stay on the MS. 
 In the $\Lambda$ cold dark matter (CDM) framework, it has long been claimed that galaxies can be fueled by cold gas thanks to cold mode accretion from cosmic web filaments, without the gas being gravitationally shock-heated \citep[]{Keres2005, Dekel2009a}, which also contributes to the acquisition of their angular momentum \citep[e.g.,][]{Pichon2011}. This cold accretion mode from the intergalactic medium is expected to be ubiquitous in the early Universe and in low-mass galaxies, while the hot accretion mode may dominate at lower redshifts for high-mass systems \citep[]{VandeVoort2011, Snedden2016}. 
 Furthermore, CO observations show that SFGs appear to contain three to ten times more gas at redshift $z=1-2 $ than their local counterparts \citep[]{Tacconi2010, Daddi2010}. These gas-rich systems show more disturbed thick disks, with more dispersion-dominated kinematics. This suggests that the disks are less settled than their local counterparts  \citep[]{Kassin2010,Kassin2012} and prone to violent disk instabilities
 that trigger enhanced SFRs \citep[]{Cacciato2012}, with SFGs moving up and down the MS following their gas accretion episodes and disk perturbations. Smooth and continuous gas flows from the cosmic web appear to be the key ingredient for conveying large quantities of cold gas at high redshifts and triggering intense star formation episodes. 

 Outflows, on the other hand, are the other components that alter the evolution of SFGs. Galactic winds from massive stars and supernova (SN) explosions can expel a fraction of the gas outside the disk, which reduces the star formation efficiency and contributes to the metal enrichment of the interstellar medium (ISM) and the circumgalactic medium \citep{Dave2011, Hopkins2014, Fontanot2017}. While these processes may be efficient for low-potential-well systems, they may not be sufficient for massive ones. At high masses, active galactic nucleus (AGN) feedback appears more effective at producing high-velocity winds, ejecting a large fraction of the gas, and preventing its cooling on a short timescale, thereby halting the star formation activity of the host galaxy \citep[]{Hopkins2006,Cattaneo2009}.
 
  All these inflow and outflow processes that govern the evolution of galaxies are incorporated into the most recent semi-analytical models (SAMs) via empirical recipes and into numerical simulations as sub-grid physics \citep[e.g.,][]{Somerville2015, Dave2011a, Dubois2016}. Their comparison with observations is crucial to constraining the influence of feedback processes. Such comparisons show that stellar winds on the low-mass end and AGN feedback on the massive end can explain the shape of the observed GSMFs and their deviation from the theoretical dark matter (DM) halo mass functions \citep[e.g.,][]{Silk2012}, and they are necessary for reproducing the stellar-to-halo ratio \citep[]{Shuntov2022}.

  The star formation rate function (SFRF) is another independent constraint. In contrast to the GSMF, which provides an integrated view of the past star formation activity, the SFRF gives an instantaneous view of the distribution of the in situ SFR and its relative evolution with cosmic time. It provides insights into the importance of stellar winds, as some observations suggest a correlation between outflow velocities and SFRs \citep[]{Heckman2015}. By implementing four different stellar wind recipes in hydrodynamical simulations, \citet[]{Dave2011a} show that models can reproduce the faint end of the SFRF, but they all fail to suppress high SFRs at low redshifts ($z\le 2$). \citet[]{Katsianis2017b}, using EAGLE simulations with different AGN and SN stellar wind implementations, show that SN winds play an essential role in reproducing the SFRFs at high redshifts and that AGN feedback becomes prominent at low redshifts. However, some discrepancies arise depending on the assumed SFRF measurements, based on UV, H$_{\alpha}$, or far-infrared (FIR) estimators. The simulations are in better agreement with the high-end SFR functions from UV/H$_{\alpha}$ and underestimate the number of high-SFR systems observed with the FIR SFRFs. 

  Accurately measuring SFR functions is a difficult task. It relies on observational SFR estimates, which are timescale dependent, subject to different dust attenuation effects, and sensitive to different selection effects. This can affect the shape of the SFRFs. All the known tracers (from the far-UV to the radio) have different benefits and drawbacks.

As the emission of galaxies in the rest-frame UV is dominated by young, short-lived ($t\sim10^8$ yr), massive stars \citep{Kennicutt1998},  UV represents a direct tracer of the SFR \citep{Bouwens2009,Schiminovich2005}.
It is easily accessible over the entire history of the Universe, but UV light is efficiently absorbed and scattered by dust grains, which heat up and re-emit the absorbed energy at FIR wavelengths. A correlation between the infrared excess (IRX; $IRX=L_{\rm IR}/L_{\rm UV}$), a measurement of the UV attenuation ($A_{UV}$), and the slope of the UV continuum ($\beta$ slope) has been observed for starburst galaxies \citep[]{Meurer1999, Calzetti2000}. This dust correction is abundantly used to derive the SFR of high redshift galaxies, as the UV slope is the only accessible quantity \citep[]{Smit2012, Katsianis2017a}. However, a large scatter in the IRX-$\beta$ relation is observed for the SFG population, spanning a range  between a Calzetti- and a Small Magellanic Could-like attenuation law \citep[]{Seibert2005, Salim2007}, which depends on galaxies' ages and metallicities \citep[]{Boquien2009, Shivaei2020}. 

The total infrared luminosity ($L_{\rm IR}$) is produced by the dust continuum emission and is a direct probe of the SFR \citep[]{Kennicutt1998}. It is defined as the integrated luminosity between 8 and 1000 $\mu m$ and can be assessed either by combining multiband photometry or via monochromatic wavelengths, where tight correlations are observed between monochromatic and total luminosities \citep[]{Bavouzet2008,Goto2011}. However, short wavelength monochromatic luminosities  ($\le 30~\mu m$) can be impacted by the presence of AGNs and the heating of the dust by old stellar populations for evolved galaxies \citep[]{Cortese2008}. A constraint from the Rayleigh-Jeans part of the FIR spectral energy distribution (SED) is required to minimize their impacts on SFR estimates. Finally, FIR observations suffer from limited instrumental sensitivity and angular resolution, restricting detections to luminous distant infrared galaxies. While sensitivity can be partly compensated for by different stacking techniques \citep{Heinis2013a}, the resolution leads to important confusion issues \citep{Bethermin2012}.

The H$_\alpha$ luminosity ($L_{H_\alpha}$) is produced from the gas ionized  by short-lived massive stars ($t\sim10^7$ yr) and is thus an excellent tracer of the instantaneous SFR \citep[]{Kennicutt1998}.
It is difficult, however, to observe at high redshifts as the line is redshifted into the  near-infrared (NIR) domain. Alternatively, narrowband imaging surveys can be efficiently used to detect $H_{\alpha}$ line emitters from their color excess \citep[]{Ly2011, Sobral2013}. However, the derived luminosity is subject to several uncertainties: 
 the contribution of the adjacent [NII] line, which is sensitive to the galaxy metallicities; and dust attenuation effects, which are based on an empirical relation with a large scatter \citep[]{Ly2011, Sobral2013}. Furthermore, the density of sources is sensitive to mismatched line contamination.  \\

 The SFR functions derived from the luminosity functions of the above tracers have recently been compiled over a wide redshift range  \citep[]{Katsianis2017b,Katsianis2017a}, each with specific caveats for converting their luminosities into dust-free SFRs. Below $z\sim 2$, all the SFRFs derived from UV luminosities show a shortage of high-SFR sources (SFR$>100$ \SFR),  while FIR selection reveals sources up to SFR$=1000$ \SFR. In the low-SFR regime, the slope of the SFRFs from $H_{\alpha}$ and UV luminosities shows a large range of values, $-1.4\le \alpha\le -1.8$, while the FIR observations do not have access to this regime. 
 
 In this work we make use of a new approach, combining UV and FIR observations to derive the dust correction to be applied to the UV luminosities of a mass-selected sample of SFGs. We then derive the SFRFs up to $z\sim 2$, based on the contribution of galaxies with \MS$\ge 10^{9}$ \MM.

\citet[hereafter A13]{Arnouts2013} found that the IRX shows a remarkable behavior in the rest-frame color-color diagram $(NUV-r)$ versus $(r-K)$.
 They identified a single vector by combining the two colors, $NrK$, which captures the behavior of the IRX over a large dynamical range with a small dispersion, $\sigma(IRX)\sim 0.2$ dex, and no mass dependence. 
%Indeed, Figure~\ref{mass_nrk_cosmos.pdf} shows for the COSMOS UV-to-FIR sample, the evolution of the IRX with respect to stellar mass and a NrK vector (a specific direction in the NUVrK color diagram) for two redshifts. It reveals that the IRX vs. NrK exhibits a tighter correlation over a larger dynamic range of $\sim 3$ order of magnitudes than with stellar mass.
By using a two-component dust model \citep[i.e., birth clouds and the diffuse ISM;][]{Charlot2000} and a full distribution of galaxy inclinations \citep{Tuffs2004, Chevallard2013}, this model can reproduce the IRX distribution in the color-color diagram, confirming that it encodes information about energy transfer between starlight and dust. 
This method, by combining the UV and infrared luminosities, provides a direct measurement of the energy budget that can be used to predict the infrared luminosity with a simple optical estimator and assess the total SFR \citep[]{Bell2005}, without any assumption on the shape of the attenuation law.

While in A13 the IRX measurement was based on the $L_{\rm IR}$ derived from the \textit{Spitzer} $24~\mu m$ observations, in this work we make use of the COSMOS2020 catalog \citep{Weaver2021}, which gathers \textit{Spitzer}/MIPS ($24~\mu m$) and \textit{Herschel}/PACS and SPIRE (100, 160, 250, 350, and 500 $\mu m$) bands \citep{Jin2018},  allowing us  to extend the calibration up to $z\sim 2$. In addition, we use the stacking technique to extend the calibration into the low stellar mass regime.  
 We then apply the new COSMOS2020 IRX versus $NrK$ calibration to the HSC-CLAUDS deep survey with very deep U ($u\sim27$) and optical imaging \citep[$i\sim 27$;][]{Sawicki2019}, where robust photometric redshifts have been estimated \citep[]{Desprez2023}.  Thanks to its depth, the CLAUDS-HSC survey is ideally suited to measure the unobscured UV luminosity functions up to $z=2$, providing the best constraints on both ends of the UV luminosity function \citep{Moutard2020a}.  In this work we restrict the analysis to the regions covered by deep NIR imaging ($\sim 5.5$ deg$^{2}$ in the COSMOS-E and XMM-LSS fields).

The paper is organized as follows. In Sect. \ref{data} we describe both the  COSMOS2020 data set used to derive the IRX calibration and the HSC-CLAUDS survey,  to which the calibration is applied to measure the SFRFs. 
In Sect. \ref{sec:physpara} we describe the estimates of the physical parameters, and in Sect. \ref{sec:calib} we perform the calibration of the IRX from direct FIR observations and extend it to low-mass systems using the FIR stacking technique.
Section~\ref{functions} presents the SFR functions derived between $0\le z\le2$ and a comparison with previous results from the literature, as well as a comparison with four hydrodynamical simulations, \tngone from the \tng project \citep{Pillepich2018,Nelson2019}, \eagle \citep{Crain2015,Schaye2015,McAlpine2016}, \hagn \citep{Dubois2014}, and \simba \citep{Dave2019}.
%(\tngone, \eagle, \hagn, \simba).  
Finally, Sect. \ref{SFRD} presents the cosmic SFRD for different stellar mass and SFR regimes, and we conclude in Sect.  \ref{discussion}.
 Appendix A describes the stacking analysis with the \textit{Spitzer}-24$\mu$m and \textit{Herschel}-SPIRE data, and  Appendix B gives a more detailed description of the four hydrodynamical simulations used in this work.  
Throughout this paper, we use a \cite{Chabrier2003} initial mass function, all magnitudes are in the AB system \citep{oke1974}, and we adopt a flat ${\rm \Lambda CDM}$ cosmology with $\Omega_m=0.3$, $\Omega_{\Lambda}=0.7$ and the Hubble constant $H_0=70 {\rm~ km~s^{-1}~Mpc^{-1}}$. %\kat{Shouldn't we add also appendices?}
%
% Other paper that could be used:\cite{Boissier2013,Ilbert2018,Best2018,Dekel2004,Lanc2018a,Peng2010,Davidzon2017,Bouche2010,Whitaker2012,Franx2008}

%
%%%%%%%%%%%%%%%%%%%%%%%%%%%%%%%%%%%%%%%%%%%%%%%%%%%%
\section{Data} \label{data}
%%%%%%%%%%%%%%%%%%%%%%%%%%%%%%%%%%%%%%%%%%%%%%%%%%%%
%
In this work we used the latest version of the Cosmic Evolution Survey (COSMOS) catalog provided by \citet[][COSMOS2020]{Weaver2021}, combined with the super-deblended FIR photometry of the \textit{Spitzer} and \textit{Herschel} data from \citet{Jin2018}. We performed the calibration of the IRX as a function of  $NrK$  
vector and stellar mass parameters to assess the SFR of individual galaxies, first based on direct measurements with detected FIR sources then using stacking techniques to extend the calibration to lower stellar masses.  This calibration was then applied to the sources detected in the HSC-CLAUDS-NIR catalog \citep[][hereafter D23]{Desprez2023} to derive the SFR functions up to $z\sim 2$.

\subsection{COSMOS and far-infrared catalogs}
\subsubsection{COSMOS2020 Catalog}

COSMOS is a major extragalactic field with a large multiwavelength photometric coverage over a 2 ${\rm deg}^2$ field. It collects ground-based optical observations with intermediate and broadband filters, NIR photometry for a total of 31 filter passbands as described in \citet[][ COSMOS2015]{Laigle2016}.
 The main improvement of the latest COSMOS2020 catalog with respect to COSMOS2015 is the depth of broadband imaging. It combines the deeper CFHT $U$-band imaging from CLAUDS survey \citep[]{Sawicki2019}, the ultra-deep data from Public Data Release 2 (PDR2) of the HSC Subaru Strategic Program \citep[HSC-SSP;][]{Aihara2019} and the latest release of UltraVISTA  \citep[DR4;][]{McCracken2012} as well as all the mid-infrared (MIR) imaging available with the \textit{Spitzer}/IRAC channels.

 COSMOS2020 provides a set of four photometric redshift catalogs. They rely on two different photometric extractions, based on SExtractor software \citep{Bertin1996} and the Farmer \citep[a package running the Tractor code on multiwavelength images, ][Sect. 3.2]{Weaver2022} and the two photometric redshift codes LePhare \citep[][]{Arnouts2002, Ilbert2006} and EAZY \citep[][]{Brammer2008}. For more consistency with our catalog, in the following, we only refer to the SExtractor-LePhare catalog for both the calibration of the $NrK$ versus $IRX$ relations and the comparisons of the photometric redshifts and physical parameters derived with our HSC-CLAUDS catalog.

%%%%%%%%%%%%%%
\subsubsection{Far-infrared catalog}
 We use the "super-deblended" FIR to millimeter photometric catalog from \citet{Jin2018}. They use the \textit{Spitzer}/MIPS 24 $\mu$m images from the COSMOS-\textit{Spitzer} survey \citep[]{LeFloch2009}, the \textit{Herschel}/PACS (100 and 160$\mu$m) images from PEP survey \citep[]{Lutz2011} and \textit{Herschel}/SPIRE (250, 350, and 500 $\mu$m) from the HerMES  survey \citep[]{Oliver2012}.
 Point spread function (PSF) prior-fitting multiband photometry was performed by adopting the prior positions of the 24 $\mu$m, radio, and  $K_s$-band mass-selected sources \citep[]{Liu2018}. The typical fluxes at S/N=5 correspond to 50 $\mu$Jy in \textit{Spitzer}/24 $\mu$m, 8.3, 22.9 mJy in PACS-100 and 160$\mu$m, 7.6, 11.0, 13.2 Jy in SPIRE-250, 350, and 500 $\mu$m.

 We restricted the FIR population to the 24$\mu$m sources with a signal-to-noise S/N$\ge$5 and rejected a few anomalous sources with m$_{AB}$(3.6 $\mu$m)$>$22.5, leading to a catalog of $\sim$26,000 galaxies.  Our sample is driven by the 24$\mu$m, and only the most luminous sources are detected with \textit{Herschel}. When adopting a S/N=3 for \textit{Herschel}, $\sim$ 12\%, 10\%, 34\%, 8\%, and 4\% are detected respectively in PACS-100 $\mu$m, PACS-160 $\mu$m, and SPIRE-250, 350, and 500 $\mu$m.

 The FIR sources are matched with the COSMOS2020 catalog, adopting a small positional uncertainty of 0.25 arcsec (as all the high S/N sources of interest are attached to a K-band counterpart in the COSMOS2015 catalog). Limiting the redshift range in between $0\le z\le2$, the final FIR catalog contains $\sim$23,200 sources. \\

%%%%%%%%%%%%%%%%%%%%%%%%%%%%%%%%%%%%%%%%%%%%%%%%%%%%
\subsection{The HSC-CLAUDS survey}
%%%%%%%%%%%%%%%%%%%%%%%%%%%%%%%%%%%%%%%%%%%%%%%%%%%%
%%%%%%%%%%%
\bfigs{1}{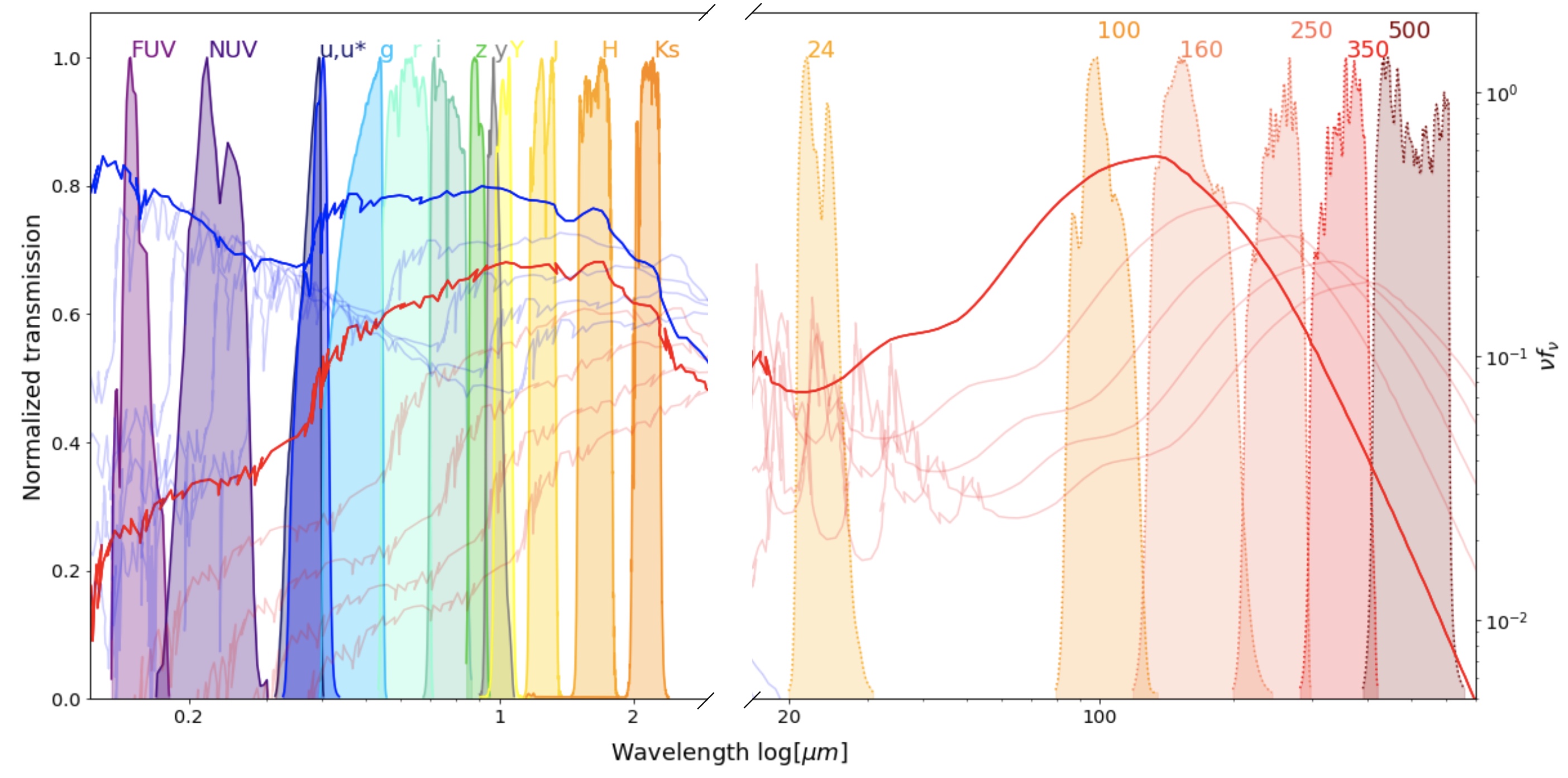}{Transmission curves of the photometric bands used to derive galaxy properties (left: UV to NIR) and infrared luminosities (right: \textit{Spitzer}/MIPS +  \textit{Herschel}/PACS-SPIRE) for the calibration of the IRX. Transmission curves are arbitrarily scaled to one. 
Two spectra of SFGs, without dust (blue) and with dust (red), are shown at $z$=0, 0.5, 1, 1.5, 2. }{test}

\bfigs{1}{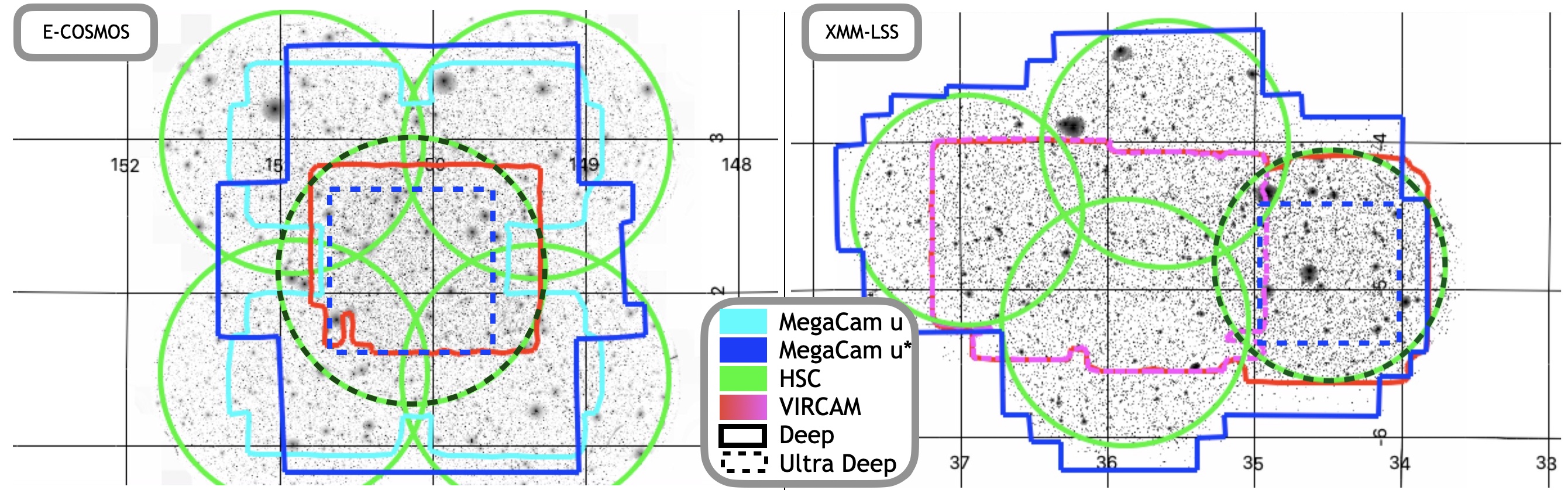}{Deep (solid lines) and ultra-deep (dashed lines) footprints of the HSC-CLAUDS and VIRCAM observations, as indicated in the inset and overlaid on the background detection images. Starting in 2015, the $u^*$ filter (blue lines) was replaced by a new filter ($u$, with a slightly bluer effective wavelength; light blue line). }{fields2.jpg}
%%%%%%%%%%%%%%%%%%%%%%%%%%%%%%%%%%%%%%%%%%%%%%%
%
\subsubsection{Observations}
HSC-CLAUDS combines the PDR2 of the deep and ultra-deep layers of the HSC-SSP  \citep[][]{Aihara2019} with the deep U-band observations carried out with the MegaCam instrument at CFHT \citep[CLAUDS, ][]{Sawicki2019}. This data set is a unique combination of depth and area, reaching 26-27$^{\rm th}$ magnitude over a total area of $\sim$20 deg$^2$, split into four separate regions (E-COSMOS, XMM-LSS, ELAIS-N1, and DEEP2-3). In the E-COSMOS and XMM-LSS fields, HSC-CLAUDS is combined with the publicly available VIRCAM  NIR observations from the VIDEO  and UltraVISTA surveys \citep[][respectively]{Jarvis2013, McCracken2018}.
The different bands are shown in Fig. \ref{newfilters2.jpg}. \\

 Two catalogs have been produced for the source extraction and flux measurements, which are described in the companion paper by D23. They discuss the processing steps to combine all the above data sets into the HSC grid and the addition of the external bands into the dedicated photometric HSC pipeline \citep{Bosch2018}. In addition to the catalog produced with the HSC pipeline, a second catalog based on SExtractor software \citep{Bertin1996} is produced using a multiband $\chi^2$ image in dual mode. In the SExtractor catalog we also include the far-UV (FUV; $\lambda\sim 1500$\AA) and near-UV (NUV; $\lambda\sim 2300$\AA) imaging from the GALEX  satellite \citep{Martin2005}. The deblending of the UV photometry was performed with the EMPHOT code \citep{Conseil2011} by using optical u-band detections as prior down to $u\sim 25$ \citep[e.g.,][]{Zamojski2007,Moutard2016a}. \\
 D23 gives all pieces of information regarding magnitude, color, depth, and photometric redshift measurements, and detailed comparisons between the two catalogs and external data sets (COSMOS2020, CANDELS, and spectroscopic samples) are discussed. The two catalogs are publicly available.\footnote{\url{https://www.clauds.net/}}
We restricted our analysis to the E-COSMOS ($\alpha\sim150^{\circ}$, $\delta\sim2.5^{\circ}$) and XMM-LSS ($\alpha\sim35.5^{\circ}$, $\delta\sim-5^{\circ}$) fields, where the deep NIR photometry is available.
Figure~\ref{fields2.jpg} shows the layouts of the CLAUDS (with $u$ and $u^\star$ filters), HSC-SSP, and VIRCAM observations in these two regions. The 5$\sigma$ depths in 2 arcsec apertures for the deep and ultra-deep components as well as their respective area are given in Table \ref{tab:survey}. 
According to those expected depths, we restrict our analysis to galaxies with $i_{\rm AB}\le 26.5$ and $K_{\rm AB}\le 24.5$, which also corresponds to the limits where the number counts in the two fields are consistent (D23). The large area covered benefits the present analysis by reducing the impact of cosmic variance in the estimate of the SFR functions and cosmic SFRD. \\
%%%%%%%%%%%

\subsubsection{Photometric redshifts}
The photometric redshifts are derived with LePhare code using all available passbands. They are described in D23 and compared to the extensive spectroscopic redshift samples available in the two fields. 
A good agreement is observed at magnitudes brighter than $i_{AB}\le 25$, with a scatter $\sigma\le$0.03  ($\sigma$ being defined as $\sigma=1.48\times \rm{Median}(|z_p-z_s|/(1 + z_s)$) and a small fraction of outliers $\eta\le 7\%$ ($\eta$ being defined as the fraction of galaxies with $\Delta z > 0.15 \times (1 + z)$).

At fainter magnitude, $i_{AB}\ge 25.5$, the scatter remains small but the outlier fraction doubles with the majority being catastrophic redshifts between low and high redshifts. 

%\begin{table}[H]
\begin{table}[]
\begin{center}
\caption{Summary of UV-optical-infrared observations with the depth (corresponding to the magnitude at $5\sigma$ in $2’'$ aperture) and area for the deep (DD) and ultra-deep (UDD) regions. The two regions are entirely covered by NIR data, except for the J band in one pointing.}
\begin{tabular}{ccccc}
\label{tab:survey}
\textbf{Telescope} & \textbf{Filter} & \textbf{Central} & \textbf{Depth} & \textbf{Surface  } \\
(Survey) &  & $\lambda$ [\AA] & DD(UDD) & deg$^2$ \\
\hhline{=====}
\textbf{GALEX} & FUV & 1526 & 25.5 & 5.3 \\
(GALEX-DIS) & NUV & 2307 & 25.5 & 4.6 \\
\hline
\textbf{CFHT} & u & 3709 & 27.2 & 1.4 \\
(CLAUDS) & u$^*$ & 3858 & 26.0 (27.5) & 5.6 (1.7) \\
\hline
\textbf{SUBARU} & g & 4847 & 27.0 (27.7) & 5.6 (2.8) \\
(HSC-SSP) & r & 6219 & 26.6 (27.5) & 5.6 (2.8) \\
 & i & 7699 & 26.5 (27.2) & 5.6 (2.8) \\
 & z & 8894 & 26.0 (26.5) & 5.6 (2.8) \\
 & y & 9761 & 25.0 (26.0) & 5.6 (2.8) \\
\hline
\textbf{VISTA} & Y & 10216 & 26.0 / 25.5 & 4.2 / 1.4 \\
VIDEO/U-VISTA & J & 12525 & 25.7 / 25.3 & 2.9 / 1.4 \\
 & H & 16466 & 25.2 / 25.0 & 4.2 / 1.4 \\
 & $K_s$ & 21557 & 24.8 / 24.8 & 4.2 / 1.4 \\
\hline

\end{tabular}
\end{center}
\end{table}

\section{Physical parameter estimates}
\label{sec:physpara}
\subsection{Stellar masses and luminosities}\label{sec:31}

 To derive the physical parameters and rest-frame luminosities we use LePhare code \citep{Arnouts2002, Ilbert2006} with the stellar population synthesis model from \cite[hereafter BC03]{Bruzual2003}, following a similar procedure as \cite{Ilbert2015}.
 We fix the redshift to its spectroscopic value if available or to the photometric redshift value based on the median of the marginalized probability distribution function.
 Our BC03 library includes six exponentially declining star formation histories, following $\tau^{-1} e^{-t/\tau}$ with $\tau$ varying between 0.1 and 30 Gyr, and two delayed star formation histories; two stellar metallicities ($Z/Z_\odot=0.4,~1$); two extinction laws with a maximum dust reddening of $E(B-V)$=0.7. We also imposed the prior $E(B-V)<0.15$ if age/$\tau>$4 (a low extinction is imposed for galaxies that have a low SFR) and include the contribution of emission lines using an empirical relation between the UV light and the emission line fluxes \citep[][]{Ilbert2009}.
 The physical parameters are derived by computing the median of the marginalized likelihood for each parameter and the errors corresponding to the 68\% confidence level. To derive the rest-frame luminosities (or absolute magnitudes), we adopted the same approach as \citet{Ilbert2005}, using the photometry in the nearest rest-frame broadband filter to minimize the dependence on the k-correction.
 We note that in this work, the SFR measurement does not rely on the 
 SED fitting ingredients, such as the dust attenuation law or the reddening excess. They are only used to best fit the observed multiband  photometry and derive the luminosities in different passbands. The SFR is then estimated from a combination of the NUV, r, K$_s$ luminosities, and redshift (see Sect. 4.1).

As we are only interested in SFGs in this analysis, we use the rest-frame color-color diagram  $(NUV_{\rm abs} - r_{\rm abs})$ versus $(r_{\rm abs} - K_{\rm abs})$  (Fig. \ref{IRX_COSMOS_CLAUDS_both3.jpg}) to classify quiescent and SFGs \citep[]{Arnouts2013}, following the redshift dependence proposed by \cite{Moutard2016b} (see their Fig. 8 and Sect. 5.1). All the galaxies not classified as quiescent are classified as star-forming.

 In Fig.~\ref{mass_nrk_cosmos.pdf} we compare the stellar mass and the $NrK$ vector  (defined in the next section) derived with our HSC-CLAUDS photometry and with the COSMOS2020 catalog, as a function of redshift and stellar mass.
 For this comparison, we restrict the sample to the SFGs with photo-$z$ within $|z_{\rm COSMOS} - z_{\rm CLAUDS}|/(1+\bar{z}) \le0.1$. This cut will always be used when comparing the physical parameters.
 Our stellar masses (top panels) appear in good agreement with COSMOS2020 estimates with a relative difference lower than 25\% (which means less than $\sim$0.1 dex) and no significant trend is observed with redshift and stellar mass. Since the K-band luminosity relies on some extrapolation of the SEDs at high redshift, in the bottom panels we compare our $NrK$ vectors with the COSMOS2020 estimates, which are better-constrained thanks to the use of the IRAC MIR photometry. The $NrK$ relative differences show no bias with redshift and stellar mass and typical variation of less than $\sim$25\% except in the highest redshift bins and lowest stellar mass bins where it increases up to 40\%.
%%%%%%%%%%%
\sfigs{1}{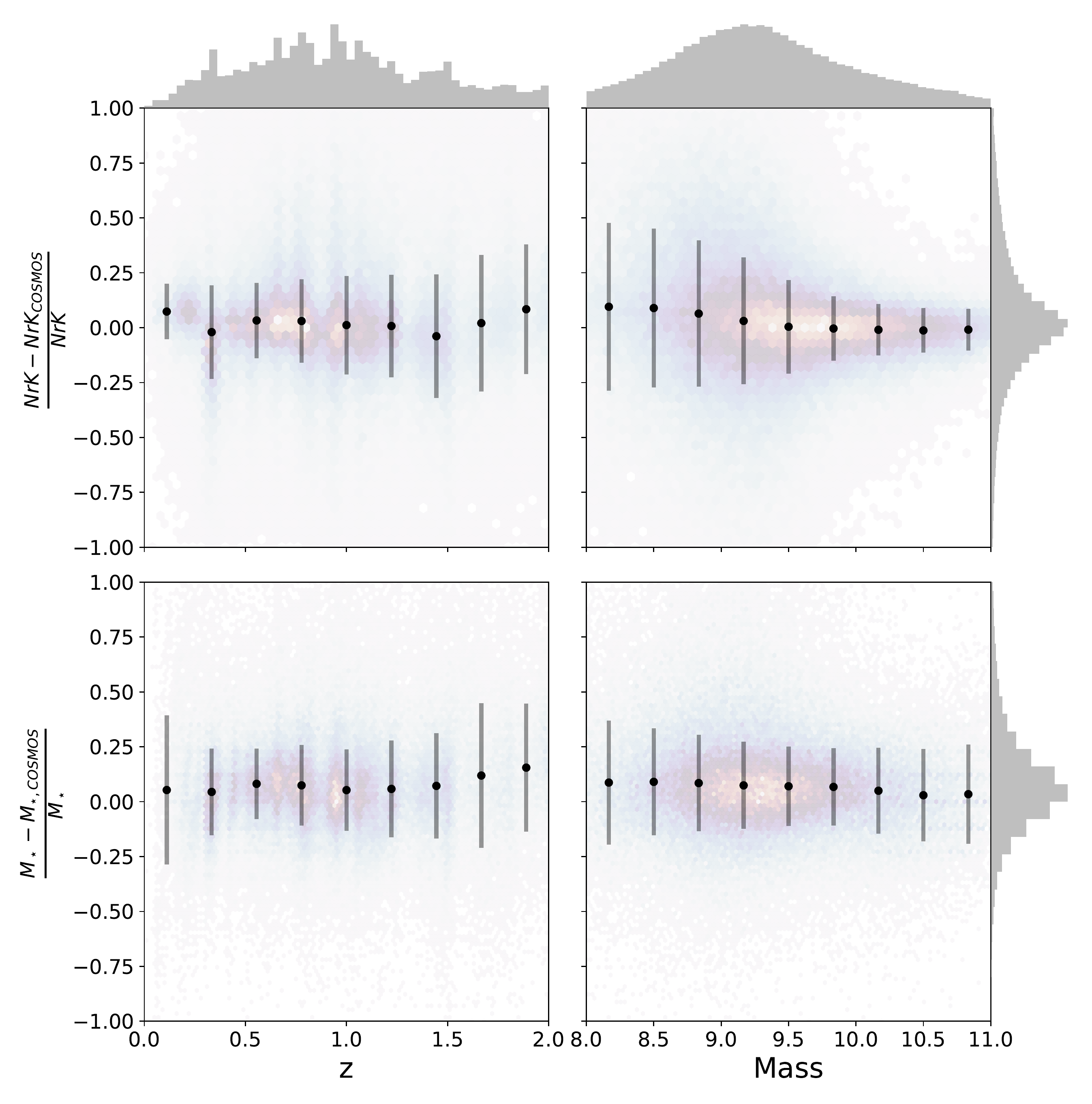}{Relative differences of the stellar masses and the $NrK$ vectors (see Sect. \ref{sec:calib}) between HSC-CLAUDS and COSMOS2020 as a function of redshift and stellar mass.
}{fig:para}
%%%%%%%%%%%%

The stellar mass completeness (smallest mass at which most of the objects would still be observable) of our SFG sample ($i_{\rm AB}<26.5$ and $K_{\rm AB}<24.5$) is empirically computed following the commonly used  method \citep{Ilbert2013, Weaver2021} developed by \citet{Pozzetti2010} based on the mass-to-light ratio $M/L$ dependence.
At each redshift, we define a minimum mass $M_{min}$ above which the stellar mass function of each subpopulation is complete. To do so, we rescale each galaxy's stellar mass computed via template-fitting to the stellar mass limit ($M_{lim}$) it would have at its redshift if its $K_{AB}$ apparent magnitude was equal to the limiting magnitude of the survey ($K_{\rm lim} = 24.5$): log$_{10}(M_{\rm lim})={\rm log_{10}}(M_{\rm med}) + 0.4(K-K_{\rm lim})$. At each redshift, to derive a representative limit, we select the $20\%$ faintest galaxies of our sample.
The stellar mass completeness can then be determined in a given redshift bin from the distribution of the rescaled masses of this faint subsample: the 95$^{th}$ percentile of the distribution defines a mass at which most of the objects would still be observable.

%%%%%%
\subsection{Far-infrared luminosities and far-infrared sample properties}\label{Far-InfraRed luminosities and FIR sample properties}
%%%%%%
The infrared luminosity ($L_{\rm IR}$) is defined as the luminosity integrated from 8 to 1000 $\mu m$:
\begin{equation}
L_{\rm IR}=\int_{\lambda=8\mu m}^{1000\mu m}L(\lambda)d\lambda
\label{LIR}
.\end{equation}
It is derived by using the code LePhare \citep{Arnouts2002, Ilbert2006} combined with the FIR SED templates of \citet{Dale2002}.
 In A13, the $L_{\rm IR}$ was derived by extrapolating the 24$\mu$m observed flux density with the FIR templates in such a way that the templates follow the locally observed dust temperature-luminosity relationship \citep{Chary2001, Goto2010}.
 By using the \textit{Herschel} observations, \citet{Elbaz2011} confirmed the tight correlation between the infrared luminosity extrapolated from 24$\mu$m photometry ($L_{\rm IR}^{24\mu m}$) and the infrared luminosity ($L_{\rm IR}$) measured with the PACS and SPIRE photometry. However, they noticed a bias at high luminosity ($L_{\rm IR}^{24\mu m}> 10^{12}~L_{\odot}$), where the $L_{\rm IR}^{24\mu m}$ systematically overestimates the $L_{\rm IR}$.
 In Fig.~\ref{Difference_Lirs.jpg} we compare the infrared luminosity measured with the six passbands (24$\mu$m to 500$\mu$m) with the one derived from the 24$\mu$m alone. All the galaxies in this sample have at least one \textit{Herschel} flux with $S/N>3$ and the scaling and SED shape are left free in the fitting procedure for the $L_{\rm IR}$ estimate. We confirm the excellent agreement for the SFGs with low infrared luminosity, with a small scatter ($\sigma\sim 0.07$ dex) in between the two estimates, while at high luminosity we confirm the overestimation of $L_{\rm IR}^{24\mu m}$ compared to $L_{\rm IR}$. We therefore included the \textit{Herschel} photometry in the analysis to improve the estimate at high infrared luminosity; it does not impact the $L_{\rm IR}$ estimate for the bulk of the 24 $\mu$m sample. \\

\sfigs{1}{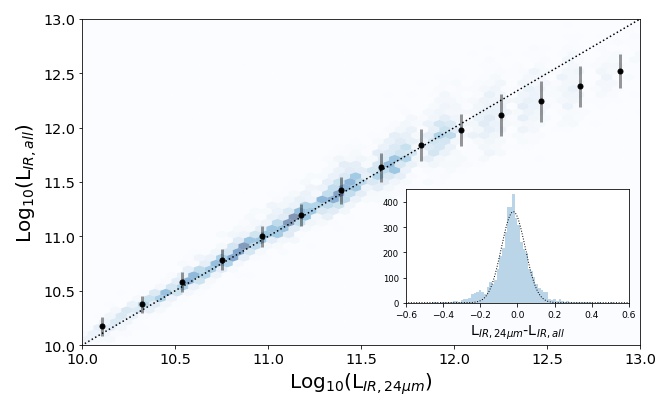}{Difference between the infrared luminosity, $L_{\rm IR}$,  estimated from the 24$\mu m$ photometry alone and by adding the \textit{Herschel} photometry 
 up to $z\sim 2$.}{fig:Lirs}

%%%%%%%%%%%%%%%%%%%%%%%
\sfigs{1}{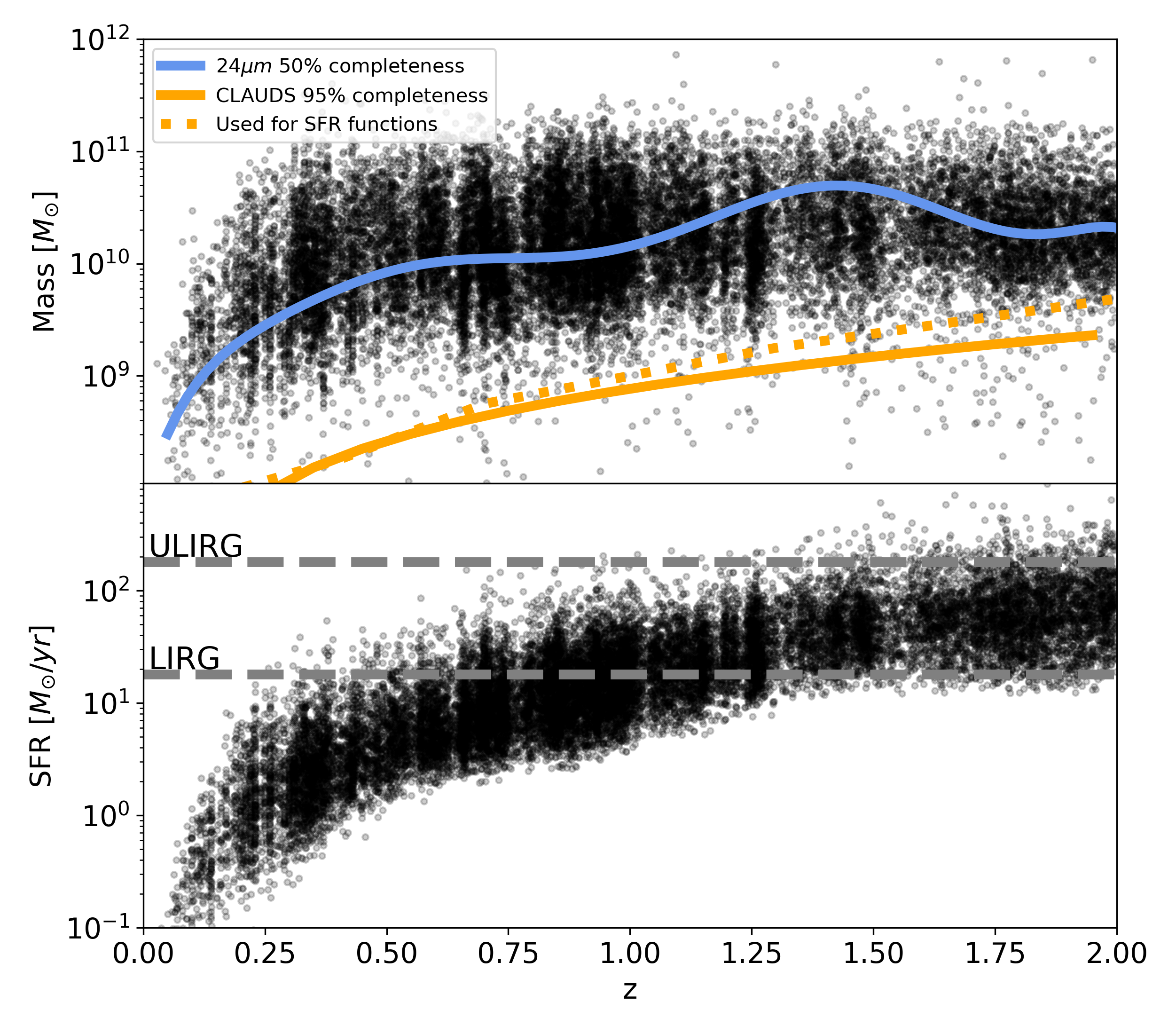}{Stellar mass distributions (top panel) and SFR distributions (bottom panel)  as a function of redshift for the 24$\mu$m sample. The solid blue line represents the 50\% mass completeness limit of the FIR sample (see text), while the orange line represents the stellar mass completeness of the HSC-CLAUDS K-selected sample ($K\le 24.5$; see text). The horizontal gray lines in the bottom panel indicate the SFR threshold that corresponds to the luminous (LIRG, $L_{\rm IR}\ge 10^{11}~L_{\odot}$) and ultra-luminous (ULIRG, $L_{\rm IR}\ge 10^{12}~L_{\odot}$) infrared galaxies.}{fig:FIRsample}
%%%%%%%%%%%%%%%%%%%%%%%
%
 The properties of the FIR population are shown in Fig. \ref{mass_sfr_complete2.png}.
 In the bottom panel, we show the SFR distribution as a function of photometric redshift. The SFR is defined as the sum of the UV and FIR contribution, as in A13: 
\begin{equation}
SFR(M_{\odot}/yr)=8.6\ 10^{-11} \times (L_{\rm IR} + 2.3 \times {\cal L}_{\rm NUV}),
\label{SFR_tot}
\end{equation}
where ${\cal L}_{\rm NUV}$ is the monochromatic NUV luminosity: ${\cal L}_{\rm NUV}/L_{\odot} = \nu L_{\nu}(2300\AA)$.
 As already shown by \citet[]{LeFloch2005}, below $z=0.5$, the population is composed of moderately SFGs with $SFR\le 10$ \SFR.
 The fraction of luminous infrared galaxies (LIRGs) gradually increases from $z=0.5$ to $z=1$ and dominates at $z\ge 1$. At all redshifts, the fraction of ultra-luminous infrared galaxies (ULIRGs) is negligible.
 In the top panel, we show the stellar mass distribution as a function of photometric redshift. The FIR sample is dominated by galaxies with \MS$\ge 5\ 10^{9}\ M_{\odot}$.  To characterize how representative the 24$\mu$m sample is with respect to the entire star-forming population at a given mass and redshift, we define a 50\% completeness stellar-mass limit with the ratio $\int_{M_{lim}}^{\infty} \Phi_{SF}^{FIR}(M_{\star},z)dM /  \int_{M_{lim}}^{\infty} \Phi_{SF}^{all}(M_{\star},z)dM_{\star}\sim$0.5;  where $\Phi_{SF}^{FIR}(M_{\star})$ and $\Phi_{SF}^{All}(M_{\star})$ are the $V_{max}$ weighted comoving volume densities of FIR ($f_{24\mu m}\ge 50\ \mu$Jy) and $K_s$-selected (K$\le$24.5) samples of SFGs, respectively.  Above this limit (shown as a solid blue line), we consider the physical properties of the FIR population to be representative of the whole star-forming sample. This corresponds to a stellar mass of M$_{\star}\sim 1-2\ 10^{10} M_{\odot}$ at $z \ge$0.5. The lower completeness near $z\sim$1.4 reflects the dip in between the polycyclic aromatic hydrocarbon features passing in the 24$\mu$m passband (Fig.~\ref{newfilters2.jpg}).

%%%%%%%%%%%%%%%%%%%%%%%%%%

\section{The infrared excess in the NUVrK diagram}\label{sec:calib}
For the majority of galaxies, the summation of the infrared and UV luminosities is a reliable indicator of the bolometric luminosity coming from young stars and therefore a good proxy for SFR \citep{Buat2002}.
 As the UV luminosity is of easy reach up to high redshift, measuring the IRX, defined as $IRX=L_{\rm IR}/{\cal L}_{\rm NUV}$, allows for assessing the obscured star-formation contribution and offers an interesting alternative to SED fitting derived SFR, which depends strongly on the adopted attenuation laws\footnote{This is not the case for SED fitting codes preserving the energy budget, which gives an SFR consistent with the addition of infrared and UV contribution, and also corrects for the contribution of the infrared emission due to dust heating by old stars and not directly connected to star formation}.

 In contrast, the IRX is weakly dependent on the age of the stellar population, dust geometry, and nature of the extinction law \citep{Witt2000}. We note that our definition of the IRX differs from the literature, which usually adopts the FUV luminosity.  Indeed, the NUV luminosity can be impacted by the dust attenuation bump at 2175Å, which falls in the blue side of the NUV passband. But we decided to use the NUV luminosity for practical reasons as it proves to be more reliable thanks to the deep GALEX NUV and CFHT u-band observations in the whole redshift range considered in this work. As shown by \citet[]{Hao2011}, this choice does not impact the reliability of the SFR estimates.

Based on a 24$\mu$m-selected sample, A13 have measured a tight correlation between the $NrK$ vector and the IRX values up to redshift $z=1.3$, with almost no dependence on the stellar mass.
 In this section we revisit this analysis and extend it to higher redshift, $z\sim2$, by using the FIR photometry (including MIPS/\textit{Spitzer} and PACS and SPIRE/\textit{Herschel}) and the latest COSMOS2020 catalog  \citep{Weaver2021}. We also extend to lower-mass populations, based on an FIR stacking technique.

\subsection{IRX calibration with detected FIR sources}\label{IRX_calib_subsect}

\bfigs{1}{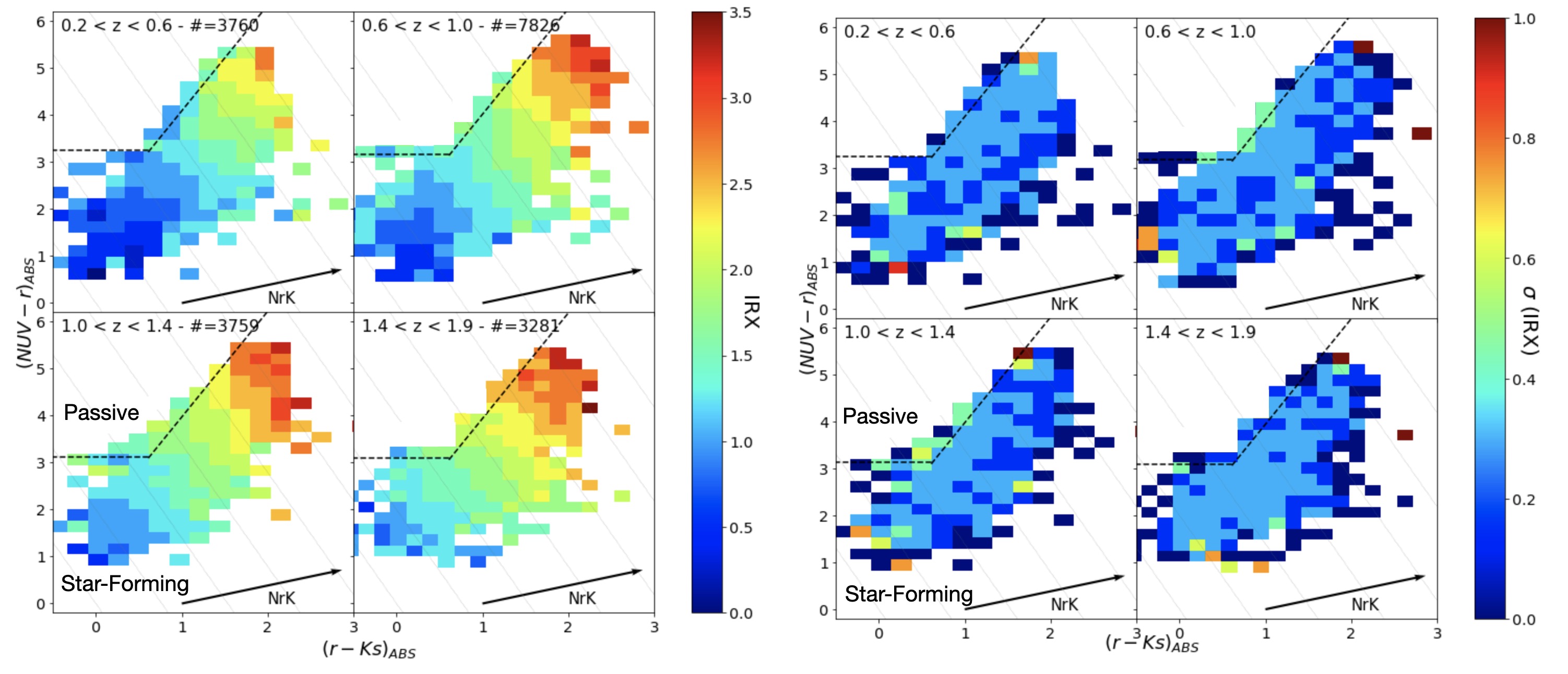}{Behavior of the IRX in the NUVrK color-color diagram. \textbf{Left:} Mean IRX ($\langle IRX \rangle$ color-coded in logarithmic scale) in four redshift bins. The thin parallel lines show the modeled evolution of the $\langle IRX \rangle$ stripes with the norm of the $NrK$ vector perpendicular to them.  In each panel, the dotted line indicates the region of passive galaxies \citep{Moutard2016b} that are not included in this analysis. \textbf{Right:} Same as the left panel but for the dispersion around the mean ($\sigma$(IRX)), color-coded in a logarithmic scale.}{}{}

Figure~\ref{IRX_COSMOS_CLAUDS_both3.jpg} shows the volume-weighted mean IRX ($\langle {\rm IRX} \rangle$) in the NUVrK diagram.
In all the redshift bins, we observe an increase of the $\langle {\rm IRX} \rangle$ by $1.5-3$ dex from the bottom left corner to the upper right one with a small scatter around the mean. We also observe constant IRX stripes, which allows us to describe the variation in IRX by a single vector perpendicular to those stripes:
\begin{equation}
NRK = sin(\phi) \times (NUV - r) + cos(\phi) \times (r - K)
\label{NRK}
.\end{equation}
Since the dispersion $\sigma[IRX(\phi)]$ reaches a minimum when the vector $NRK (\phi)$ is perpendicular to the stripes, we minimize $\sigma[IRX(\phi)]$ to find the best angle.
We obtain an angle of $\phi=21^\circ$, consistent with A13 ($\phi=18^{\circ}$), but we note that a difference up to 10 degrees does not impact the final calibration. The right side of the Fig.~\ref{IRX_COSMOS_CLAUDS_both3.jpg} shows that the scatter at all redshift stays below 0.2 dex over the whole color diagram.

With the definition of the $NrK$, we can now derive the relationship between $\langle {\rm IRX} \rangle$ and $NrK$.
In Fig.~\ref{IRX_vs_nrk_CLAUDS.jpg} we show the mean $\langle {\rm IRX} \rangle$ and the associated scatter for the FIR-detected sources as a function of $NrK$ and redshift.
In all panels, a tight linear correlation is observed with a small scatter ($\sigma \leq  0.3\,$dex) compared to the dynamical range covered by $\langle {\rm IRX} \rangle$ ($\sim$2 dex). The entire calibration does not depend on the stellar mass, and while the slope of the relation does not change with redshift, the normalization increases with redshift. At a fixed redshift, the populations in different stellar mass bins follow the same relation but occupy different regions as their $NrK$ distributions shift to higher values at higher stellar mass. 

We first adopted the same parametrization for the $\langle {\rm IRX} \rangle$ versus $NrK$ relation as A13, with two separated quantities $\langle {\rm IRX} \rangle = f(z) + \alpha \cdot NrK$, where f(z) is a third-order polynomial function describing the redshift evolution, and $\alpha$ is a constant for the $NRK$ dependence. We performed a linear least-square fit to derive the four free parameters and their standard deviation uncertainty.
For the redshift evolution $f(z)= a_0 + a_1\cdot z + a_2\cdot z^2 + a3\cdot z^3$ we derived  $a_0 = -0.73 \pm 0.02$; $a_1 =2.02 \pm 0.06$; $a_2= -1.33 \pm 0.06$ ; $a_3 = 0.33 \pm 0.02$ ; for the $NrK$ term, $\alpha = 0.63 \pm 0.003$.

The result of this calibration is added as dashed black lines in Fig. \ref{IRX_vs_nrk_CLAUDS.jpg}, and the predicted $L_{\mathrm{IR}}$ ($L_{\mathrm{IR}}^{NrK}$) that comes from it are compared to the reference $L_{\mathrm{IR}}$ in Fig. \ref{perf_lir.pdf}. The predicted and reference $L_{\mathrm{IR}}$ are in good agreement with almost no bias with redshift and a global dispersion $\sigma\sim0.24$ dex. While this scatter is slightly larger than in the analysis by A13, it covers a larger redshift range by extending the analysis up to $z=2$. Finally, the comparison as a function of infrared luminosity shows almost no bias except in the extreme regimes (for ULIRGs and low infrared luminosity galaxies), where our calibration slightly underestimates or overestimates the luminosity by less than a factor of 2.

\bfig{1}{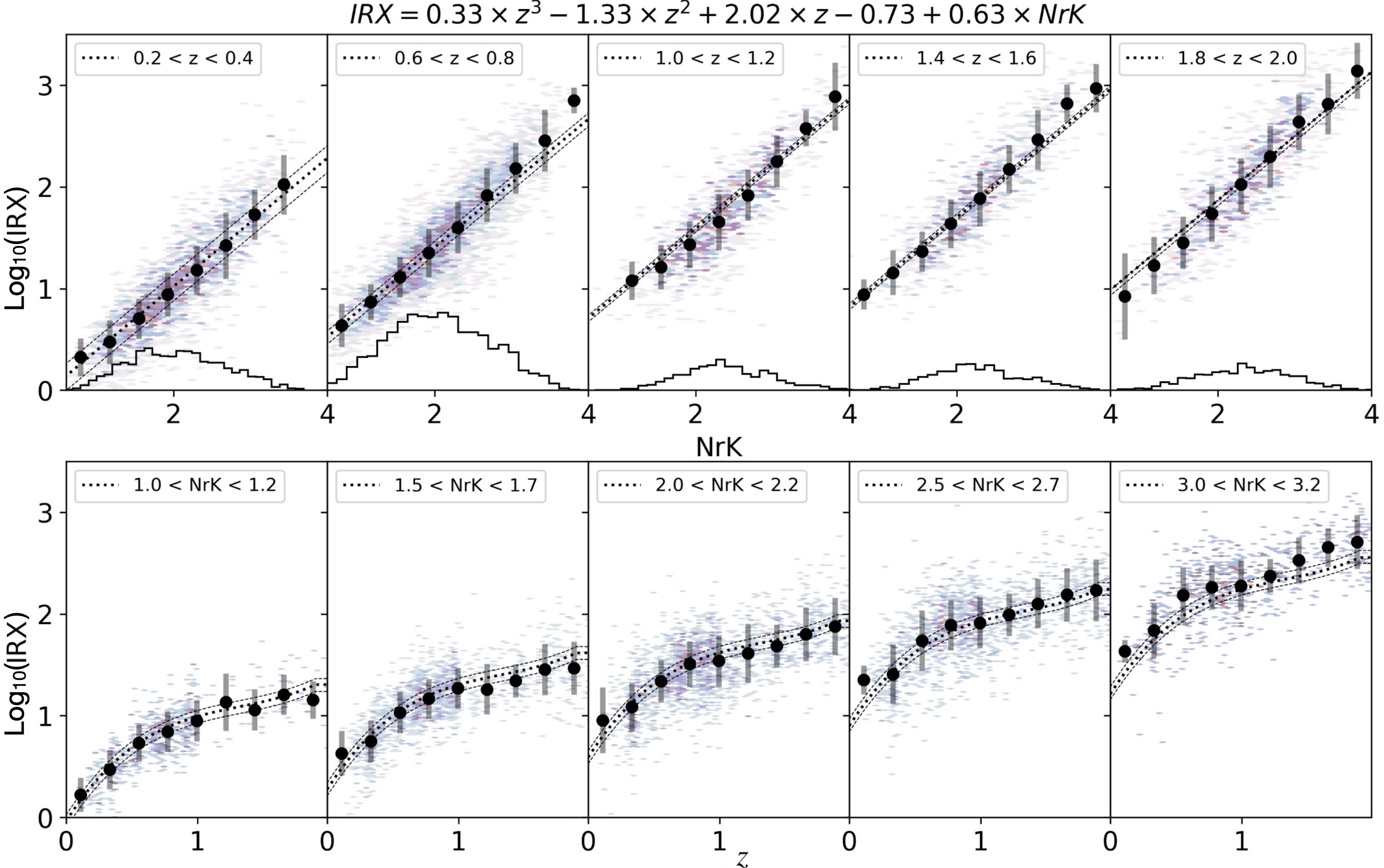}{Evolution of the IRX with $NrK$ (upper plots) and redshift (bottom plots) in different redshift and $NrK$ bins. The back dots represent the IRX computed on the 24$\mu m$-selected sample, and the dotted line corresponds to the minimum least square fitting at the center of each bin. The back lines around each dotted line represent the calibration for the edge of each bin.}

\sfig{1}{perf_lir.pdf}{Comparison of the infrared luminosity, $L_{\rm IR}$, estimated from the COSMOS2020 FIR sample with that estimated with the IRX-NrK calibration ($L_{\rm IR}^{NrK}$) as a function of redshift and infrared luminosity. }

\subsection{IRX calibration based on FIR stacking}

Surveys conducted in the thermal infrared regime are generally not sensitive enough for detecting low-mass sources individually. To explore the validity of the $IRX -NrK$ relationship in the low-to-intermediate stellar-mass regime we resort to the FIR stacking technique \citep{Bethermin2010}.
However, while the stacking technique is often applied to stellar mass-selected samples \citep{Whitaker2014}, at a fixed stellar mass the $\langle {\rm IRX} \rangle$ spans a large dynamical range, which can be better constrained by dividing the sample as a function of the unobscured UV luminosity \citep{Heinis2014} where the most extinguished galaxies are expected to be the less UV luminous in a fixed stellar mass bin, or alternatively (and hereafter), as a function of $NrK$ values as it best follows the evolution of $\langle {\rm IRX} \rangle$ in order to minimize the dispersion around the mean.    We thus split the sample into $NrK$, stellar mass, and redshift bins and co-add the images from the \textit{Spitzer}-24 $\mu$m and SPIRE-\textit{Herschel} FIR channels for all the selected sources to get their average FIR emissions. Some stacks are shown in Fig. \ref{stacking_both.jpg} and the details about the procedure are given in Appendix~\ref{App:stacking}.

The results of the stacking procedure are shown in Fig. \ref{IRX_stacking.pdf} with the mean infrared luminosity derived from the mean 24 $\mu$m flux (blue diamonds) or by combining the 24, 250, 350, and 500 $\mu$m fluxes (red circles).
Despite all the potential, hard-to-control biases that such a technique can introduce, as discussed in the appendix,
we observe an overall good agreement between the stacking results and the individual calibration (black circles). We obtain the same trend with the stacking technique, namely the tight correlation between $\langle {\rm IRX} \rangle$ and $NrK$, with a similar slope and small scatter around the mean values. However, noticeable changes in the normalization in the different bins are observed.
In the regime where the FIR-detected sources are the dominant population (panels below the thick black line), the stacking is in agreement within the respective uncertainties. For the panels above the black line, the individual detections lead to a systematically higher $\langle {\rm IRX} \rangle$ at a fixed $NrK$. By nature, the FIR detections biased the samples toward the most obscured ones and did not reflect the behavior of the whole population in a given stellar mass and $NrK$ bin.

We also observe that at high redshift and stellar mass bins (bottom-right panels), the $L_{\mathrm{IR}}$ measured from the 24$\mu$m mean flux alone tends to be slightly higher than the one estimated with 24+SPIRE mean fluxes. This is consistent with the expected overestimation of the $L_{\mathrm{IR}}$ at high luminosity discussed in Sect. \ref{Far-InfraRed luminosities and FIR sample properties}.

To take the stacking results into account in the following analyses, we considered two cases. In the first case, we disregarded the stacking and kept the original $IRX-NrK$ calibration as derived with the FIR-detected sources. In the second case, we used a mixed calibration. For galaxies below the FIR stellar mass completeness, we applied a linear correction with stellar mass and redshift (shown as red lines in Fig. \ref{IRX_stacking.pdf}) and parameterized as

\begin{equation}
IRX_{\mathrm{stack}}= IRX_{NrK} - (0.34 z  - 0.10 M_{\star} + 0.69),  \;\;\mathrm{if} \;\;M_{\star}<M_{\mathrm{lim}}(z)
\label{eq:offset}
.\end{equation}

In each (\MS, $z$) panel of Fig. \ref{IRX_stacking.pdf}, we report the corresponding shifts between the original and stacking calibrations. While the former will lead to an overestimate of the predicted SFR, the latter will lead to lower SFR values but maybe yet more appropriate at low mass or high redshift for the whole population.

\bfig{1}{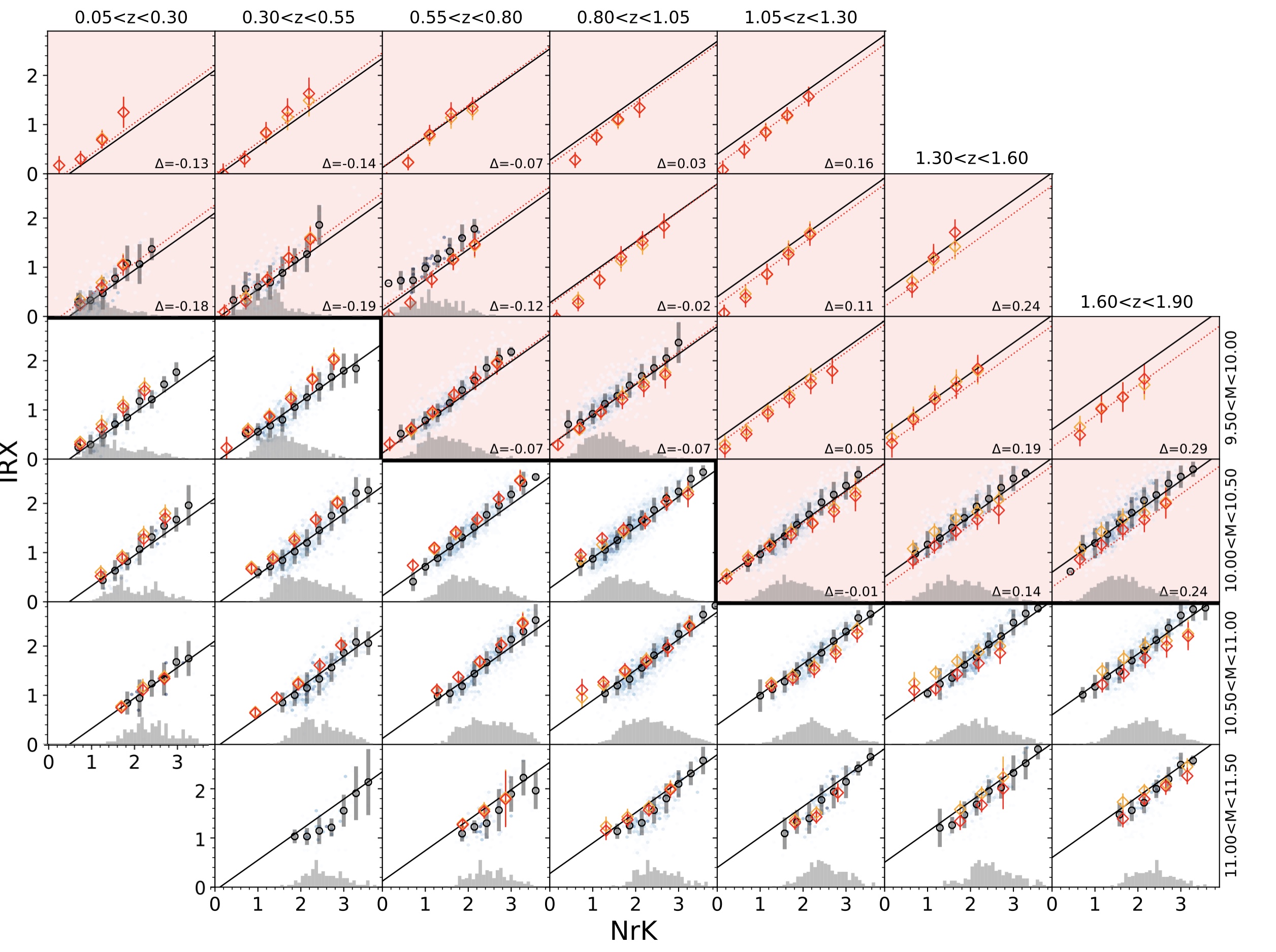}{
Mean IRX as a function of $NrK$ for the detected FIR sources (black circles and gray shaded histograms) and for the stacked populations (open blue diamonds  with 24$\mu$m alone and red circles with 24$\mu$m+SPIRE images) in different redshift and mass bins. The IRX-$NrK$ calibration based on the detected sources is shown as solid black  lines, while the dashed red lines show the calibration based on the stacking results, which are considered only in the panels where the FIR-detected sources are not representative of the whole population (limits are established with the blue line of Fig. \ref{mass_sfr_complete2.png}, and corresponding panels are colored in pale red) and delineated by the thick solid black line. The shift between the two calibrations is reported in each panel. }

\subsection{Dust attenuation evolution with  stellar mass and redshift}
\label{sec:dust_attenuation} 

 The IRX, or equivalently the dust attenuation, has been found to be a strong function of stellar mass at low \citep[]{Garn2010} and high  \citep[]{Pannella2009, Reddy2010, Whitaker2012, Heinis2014, Shapley2022} redshift, with the most massive galaxies being more highly obscured. 
  Since our $NrK$ calibration does not rely explicitly on the stellar mass, in Fig. \ref{whitaker4.pdf}, we show the behavior of the mean $IRX$ as a function of stellar mass in four different redshift bins. The mean IRX values are measured by using the FIR-detected sources (solid dots), the stacked sample (empty diamonds) both from the COSMOS2020 sources and by using the $NrK$ prediction with or without the stacking correction (solid and dashed lines, respectively) for the HSC-CLAUDS sample. In each redshift panel, the stellar mass distributions are  shown in the bottom part. The bottom right panel combines  the  $\langle {\rm IRX} \rangle$ versus stellar mass relationship for the four redshift bins.
\bfig{1}{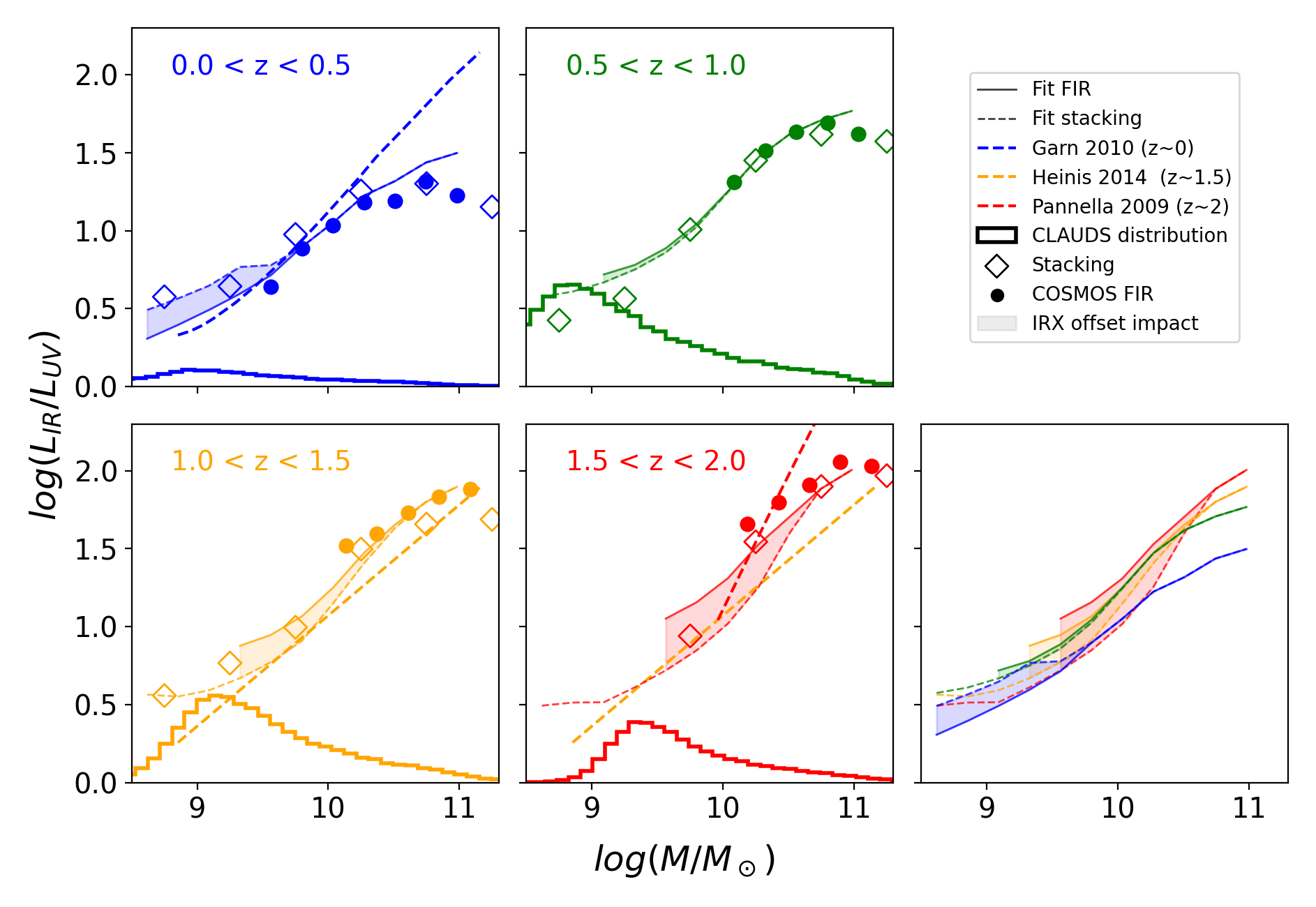}{Evolution of the mean IRX with stellar mass in four redshift bins based on the measurements from the COSMOS2020 FIR-detected sources (empty circles), FIR stacked  sources (filled diamonds), and the HSC-CLAUDS sample with or without the IRX stacking correction (solid and dashed lines, respectively). 
In the bottom right panel, we show the behaviors for the COSMOS2020 FIR stacking in the four redshift bins.}

The mean IRX derived from the different approaches is consistent with each other. Despite the absence of calibration with stellar mass, the $NrK$-based estimates  reproduce the same behavior as a function of stellar mass and redshift. We note that in Fig. \ref{whitaker4.pdf} we do not show results from the literature as some assumptions regarding the dust attenuation law \citep[which vary across the $NUVrK$ plane and the specific SFR of galaxies,][]{Arnouts2013}  is required to convert $A(H_{\alpha})$ \citep{Garn2010,Shapley2022} or $L_{\rm IR}/L_{\rm FUV}$ \citep{Heinis2014, Whitaker2014} into our $L_{\rm IR}/L_{\rm NUV}$ measurements, making the correction uncertain. \\

Two main features are observed. First, at low stellar mass ($M_{\star}\le 10^{10.3}~M_{\odot}$), almost no evolution of the IRX versus stellar mass relationship is observed between $z=0$ and $z=2$ \citep[in agreement with ][]{Whitaker2014, Shapley2022}. 
Considering the well-established evolution of the atomic/molecular gas, dust content, and metallicity at fixed stellar mass with redshift, the lack of evolution of the attenuation with redshift is rather puzzling \citep{Bogdanoska2020} and could underlie some changes in the dust properties such as grain size distribution and composition \citep[e.g.,][]{Shapley2022}.  \\

Second, at high stellar mass ($M_{\star}\ge 10^{10.3} $ \MM), a $\langle IRX \rangle$ plateau is observed despite the large scatter in the measurements. It appears at different stellar mass with redshift, near $M_{\star}\sim 10^{10.7}$ \MM{} at $z\sim 2$ and $M_{\star}\sim10^{10.3}$ \MM{} at $z=0.25$. This result is in qualitative agreement with the trend reported by \citet[][their Fig. 5]{Whitaker2014} based on the stacking of \textit{Spitzer}-24 $\mu m$ data.

The origin of the plateau and its redshift dependence is also unclear. It may be related to a change in the balance between the formation and destruction of dust grains in massive galaxies. 
SNe and AGB stars are efficient sources to produce and inject dust grains in the ISM \citep[]{Gehrz1989} while accretion growth in dense molecular clouds \citep[]{Draine2009} contributes to increasing the dust mass in galaxies. On the other hand, hot gas ISM and SN shock waves can destroy dust grains (or modify dust grain size) by shattering or sputtering processes \citep[e.g.,][]{Inoue2011}.

Alternatively, high redshift galaxies are more gas-rich, with more turbulent disks than their low-$z$ counterparts \citep{Kassin2010, Schreiber2020}, and star formation is potentially more concentrated in long-live giant clumps as suggested by hydrodynamical simulations \citep{Fensch2021}.
At a fixed stellar mass and lower redshift, galaxies have a lower SFR and are less clumpy. This contributes to reducing the dust production by SNe and growth by accretion while keeping efficient dust destruction in the hot gas ISM by sputtering and SN shock waves, as well as the presence of X-ray feedback from AGNs \citep[]{Choi2012, Dave2019}, which may lead to a reduced dust mass and thus produces a lower mean dust attenuation. While it is beyond the scope of this paper to propose a coherent model to explain the behavior of the $IRX$ versus stellar mass relation, it will be of interest to know if the origin of the plateau is related to the survival/destruction of dust  and/or to the disk properties with cosmic time in massive galaxies.
% \end{itemize}
 
%
\subsection{Comparison of individual SFR estimates with literature values} 
\label{section4.4}
The two previous $\langle {\rm IRX} \rangle-NrK$ calibrations have been applied to the HSC-CLAUDS sample. The total SFR is derived by computing respectively $L_{\mathrm{IR}}^{NrK}$ ($L_{\mathrm{IR}}^{NrK}=\langle {\rm IRX} \rangle \cdot L_{\mathrm{NUV}}$) and $L_{\mathrm{IR}}^{NrK, \mathrm{stack}}$ and by applying Eq. \ref{SFR_tot}.

In Fig.~\ref{comparaison_tot_irx4.jpg} we compare the SFRs for the HSC-CLAUDS samples based on the $NrK$ methods with the FIR-based SFR  measured with the COSMO2020 data set at high masses ($M_{\star}\ge 10^{10}$ \MM), while at low mass we compare with the SFRs measured in the CANDELS  fields \citep[]{Barro2019}, for the two fields included in the HSC-CLAUDS regions, to extend our comparison to low-mass galaxies. For the latter, we use their SFR-ladder, estimated from a combination of three tracers. They use FIR and MIR photometry when available to estimate the $L_{\mathrm{IR}}$ and the $\beta$-slope when infrared data are not available.
For extreme galaxies detected by both MIPS and \textit{Herschel}, total SFRs are calculated with dust emission models fitting the infrared data points and adding the unobscured star formation. For sources having only 24 um flux measurements (undetected at longer wavelength), the obscured SFR is computed following \citet[][]{Wuyts2008}. %, Sect 8.2.2.
For the remaining galaxies undetected by MIR and FIR surveys, the SFRs are estimated using the IRX-$\beta_{UV}$ relations (85\% of the $\sim10$K matched sources).
In all the mass bins the different SFR estimates are consistent with each other within their relative uncertainties.
At low mass, $M_{\star}\le 10^{10}$ \MM , our results are in overall good agreement with CANDELS SFRs, with a small scatter ($\sigma<0.2$) and bias. Adopting the calibration based on the stacking results slightly reduces the bias, suggesting that this calibration may be more appropriate in the low-mass regime. 
%
%%%%%%%%%%
At high masses, with the COSMOS2020 FIR sample, we still have an overall good agreement. However, we observe a wavy shape of the mean difference with redshift, which contributes to enlarging the scatter reported in each panel, up to $0.3$ dex. We checked that this was not due to the estimates of the $NrK$ vector between the HSC-CLAUDS and COSMOS2020 data set. It is rather due to the COSMOS2020 calibration, where our parametric form with redshift does not capture this residual variation at all masses\footnote{Note that the comparison of our SFR-NrK with the SFR from CANDELS at low stellar mass is essentially based on SFRs derived with the beta slope in UV and not FIR measurement}, which does not exceed $\pm$0.1dex. Since it affects all masses it does not modify the SFR functions but we account for this redshift residual in the cosmic SFRD measurements (Fig~\ref{SFR_density_sfr_new6_sfr.pdf}) by adding them in quadrature in the error bars. In the highest stellar mass bin, \MS$>10^{11}$ \MM{}, the global shift is due to the fitting procedure, which is slightly above the measurements in Fig. \ref{IRX_stacking.pdf}. We checked that changing the normalization in the highest mass range does not impact the bright end of the SFR functions and then the results discussed in Sect. \ref{functions}.          
 
%%%%%%%

\bfig{1}{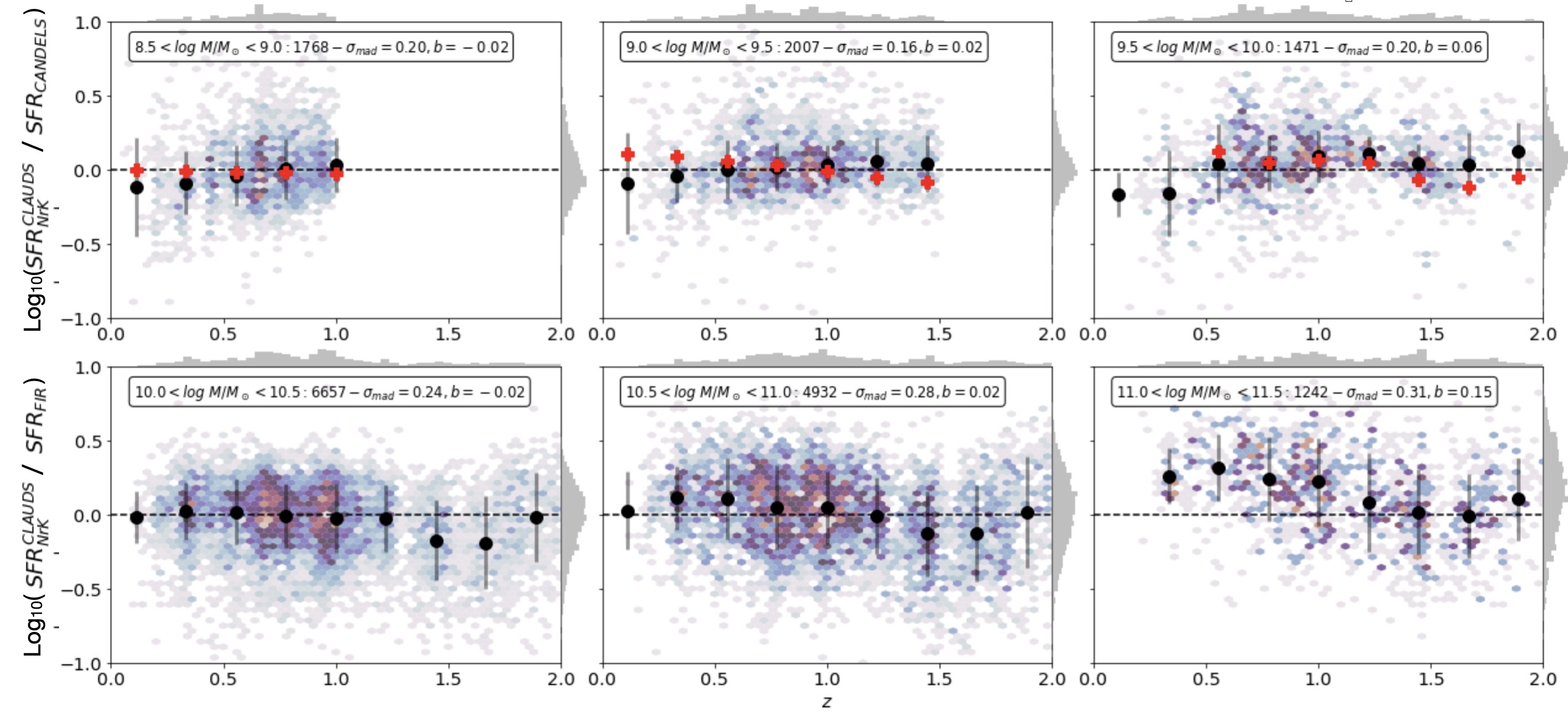}{Comparison of the $NrK$-based SFR of HSC-CLAUDS with  that of CANDELS based on the SFR-ladder estimates at low masses (top panels) and with COSMOS2020 based on the FIR SFR estimates at high masses (bottom panels). The $NrK$-based SFRs are derived from the calibration without (black circles) or with (red crosses) corrections based on the staking results.}

 In Fig.~\ref{main_sequence.png} we show the behavior of SFR$-$stellar mass relation as a function of redshift. This correlation is known to be tight and to evolve with redshift \citep{Elbaz2007, Noeske2008, Speagle2014}. The mean SFRs per stellar mass bins are estimated for the two $\langle {\rm IRX} \rangle-NrK$ calibrations discussed above. We compare our estimates with the local measurements from GALEX-SDSS  survey \citep{Salim2007}, the \textit{Herschel} Reference Survey \citep{Ciesla2016},  deep \textit{Herschel} observations \citep{Rodighiero2010, Magnelli2014, Heinis2014, Schreiber2015} and radio 3 GHz stacking \citep{Leslie2020}.
Our local, z$\sim$0, estimate is in excellent agreement with GALEX \citep{Salim2007} and HRS  \cite{Ciesla2016}  surveys.

At higher redshift, the normalization of the MS agrees with previous studies up to $z\sim 2$ and over the whole stellar mass range. As previously noticed \citep[]{Salim2007, Ilbert2015,Magnelli2014,Leslie2020, Schreiber2015}, a single slope does not provide the best fit as a flattening is observed around $M_{\star}\sim 10^{10}M_{\odot}$.  Indeed, the double slope fit of the MS proposed by \citet{Magnelli2014} or \citet{Schreiber2015} provides a good fit for our SFR-\MS measurements.

\bfig{1}{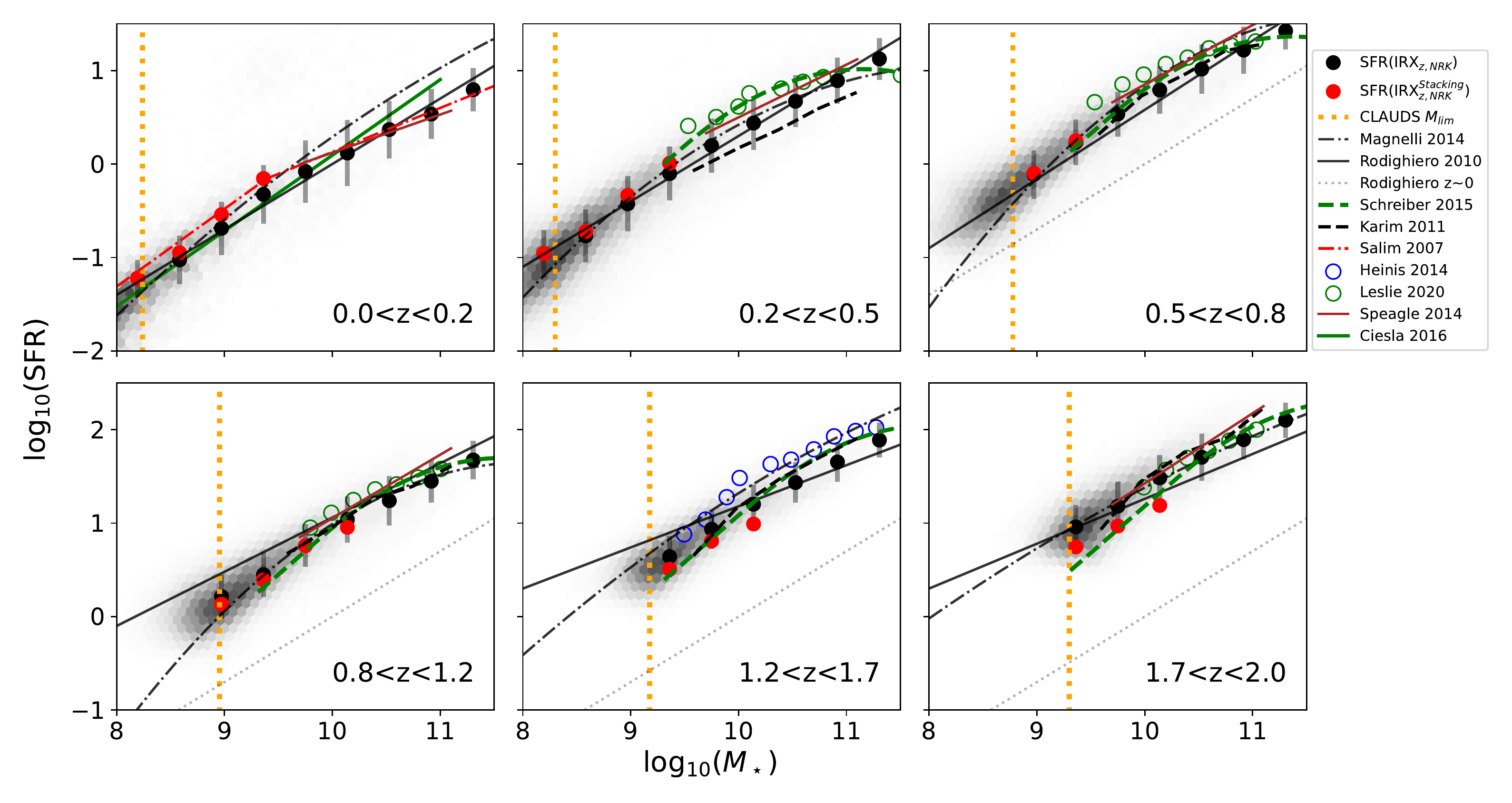}{Mean $NrK$-based SFR versus stellar mass relation as a function of  redshift. The two SFRs derived from the $IRX-NrK$ relations with (red dots) and without (black dots) the stacking results are shown. The completeness limits in stellar mass are shown as vertical orange lines. Comparison with the literature is also shown, with the symbols specified in the inset. 
The MS at $z\sim 0$ from \citet{Rodighiero2010} is also reported in each panel.}
%
%%%%%%%%%%%%%%%%%%%%%%%%%%%
\sfig{1}{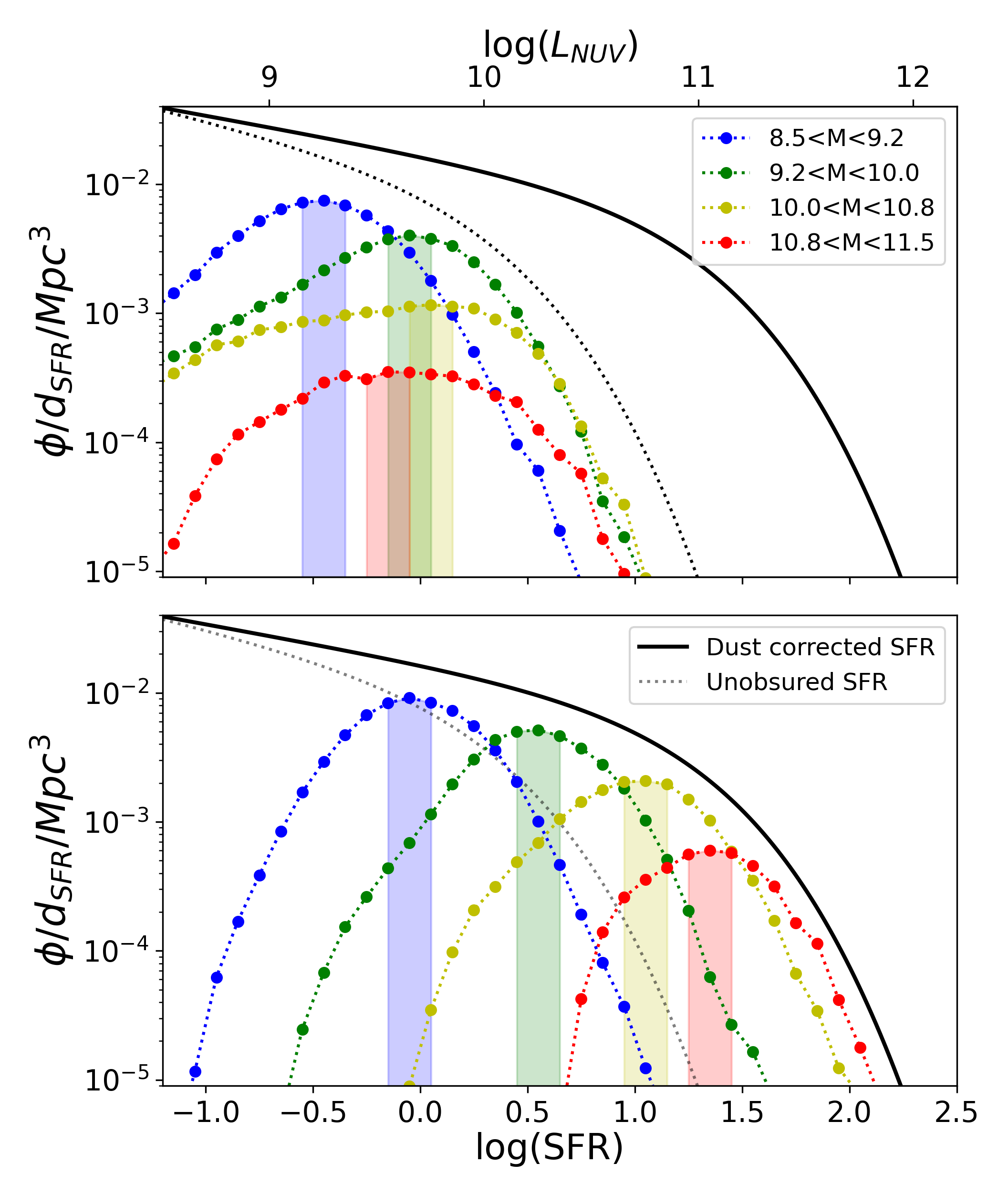}{Impact of dust correction on the shape of the SFR functions. The unobscured (upper panel) and dust corrected (lower panel) SFR functions are shown for different stellar mass bins (in color) and the whole sample (solid and dotted black lines) in the range $0.6\le z\le 1.0$. One noticeable feature of the dust correction is the ability to dissociate the SFR functions for each stellar mass bin with a larger correction at high masses and to modify the SFR functions per stellar mass bin from an asymmetric to a nearly Gaussian distribution.}
\section{Total SFR functions}
\label{functions}
In this section we present the construction of the total SFR functions and compare their evolutions to observations and simulations. 
We use the results from the previous section to correct for dust attenuation.
\subsection{From the unobscured to total SFR functions}

Figure~\ref{sfr_functions_obscured_masses.png} illustrates the impact of our dust correction on the SFR function. In the top panel, we show the unobscured SFR function (based on the observed NUV luminosity function) for different stellar mass bins (colored lines) and for the whole sample (black solid line) in the redshift range $0.6\le z\le 1$. One noticeable feature is that the unobscured SFR distributions for the three high-mass bins saturate in SFR around a similar value ($L_{\rm NUV}\sim 10^{9.7} L_\odot$). This saturation effect was already reported by \citet{Martin2005a} in their analysis of the bi-variate FIR and UV luminosity function in the local Universe. They showed that at low luminosity, the FIR and FUV luminosities track each other. At high luminosity, a saturation effect happens around $L_{\mathrm{UV}}\sim 10^{10} ~L_{\odot}$, while the $L_{\mathrm{FIR}}$ continues to increase by a factor of $\sim$100.
The wide range of dust attenuation at a fixed UV luminosity prevents the derivation of a statistically meaningful dust correction that could be used to translate the unobscured SFR functions into dust-free SFR functions.

By splitting it into stellar mass bins, \citet{Heinis2014} showed that in addition to the trend between dust attenuation (or IRX) and stellar mass, a dependence with $L_{\mathrm{UV}}$ is observed, so that fainter $L_{\mathrm{FUV}}$ have higher $\langle {\rm IRX} \rangle$ than brighter ones at fixed stellar mass, providing a better perspective to derive the SFR function by estimating the $\langle {\rm IRX} \rangle$ in bins of $L_{\mathrm{UV}}$ and stellar mass \citep[see also ][]{Bourne2017}. 
The $IRX-NrK$ calibration incorporates those dependences at once. The dust-free SFR function is shown in Fig. \ref{sfr_functions_obscured_masses.png} bottom panel. While the main effect is a clear separation with stellar mass, the residual correction at fixed stellar mass brings the asymmetric shapes into nearly Gaussian SFR functions per stellar mass bin at the origin of the MS and its moderate scatter. Adopting a dust correction based only on $\langle {\rm IRX} \rangle-M_{\star}$ relationship would translate the unobscured SFR per stellar mass bins but would fail to produce the right shape of the SFR function.
Finally, the dust-free SFR function leads to a flatter faint-end slope resulting from a larger correction for the higher stellar mass bin and a more concentrated SFR distribution in a fixed stellar mass bin.
\subsection{Measurement of the SFR functions}
After applying the above dust correction to each individual galaxy, we can derive the SFR functions in different redshift bins up to $z=2$. We restrict our sample to SFGs with $i<26.5$, $K<24.5$.

The SFR functions are then measured using a $V_{\mathrm{max}}$ estimator \citep{Felten1976, Ilbert2004} combining the $i$ and  $K_s$ band selections. Each galaxy is then weighted by $1/V_{\mathrm{max}}(z=\mathrm{min}(z_i,z_k))$,  the maximum comoving volume within which the galaxy could have been observed in both filters given the limiting magnitudes of our sample and its best-fit template.

To estimate the SFR function uncertainties, we take into account the contribution from the Poissonian errors ($\sigma_{\mathrm{poisson}}$) and the cosmic variance due to large-scale density fluctuations ($\sigma_{\mathrm{cv}}$):

  \begin{align}
  \sigma_{\mathrm{tot}}^2 &=  \sigma_{\mathrm{poisson}}^2+  \sigma_{\mathrm{cv}}^2\\
  &=\langle N \rangle +  \langle N \rangle^2 \times \frac{1}{V^2}\int_V dV_1dV_2 \xi_{\mathrm{DM}} b^2(m,z)\\
  &= \langle N \rangle + \langle N \rangle^2\times b^2(m,\bar z)\times\sigma^2_{\mathrm{DM}}(\bar z).
\end{align}

The cosmic variance term, $\sigma_{\mathrm{cv}}$, is estimated by using the \citet{Moster2011} cookbook.  The DM cosmic variance,  $\sigma_{\mathrm{DM}}(\bar z)$, is given by their Eq. 10 and depends on the redshift and the considered surface. To scale this term to the surface of our survey, we linearly extrapolated their estimate predicted for 2 deg$^2$ to  5.5 deg$^2$. The galaxy bias, $b(m,\bar z)$, is parameterized in their Eq. 13 for different stellar mass bins that we converted into SFR bins assuming the MS relationship.

Figure~\ref{sfr_functions4.pdf} shows the SFR functions in ten redshift bins between $0.05\le z\le 2.05$ and derived with the $IRX-NrK$ calibration  without (black dots) and with (red dots) the stacking correction. \\
The orange shaded area shows the Vmax weighted SFR functions derived for the COSMOS2020 FIR-selected catalog ($f_{24\mu m}\ge 50\mu$Jy), with the total SFR derived according to Eq. \ref{SFR_tot}.
As can be seen, the high-SFR regimes of the HSC-CLAUDS SFR functions are in excellent  agreement with the FIR-selected SFR functions. The $IRX-NrK$ calibration does not over/under-predict the density of galaxies with high SFRs, validating our dust correction procedure. On the other hand, since our sample is optical/mass-selected, it allows us to probe the slope of the SFRFs in the low-SFR regime beyond what is currently reached by FIR observations. 
\subsection{Parametric fits of the SFR functions}
\label{Fitting}
% \newpage
 We fit the SFR functions with a three-parameter Schechter function \citep{Schechter1976}:
\begin{equation}\label{schechter}
\phi(SFR)dSFR = \phi^\star_{\rm SFR}    e^{-\frac{SFR}{SFR^\star}}\left(  \frac{SFR}{SFR^\star} \right)^\alpha \frac{dSFR}{SFR^\star}
,\end{equation}
where $\alpha$ is the faint-end slope of the power-law regime and $\phi^\star_{SFR}$ and $SFR^\star$ the characteristic density and SFR separating the power-law and exponential regimes.

Before performing the fit, we account for the Eddington bias \citep{Eddington1913}, due to the uncertainty in the SFR estimate and the exponential cutoff of the SFR function. As a result, more galaxies are shifted toward high SFRs than the reverse, producing a shallower decline in the high-SFR regime than the intrinsic one. To correct for it we adopt the same procedure as 
\citet[]{Ilbert2013}. We first estimate the uncertainty according to Fig \ref{comparaison_tot_irx4.jpg}. We adopt an uncertainty of $\sigma=0.15$  at $z\le 1$ and $\sigma=0.2$ at $z\ge 1$, assuming that the two SFR estimators used in the comparison have similar errors (i.e., by dividing the observed error by $\sqrt 2$). We then convolve the Schechter function (Eq.~\ref{schechter}) by the SFR uncertainty:

\begin{equation}
\phi(SFR)^{\mathrm{conv}} = \phi(SFR)  \star G(\sigma=0.15,0.2)
.\end{equation}
A least square fit procedure is performed between $\phi(SFR)^{conv}$ and the $V_{\mathrm{max}}$ weighted SFR functions up to the completeness limit derived by converting our stellar mass limit into an SFR limit by assuming the MS $SFR-M_{\star}$ relations. The SFR limits are illustrated by a change in symbol size in the different panels. 
 
To estimate the slope of the SFRFs, we fit the lowest redshift SFRF where we get the best constraint. We derive $\alpha=-1.3\pm 0.1$. Since there is no evidence for the evolution of $\alpha$ up to redshift $z=2$ in our data within the fit uncertainties (which is also consistent with \cite{Ilbert2015, Mancuso2015}), we simply assume a fixed slope at all redshift. The slope uncertainty $\sigma({\alpha})=0.1$ is then propagated into the cosmic SFRD measurements (next section).

\begin{figure*}[b]
% \begin{strip}
  \thisfloatpagestyle{empty}
  % \vspace*{-2\baselineskip}
  \makebox[\textwidth][c]{\includegraphics[width=1\textwidth]{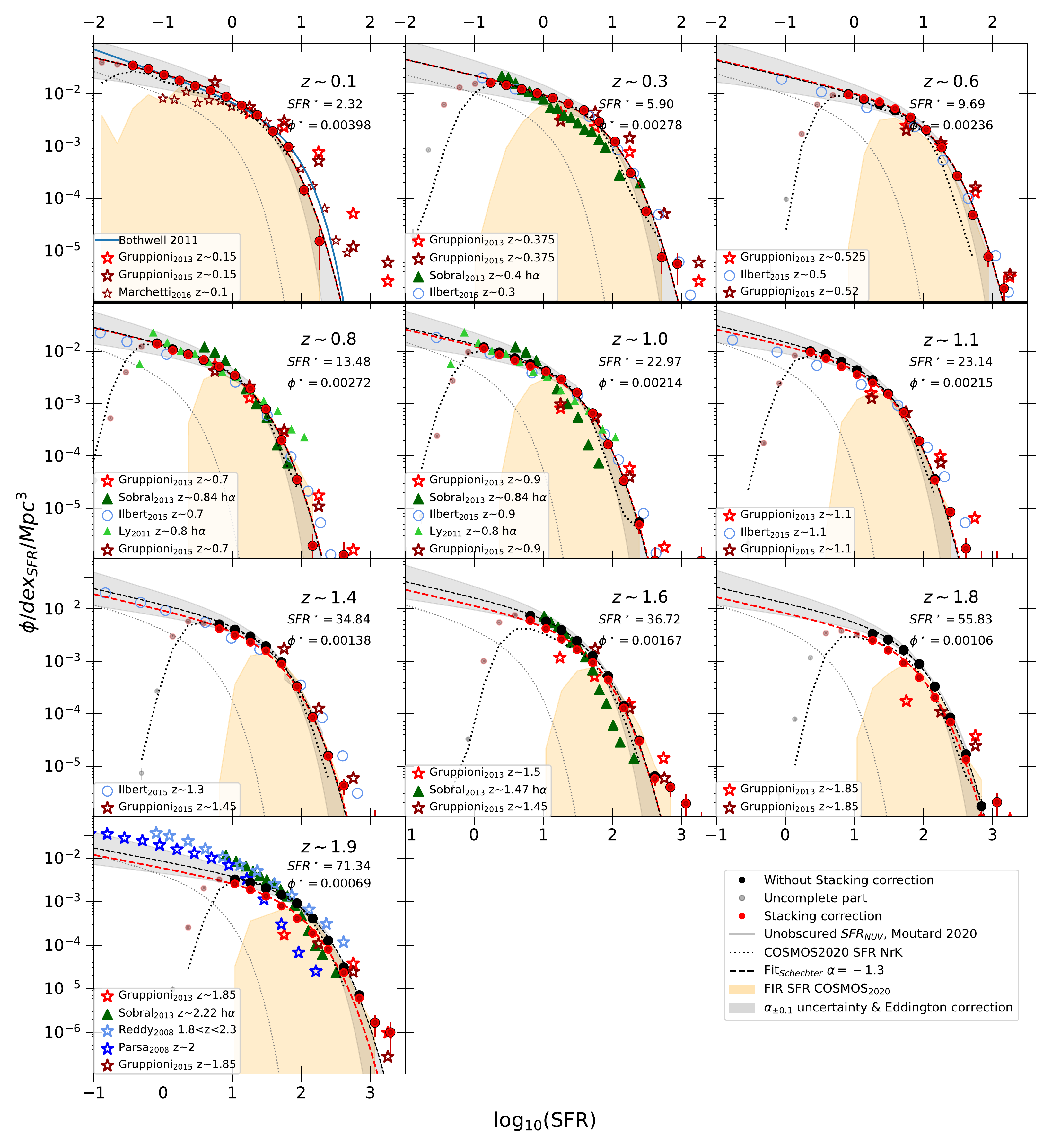}}%
  \caption[Evolution of the total SFR functions between redshift 0 and 2]{SFR functions per redshift bin from $z=0.05$ to $z=2.05$. The black- and red-filled circles correspond to the HSC-CLAUDS data with and without the stacking correction. The short-dashed black and long-dashed red lines correspond to the best-fitted Schechter functions assuming a fixed slope parameter ($\alpha=-1.3$). The gray area corresponds at the bright end to the Eddington correction and at the faint end to the slope uncertainty ($\Delta \alpha=0.1$). The orange areas represent the SFR functions based on the SFR derived from the  COSMOS2020 FIR data set. In the top panels, the SFR range is translated by 1 dex to make the low-SFR regime of the SFRFs visible.}
  
  \label{sfr_functions4.pdf}
\end{figure*}

The best-fit parameters of the SFRFs are reported in Table~\ref{sfrtable} (before convolution by the SFR uncertainty) and the final SFRFs for the two calibrations are shown in Fig \ref{sfr_functions4.pdf} (black and red lines). The gray shaded area reflects the slope uncertainties ($\delta\alpha=\pm 0.1$) at the faint end and the Eddington correction at the bright end. 

Finally, we also fit the SFR functions by a double power law with the same fixed faint-end slope:
\begin{equation}\label{exp}
\Phi(SFR) = \Phi^\star\left ( \frac{SFR}{SFR^\star} \right )^{(1-\alpha)} \mathrm{exp}\left ( -\frac{\mathrm{log}_{10}^2 (1+\frac{SFR}{SFR^\star})}{2\sigma^2} \right )
.\end{equation}
The double-power-law fits give very similar results to the Schechter functions down to  $\Phi(SFR)\sim10^{-5}$. The parameters are reported in Table~\ref{sfrtable} but are not shown in Fig \ref{sfr_functions4.pdf} for clarity. As seen in Table~\ref{sfrtable}, the characteristic SFR ($SFR_{\star}$) exhibits a monotonic decline from $\sim 60$ to $\sim 2$ \SFR from $z=2$ to $z=0$ and a smooth, though a noisier, increase of the normalization $\Phi_{\star}$ from $\sim 0.001$ to $\sim 0.004$ Mpc$^{-3}$.
% \end{strip}
\clearpage

% \begin{center}
% \input{sfr_functions3}
% \end{center}
\begin{table*}
\caption[Parameter values of the SFR function Schechter fit]{Parameter values of the SFR function Schechter fit for both individual FIR calibration and the verified calibration. The fit parameters of the double power law (noted DP) are also given for the original IRX calibration.}%We also add the Cosmos SFR density for both calibration
\resizebox{\textwidth}{!}{\begin{tabular}{ccccccccc}
\hhline{=========}
$z_{\mathrm{min}}$ & $z_{\mathrm{max}}$ & $\Phi^\star_{\mathrm{No-Stacking}}$ & $SFR^\star_{\mathrm{No-Stacking}}$ & $\Phi^\star_{\mathrm{Stacking}}$ & $SFR^\star_{\mathrm{Stacking}}$ & $\Phi^\star_{\mathrm{DP}}$ & $SFR^\star_{\mathrm{DP}}$ & $\sigma_{\mathrm{DP}}$ \\
 &  & $(10^{-3}Mpc^{-3})$ & \SFR & $(10^{-3}Mpc^{-3})$ & \SFR & $(10^{-3}Mpc^{-3})$ & \SFR  &  \\

\hhline{=========}
$0.05$ & $0.25$ & $3.98 \pm 0.20$ & $2.32 \pm 0.18$ & $3.98 \pm 0.20$ & $2.32 \pm 0.18$ & $12.32 \pm 1.60$ & $0.73 \pm 0.25$ & $0.46 \pm 0.06$ \\
$0.25$ & $0.45$ & $2.78 \pm 0.16$ & $5.90 \pm 0.42$ & $2.78 \pm 0.16$ & $5.90 \pm 0.42$ & $5.67 \pm 0.24$ & $5.97 \pm 0.71$ & $0.27 \pm 0.02$ \\
$0.45$ & $0.65$ & $2.36 \pm 0.10$ & $9.69 \pm 0.42$ & $2.27 \pm 0.09$ & $9.81 \pm 0.38$ & $4.90 \pm 0.23$ & $7.13 \pm 0.81$ & $0.33 \pm 0.02$ \\
$0.65$ & $0.85$ & $2.72 \pm 0.13$ & $13.48 \pm 0.65$ & $2.77 \pm 0.12$ & $13.31 \pm 0.58$ & $6.14 \pm 0.43$ & $9.16 \pm 1.51$ & $0.34 \pm 0.03$ \\
$0.85$ & $1.05$ & $2.14 \pm 0.10$ & $22.97 \pm 1.11$ & $2.35 \pm 0.09$ & $22.09 \pm 0.86$ & $5.75 \pm 0.40$ & $11.55 \pm 1.91$ & $0.39 \pm 0.03$ \\
$1.05$ & $1.25$ & $2.15 \pm 0.13$ & $23.14 \pm 1.31$ & $2.59 \pm 0.10$ & $21.72 \pm 0.80$ & $6.91 \pm 0.50$ & $8.61 \pm 1.28$ & $0.43 \pm 0.02$ \\
$1.25$ & $1.45$ & $1.38 \pm 0.08$ & $34.84 \pm 1.75$ & $1.82 \pm 0.10$ & $32.04 \pm 1.63$ & $3.73 \pm 0.22$ & $21.57 \pm 2.70$ & $0.35 \pm 0.02$ \\
$1.45$ & $1.65$ & $1.67 \pm 0.13$ & $36.72 \pm 2.41$ & $2.43 \pm 0.12$ & $33.80 \pm 1.44$ & $6.41 \pm 0.52$ & $13.23 \pm 2.02$ & $0.44 \pm 0.02$ \\
$1.65$ & $1.85$ & $1.06 \pm 0.12$ & $55.83 \pm 4.74$ & $1.64 \pm 0.07$ & $56.02 \pm 1.86$ & $2.90 \pm 0.22$ & $42.86 \pm 5.90$ & $0.34 \pm 0.02$ \\
$1.85$ & $2.05$ & $0.69 \pm 0.07$ & $71.34 \pm 6.88$ & $0.95 \pm 0.03$ & $83.77 \pm 2.34$ & $2.15 \pm 0.14$ & $47.90 \pm 6.62$ & $0.37 \pm 0.02$ \\
\hline
\end{tabular}}
\label{sfrtable}
\end{table*}

\subsection{Comparison with observations}

In Fig.~\ref{sfr_functions4.pdf} we compare our results with previous SFR functions from the literature based either on the FIR+UV luminosities or the dust-corrected H$_{\alpha}$ emission lines, and we convert them to a Chabrier initial mass function if needed \footnote{Most of them were already compiled in \citet{Katsianis2017a}.}.

\citet{Bothwell2011} used the IRAS  Faint Source Catalogue and the GALEX All-Sky Imaging Survey (AIS) to perform a combined weighted analysis to derive the SFR function in the local Universe. They also include deep \textit{Spitzer} and GALEX imaging of galaxies in the Local Volume Legacy (LVL) survey ($\le$11 Mpc) to constrain the faint end of the local SFR function down to SFR $<0.01$ \SFR.

\citet{Gruppioni2013,Gruppioni2015} used the PEP and HerMES surveys of the \textit{Herschel} mission, covering the passbands at 70, 100, and 160 $\mu$m (PACS) and 250, 350, and 500 $\mu$m (SPIRE), in the COSMOS and GOODS-South fields to measure the infrared luminosity functions up to redshift $z=4$ with a flux-limited sample at 160$\mu$m. In \citet{Gruppioni2015} they also perform a SED fitting from the UV to the submillimeter to subtract possible contributions from AGNs to the infrared luminosity and derive the SFRs by summing up the UV and infrared luminosities. 

\citet{Reddy2008} used a sample of spectroscopically confirmed Lyman-break galaxies  and \textit{Spitzer} MIPS 24 $\mu$m observations to derive the SFR (UV+infrared) functions in between $1.9\le z\le 2.7$ after correction for incompleteness effects. \citet{Marchetti2016} used \textit{Herschel} observations to infer the infrared luminosity down to $L_{IR}=10^9$ L$_\odot$ at $z<0.2$. Similarly, \citet{Wang2016} derive a luminosity function  at 250 $\mu$m up to $z=0.5$. Their faint-end slope is consistent with \citet{Marchetti2016}. In the following, we only refer to the \citet{Marchetti2016} value.

\citet{Ilbert2015} use a 24 $\mu$m-selected sample in the COSMOS and GOODS surveys up to $z=1.4$. The SFR is estimated by combining the UV+infrared luminosities, and SFR functions are derived by summing up their sSFR functions split per stellar mass bins. 

\citet{Ly2011} used emission line galaxies from narrowband imaging at 1.18$\mu$m from the New H$\alpha$ Survey corresponding to $H_{\alpha}$ at $z=0.85$. They correct for incompleteness and [N II] flux contamination. The SFR is derived by applying a luminosity-dependent dust correction following \citet{Hopkins2001}. 

\citet{Sobral2013} used four narrowband imaging observations in the UDS  and COSMOS fields to select $H_{\alpha}$ emitters at $z$ = 0.40, 0.84, 1.47 and 2.23. 
The H$_{\alpha}$ luminosity functions were then corrected for incompleteness, [NII] contamination, and dust extinction assuming an average attenuation of A($H_{\alpha}$)=1 mag. 

\citet{Parsa2016} used the deep fields (HUDF, CANDELS, and UltraVista-COSMOS) to measure the UV luminosity functions at $1.5\le z\le 2.5$. To convert it into an SFR function, \citet[]{Katsianis2017a} adopt a luminosity-dependent  evolution of the $\beta$-slope as proposed by \citet[]{Smit2012}. 
These data are the only ones based on a UV selection in this compilation as the authors claimed that this data set provides the best constraint on the slope of the UV luminosity function at this redshift. 

As a sanity check, we first compared our SFRFs with the COSMOS2020 ones derived with the same $NrK$ method and adopted their photometric redshifts and luminosity estimates. The COSMOS2020$-NrK$ SFRFs are shown as black dotted lines. At all redshifts, they are in excellent agreement with our HSC-CLAUDS SFRFs and within our uncertainties. 

 The COSMOS2020$-$(UV+FIR) SFRFs derived with the 24$\mu$m flux-limited sample are shown as shaded orange histograms. The two SFRFs are also in good agreement up to the completeness limit. Even though the $NrK$ method relies on the COSMOS2020 FIR data for the calibration, when applied to the entire HSC-CLAUDS sample, our $NrK$ method reproduces well the SFRF at high SFRs at all redshifts.  
On the other hand, the UV-optical selection of the HSC-CLAUDS sample spans a wider range of SFRs, extending down to at least a factor of 10 in the low-SFR regime, allowing us to explore the faint end slope.

At low redshift, $0\le z\le 0.5$, our $NrK$ SFR function is in good agreement with the FIR+UV SFRF obtained in the local volume by \citet[][blue line]{Bothwell2011} and exhibits a comparable faint-end slope ($\alpha=-1.41$ with the Vmax estimator). 
 It is also in good agreement with \citet[][thin red stars]{Marchetti2016} after converting their FIR luminosity function into a SFR function. While we observe a good match at the bright end, their faint-end slope is flatter, as expected since it neglects the  contribution of faint UV sources.
The 160$\mu$m-selected SFRFs \citep[red light and dark stars, respectively;][]{Gruppioni2013,Gruppioni2015} do not probe the faint end\footnote{Due to the lower sensitivity at $160\mu m$, their flux-limited sample introduces a brighter SFR cutoff than the $24\mu m$-selected samples} but their normalizations appear consistent with us around  SFR$\sim 1- 3$ \SFR. However, they overpredict the high-SFR end with respect to all the other FIR+UV and $NrK$ measurements.  This excess could be due to the FIR photometric extraction, where we adopt the super-deblended FIR photometry in the COSMOS field \citep[][see their Sect. 2.1.2]{Jin2018}, and/or to the photometric redshift estimates. We note that the \citet[]{Gruppioni2015} SFRF leads to a high SFRD (see Fig. \ref{SFR_density_sfr_new6_sfr.pdf}) due also to a steep faint-end slope. This is not the case for the \citet[]{Gruppioni2013} SFRF, based only on the FIR SFRF with a slope $\alpha=-1.2$.  
The SFRF from $H_{\alpha}$ by \citet[][ dark green triangles]{Sobral2013} at $z\sim0.4$ shows a steeper slope and a deficit at high SFRs, which could be attributed to a unique and averaged dust correction factor ($A(H_{\alpha})=1$mag) applied (see below).

At intermediate redshifts, $0.5\le z\le 1.5$, all the FIR+UV SFRFs are consistent with each other as do our $NrK$ SFRFs. To extend the SFRFs in the low-SFR regime, \citet[][open blue circles]{Ilbert2015} have included the contribution of low-mass galaxies by assuming that the shape of the sSFR function at low masses is the same as for their lowest stellar mass measurement (log$_{10}(M_{\star})=9.5-10$) and by normalizing the sSFR with the density of the star-forming GSMFs at the appropriated redshift. This leads to a slope of the SFRF in excellent agreement with our estimate ($\alpha\sim -1.3$) with no sign of evolution up to $z\sim 1.5$. The $H_{\alpha}$ SFRF from \citet[][ light green triangles]{Ly2011} at $z\sim 0.85$ (shown in $z=0.8$ and $z=1.0$ panels) shows a good agreement with the other measurements. They reproduce the high-SFR distribution and the faint end slope remarkably well compared to the one from \citet[][ dark green triangles]{Sobral2013}. This comes from the different dust correction treatments. \citet{Ly2011} adopted a luminosity-SFR-dependent correction as proposed by \citet{Hopkins2001} with large/small correction at high/low luminosity (similar to what we observe with stellar mass in Fig. \ref{whitaker4.pdf}), changing significantly the shape of the original $H_{\alpha}$ luminosity function.

At high redshifts, $1.5\le z\le 2$, our SFRFs are still in good agreement with \citet{Gruppioni2013,Gruppioni2015} at the bright end, while $H_{\alpha}$ SFRF slightly underestimate the high star-forming population. At $z\sim 2$, the SFRFs start to differ in different regimes. 
The UV-selected sample \citep[][dark blue stars]{Parsa2016} shows a significant shortage of high SFRs. This shortage is most likely a consequence of the uncertainty in the dust correction, especially for the most luminous as discussed at the beginning of Sect. 5. Adopting a $\beta$-slope varying only with luminosity cannot properly capture the wide scatter of dust attenuation at high UV luminosities \citep{Martin2005} and then cannot properly reproduce the high end of the SFR function. On the other hand, the UV sample explores the low-SFR regime. They derive a faint end slope consistent with our value despite a higher density normalization. 
 The Lyman break galaxies (LBG) selected sample \citep[][light blue stars]{Reddy2008}, with SFR derived from UV+FIR, also shows a higher normalization of the SFRF. In contrast to the UV and LBG samples, our HSC-CLAUDS sample has a stellar mass limit above Log$_{10}(M_{\star}/M_{\odot})=9.5$ at this redshift. We thus can miss the potential contribution of lower-mass galaxies in the SFRF around SFR$\sim 10$ \SFR and produce a flattening of the faint-end slope.  We also note that the LBG and UV SFR functions are derived in a much wider redshift bin ($\Delta z\sim 1$), which can impact the comparison, and a nontrivial correction for incompleteness is required for the LBG sample to assess the global SFR function.     
 %%%%%%%%%%%%%%
Finally, despite the very deep data set used in this work, our optical/NIR-selected sample can potentially miss heavily obscured galaxies such as submillimeter galaxies \citep[]{Chapman2005} or dark-HST galaxies \citep[]{Wang2019}. This population can contribute to the cosmic SFRD at high redshifts, but its comoving density is expected to be low in our redshift range of interest \citep[$z\le 2$,][]{Chapman2005}. It could help us better match our SFR functions above $SFR\ge 100 M_{\odot}/yr$ with those of \cite{Gruppioni2015}. Considering our bright-end SFR functions as lower limits, it  is an even more stringent test for the comparison with simulations discussed in the next section.  

%%%%%%%%%%%%%

 In conclusion, the $NrK$ method presented in this work allows, for the first time, the SFR functions to be measured for a wide range of SFRs. The derived SFR functions can reproduce the number density of high-SFR galaxies as observed with FIR samples as well as the slope at low SFRs, which is found to be relatively shallow ($\alpha\sim -1.3$), with no evolution at least up to $z\sim 1.6$ and potentially $z\sim 2$, according to the UV-selected sample. This method overcomes the current limitations of the other approaches (i.e., at the faint end for the FIR samples due to instrumental sensitivity and the dust treatment especially at the high-SFR end for the $H_{\alpha}$- and UV-selected samples).

\subsection{Comparison with simulations}\label{simu}

Star formation rate functions give an instantaneous view of the distribution of the in situ star formation at different epochs. It is a more stringent test for the models than the GSMF since the latter captures an integrated view of the past star formation activity. 

In terms of SAMs, \citet{Gruppioni2015} already made a comparison of their observed SFR functions with several SAMs and found an overall good agreement with the bright end of the SFRFs up to $z=2$, while the models fail to reproduce high star-forming systems at $z>2$. For hydrodynamical simulation, \citet{Katsianis2017a} used \eagle simulation and observed a deficit of high star-forming simulated galaxies at $z<2$.

We aim here to make a broader comparison with several  state-of-the-art cosmological hydrodynamical simulations.
In this section we confront our SFR functions with four hydrodynamical simulations: \simba \citep{Dave2019}, \hagn \citep{Dubois2014}, \eagle \citep{Crain2015,Schaye2015,McAlpine2016}, and \tngone from the \tng project \citep{Pillepich2018,Nelson2019}.

\subsubsection{Main ingredients in the simulations}
All these simulations incorporate different prescriptions to form stars, treat the stellar and black hole (BH) feedback, and adopt different observables at $z=0$ to fine-tune the sub-grid physics models. 
The simulations are described in more detail in Appendix \ref{simulation} and their main features are summarized in Table \ref{tab:simus}. 
%
%Several differences can be emphasized.
Here we highlight some of the main differences that can have an impact on the SFR functions, which is the main topic of this paper.

\simba is the only one to model on the fly the formation, growth, and destruction of dust, and to introduce X-ray heating from BHs in addition to the regular AGN feedback (see Appendix \ref{simulation}). The importance of X-ray heating has been explored in zoom simulations by \citet{Choi2012}, showing that it can potentially drive the quenching of massive galaxies. 
Indeed, as shown in \cite{Dave2019, Dave2020}, while the X-ray feedback 
has a minimal effect on the galaxy mass function, it represents an important additional energy input to fully quench massive galaxies. This leads to a quenched galaxy population \citep{RodriguezMontero2019} with reduced central
molecular gas \citep{Appleby2020}, and the galaxy–BH co-evolution  in good agreement with observations \citep{Thomas2019}.

While all of these simulations successfully reproduce the bulk massive galaxy properties, this is achieved with very different AGN feedback models. \simba employs kinetic two-mode feedback: The {radiative mode}, or the so-called  {quasar mode}, is designed to model outflows of cold molecular and warm ionized gas. The jet mode, often referred to as the {radio mode}, drives high-velocity collimated jets of hot gas in a direction given by the angular momentum of the inner disk.

The energy used to drive the AGN feedback and that serves to quench galaxies originates from the accretion energy onto BHs. \simba is unique in using a two-mode accretion model consisting of the torque-limited accretion from cold gas and the more standardly used Bondi accretion from hot gas.   
Similarly, both \hagn and \tngone employ Bondi accretion with a two-mode feedback model; however, they adopt spherical thermal feedback at high Eddington growth rates. In \hagn, the radio mode deposits kinetic energy into a bipolar jet along the accreted gas angular momentum (similar to \simba),
while in \tngone it is in a random direction.
In contrast, \eagle does not make a distinction between the two modes, it assumes Bondi accretion model onto BHs and thermal energy injection follows the quasar-mode feedback scheme.  

Finally, while all simulations rely in one way or another on the Schmidt relation to form stars (i.e., SFR proportional to the gas density over the
dynamical or free-fall timescale), the models differ in detail. While \hagn and \tngone use the density of atomic hydrogen to trigger star formation, \simba relies on the molecular hydrogen, with comparable density thresholds and efficiencies,
and \eagle adopts a metallicity-dependent density threshold for star formation.

\subsubsection{Descriptions of the snapshots}
To perform the comparison with the observed SFRFs, we analyzed the snapshots 
at five different redshifts, $z~=0,0.5,1,1.5,2$.
We restricted the analysis to galaxies identified by the galaxy finders used for different simulations, and therefore composed of at least 30 star particles (50 for \hagn). This corresponds to a stellar-mass limit of $\log_{10}$(\MS/\MM) $=$ 8.7, 8.2, 7.7, and 7.6 for \simba, \hagn, \eagle, and \tngone, respectively. 
The SFR is typically estimated based on the number of stars formed over a certain period of time  \citep[frequently 100 Myr;][]{Dubois2014}.

We explicitly verified for the \hagn and \tngone simulations that using shorter timescales (e.g., 10 Myr or 50 Myr) does not alter our conclusions. For \simba, SFR is computed from the gas particles, 
corresponding to the SFR computed from the stellar particles averaged over a timescale between 50 and 100 Myr \citep{Dave2019}.
The choice of 100 Myr roughly corresponds to a minimum measurable SFR ($M_{\star, {\rm init}}/100$ Myr) of 0.18 \SFR for \simba and 0.02 \SFR for the other simulations.
In \eagle, stellar mass and SFR are computed using particles within a fixed spherical aperture with a radius of 30 proper kpc \citep[see, e.g.,][]{Crain2015}. 
Given that the majority of star formation occurs in the central 30 kpc, even for massive galaxies, this aperture constraint has only a minimal effect on the SFRs \citep[see, e.g.,][]{Furlong2015}. The effect of the aperture on the GSMF is negligible for galaxies with \MS $< 10^{11}$ \MM; however, for more massive galaxies the aperture reduces the stellar masses \citep[see, e.g.,][]{Schaye2015}.  
Finally, we note that the SFR considered in each simulation is not corrected for the mass loss due to winds and SNe. However, we checked that taking this correction into account has a minimal impact on SFRF.

Before performing our comparisons, we exclude the quiescent population by using a unique redshift-dependent criterion for all four simulations as defined in \cite{Dave2019}, and based on the sSFR:  $\log_{10}(\rm{sSFR/yr}^{-1})=-10.8+0.3 z$. This is equivalent to removing galaxies with $\Delta$SFR $\sim$1 dex below the MS. We verified that this cut is consistent with the selection of SFGs based on the NUVrK color-color diagram. We note that, because the impact of passive galaxies is much less significant on the SFR functions than on stellar mass functions, the exclusion of simulated passive galaxies is often not applied when performing the comparison of the SFR functions with observations \citep{Katsianis2017b}.

\begin{table*}
  \caption{Summary of the different simulations used in this work: \simba, \hagn, \eagle, and \tngone. 
  %The white, light-grey, grey, and dark-grey background part recap 
  The three blocks respectively recap the cosmological parameters, general simulation parameters, and baryonic physics, including models for feedback and star formation.
  %, and the SFR function comparison.
  }

    \centering
    \setlength{\tabcolsep}{2pt}
    \renewcommand{\arraystretch}{1.5}
    \begin{tabular}{
      |
      %p{0.15\textwidth}
      >{\centering}p{0.15\textwidth}|
      >{\centering}p{0.2\textwidth}
      >{\centering}p{0.2\textwidth}
      >{\centering}p{0.2\textwidth}
      >{\centering\arraybackslash}p{0.2\textwidth}
      |
    }
\hline
\textbf{Parameters} & \textbf{\simba} & \textbf{\hagn} & \textbf{\eagle} & \textbf{\tngone} \\ 
\hline
$\mathbf{\Omega_m}$ - $\mathbf{\Omega_b}$ & 0.3 - 0.048 & 0.272 - 0.045 & 0.307 - 0.048 & 0.309 - 0.048 \\
$\mathbf{H_0}$ & 68 $\rm km~s^{-1}~Mpc^{-1}$ & 70.4 $\rm km~s^{-1}~Mpc^{-1}$ & 67.77 $\rm km~s^{-1}~Mpc^{-1}$ & 67.74 $\rm km~s^{-1}~Mpc^{-1}$\\
$\mathbf{\sigma_8}$ - $\mathbf{n_s}$ & 0.82 - 0.97 & 0.81 - 0.97 & 0.83 - 0.96 & 0.82 - 0.97 \\
\hline
\textbf{Code} & \gizmo & \ramses & \gad & \arepo\\
\textbf{Box size} & 100 $h^{-1}$ Mpc & 100 $h^{-1}$ Mpc  & 67.77 $h^{-1}$ Mpc & 75 $h^{-1}$ Mpc \\
\textbf{Particles} & 1024$^3$ & 1024$^3$ & 1504$^3$ & 1820$^3$ \\

\textbf{Star particle mass$^\alpha$} & $1.82 \times 10^7$ \MM & 2 $\times 10^6$ \MM & $1.81 \times 10^6$ \MM & $1.4 \times 10^6$ \MM\\
\textbf{Gas particle mass} & $1.82 \times 10^7$ \MM & -- & $1.81 \times 10^6$ \MM & $1.4 \times 10^6$ \MM\\
\textbf{DM particle mass} & $9.6 \times 10^7$ \MM & $8 \times 10^7$ \MM & $9.7 \times 10^6$ \MM & $7.5 \times 10^6$ \MM\\
\textbf{Gravitational softening/resolution} & 0.5 $h^{-1}$kpc down to 1 proper kpc & 0.8 proper $h^{-1}$kpc & 1.8 $h^{-1}$kpc down to 0.7 proper kpc & 0.5 $h^{-1}$kpc\\
\hline
\textbf{Galaxy finder ($n_{p}$)$^\beta$} & 6D FoF (30)   & ADAPTAHOP (50) &FoF \& SUBFIND (30) & FoF \& SUBFIND (30) \\\textbf{Mass completeness$^\gamma$} & log$_{10}(M_{\rm{comp}}/M_\odot)=8.7$ & log$_{10}(M_{\rm{comp}}/M_\odot)=8.2$ &  log$_{10}(M_{\rm{comp}}/M_\odot)=7.7$ &  log$_{10}(M_{\rm{comp}}/M_\odot)=7.6$\\
\textbf{SNe feedback} & winds-kinetic   & winds-kinetic & thermal & winds-kinetic\\
\textbf{AGN feedback} & radio mode-kinetic \& X-ray heating, quasar mode-kinetic  & radio mode-kinetic and quasar mode-thermal & thermal & radio mode-kinetic, quasar mode-thermal   \\
\textbf{Feedback calibration} & 
$z=0$ galaxy stellar mass function
& $z=0$ black hole-galaxy scaling relation  
& $z=0$ GSMF, galaxy sizes, \MS--M$_{\rm BH}$ relation 
& SFRD evolution, GSMF, SHMR at $z=0$ \\
\textbf{Star formation} & Proceeds in $n_{\rm{H}_2}$> 0.13 cm$^{-3}$  regions, following a Schmidt relation with 2\% efficiency &Proceeds in $n_{\rm H}> 0.1$ cm$^{-3}$ regions, following a Schmidt relation with 2\% efficiency & Metallicity dependant density threshold reproducing KS relation& Proceeds in $n_{\rm H}> 0.1 $ cm$^{-3}$ regions, following a Schmidt relation with 2\% efficiency\\
\textbf{SFR estimator} & $\sim$50-100 Myr & 100 Myr & 30 kpc aperture & 100 Myr$^\delta$\\
%\hline

\hline
\end{tabular}
\caption*{$\alpha$: Initial mass particle. $\beta$:  Minimal number of particles for a galaxy to be found. $\gamma$: Galaxy stellar mass completeness based on the initial stellar mass particle and the minimum number of particles used by the galaxy finder. $\delta$: \citet{ Donnari2019,Pillepich2019}.}

\label{tab:simus}
\end{table*}
\bfig{1}{sfr_functions_simu.pdf}{SFR functions, stellar mass functions, and MS predicted by simulations:
{\bf Top row:} Comparison of the SFR functions for the SFGs (gray-shaded area based on the faint end slope uncertainty) with those from the four simulations in five redshift bins: \tngone (blue), \hagn (red), \simba (green), and \eagle (orange). On the top axis, the vertical marks correspond to the SFR limits used to fit the observations (black) and the simulations (colored marks) as described in Sects. \ref{Fitting} and \ref{simu}. The data points from the simulations are fitted by a Schechter or double-power-law function (thick lines; see text). {\bf Second row from top:}  GSMF for the star-forming population of the four simulations (colored lines) with a unique star-forming criterion from \citet[][ see text]{Dave2019}. The gray-shaded area is based on the published star-forming GSMFs in the COSMOS field (see text). {\bf Second row from bottom:}  Same as above but for the whole (star-forming + quiescent) population. The gray area shows the observed total GSMF from the COSMOS field \citep{Weaver2022} except at $z=0,$ where we use \citet{Bernardi2013} and \citet{Li2009}.  %
{\bf Bottom row:} Evolution of the star-forming MS for the COSMOS2020 data set with the SFR based on the $NrK$ method (gray shaded area; see text) and the simulations (colored dots). 
}
\sfig{1}{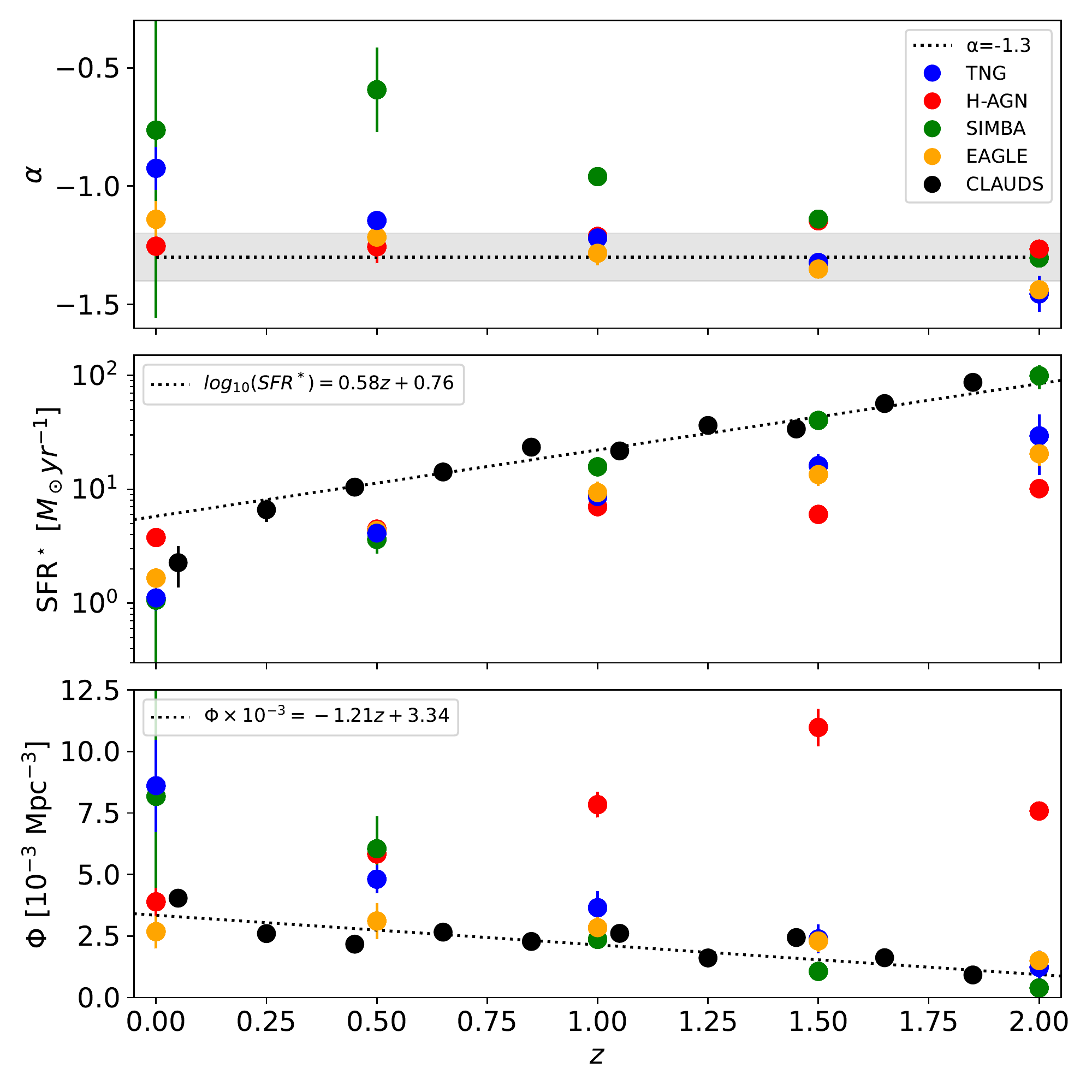}{Redshift evolution of the Schechter parameters  -- $\alpha$ (top panel), $SFR^{\star}$ (middle panel), and $\Phi^{\star}$ (bottom panel) -- of the SFR functions for the observations (black dots) and the four hydrodynamical simulations (color dots as indicated in the top panel). The fits of the redshift evolution for each parameter from the observations are shown as dotted lines.
} 
%
%%%%%%%%%%%%%%%%%%%%%%%%%%%%% COMPARISON
\subsubsection{Comparison of the simulated SFRFs with observations}

In Fig. \ref{sfr_functions_simu.pdf} we compare the SFR functions of SFGs obtained for the four simulations with the Schechter fits  of our HSC-CLAUDS sample %(which are corrected for the Eddington bias)
in five redshift bins. 
The differences between the simulations are quite noticeable, especially at increasing redshift. To quantify those differences, we fit each SFRF by a Schechter function, leaving the slope as a free parameter. We perform the fit down to the SFR limit (indicated as thick marks in each panel) as implied by the mass limit and MS relation of each simulation (see Fig. \ref{sfr_functions_simu.pdf}, bottom row). The data points and Schechter fits of the simulations are shown with dots and solid lines, respectively. 
The parameters of the Schechter fits are shown in Fig. \ref{sim_parameters.pdf}, along with those derived from the observations. 
We note that for \tngone and \simba simulations at $z=0$ and $z=0.5$, we adopt a double-power-law fitting function to better reproduce their bright ends, even if the parameters derived with a Schechter function are used for consistency in Fig. \ref{sim_parameters.pdf}.
  
Figures~\ref{sfr_functions_simu.pdf} and~\ref{sim_parameters.pdf} reveal three trends regarding the general and/or specific behaviors of the  simulations with respect to the observations. First, the SFRFs from \tngone and \eagle simulations show very similar behaviors at all $z$ and are in between the most extreme behaviors of the \hagn and \simba simulations. 

Second, while all simulations reproduce the high-SFR side at low $z$ ($z\le 0.5$), \simba is the only one able to reproduce its evolution with redshift up to $z \sim 2$, as can be seen in the evolution of $SFR^{\star}$ in Fig. \ref{sim_parameters.pdf}. This is consistent with the analysis from \cite{Lovell2021}, which shows that \simba is the only simulation able to reproduce the submillimeter galaxy number counts at $z\sim2$.
In contrast, \hagn shows a very mild evolution of $SFR^{\star}$ with an increase by less than a factor of 2, while the observations suggest a factor of $\sim$15 between $z=0$ and $z=2$. \tngone and \eagle are in between and show a shortage of high-SFR galaxies, as reflected by a low $SFR^{\star}$, at $z\ge 1$. 

Finally, the comoving density of galaxies with low to intermediate SFRs varies significantly between simulations.
At $z\sim2$, they all agree with a slope $-1.5\le \alpha\le -1.3$, consistent with the observations. At later times, while the slope in \hagn shows almost no evolution, in \eagle and \tngone it gradually flattens, but yet within the range of the observations (except at $z\sim 0$, where $(\alpha+1)>0$ for \tngone). In contrast, the evolution is much more pronounced for \simba, with $(\alpha+1)=0$ already at $z=1$ and ends up with a positive slope ($(\alpha+1)>>0$), that is to say, with a shortage of galaxies with low SFRs at $z<1$. We note that the lower mass resolution of \simba should not be responsible for this effect at least at $z\ge 0.5$.

In summary, none of the simulations analyzed here is capable of reproducing the observed SFR functions of SFGs at all redshifts from $z=2$ down to $z=0$.
Three of them, \hagn, \tngone, and \eagle, fail to reproduce the density of highly SFGs at $z\ge 0.5$, but reproduce reasonably well the low to intermediate SFR regime. On the contrary, \simba nicely reproduces the high-SFR regime but fails in the low-SFR regime below $z=1.5$.
This highlights that it remains challenging to achieve the balance between the star-forming ``MS'' galaxies and the population of quenched galaxies over cosmic time despite recent attempts at using a more physically motivated prescription for star formation and various feedback processes.
To clarify the differences in the SFR functions, in   Fig. \ref{sfr_functions_simu.pdf} we also include additional information with the comparisons of the GSMFs for the SFGs (second row from top), the whole population (second row from bottom) as well as the evolution of the MS (bottom row).

\paragraph{\textbf{The global GSMF}}
To ease the comparison with the simulations in Fig. \ref{sfr_functions_simu.pdf} (2$^{\rm nd}$ row from bottom), the observed total GSMFs are represented as a gray-shaded region. At $z\sim 0$, the gray area encloses the GSMF from \citet{Li2009} (bottom limit) and \citet{Bernardi2013} (top limit), while at higher redshifts it is based on the total GSMFs in the COSMOS survey \citep[]{Weaver2022}. The same color code is used for the simulations.  The simulations are often tuned to reproduce the observed GSMFs
 %over a broad range of cosmic history, 
 in particularly at $z \sim 0$. This is indeed the case of the simulations analyzed here. The only exception is the \hagn simulation, which is not calibrated in the local Universe apart from the choice of BH feedback parameters that reproduce the local BH mass versus stellar velocity dispersion relation.
 This possibly explains why the \hagn overpredicts at all epochs the GSMFs density \citep{Kaviraj2017}, and particularly below $10^{11}$ \MM. At $z \gtrsim 0.5$, \eagle underpredicts the density of high stellar mass galaxies \citep[see also][]{Furlong2015} and slightly overpredicts (similarly to \tng) the density of low stellar mass galaxies. 
 At all stellar masses, \simba agrees reasonably well with the observations.

The slight over-prediction of massive galaxies ($M\ge 10^{11}$ \MM) by \hagn and \tng simulation below $z=0.5$ may be due to an inefficient AGN feedback that does not quench star formation enough in these massive systems, while the over-prediction of galaxies at the low-mass end and $z\ge 0.5$ for all simulations but \simba can instead be attributed to an insufficient SN feedback model.

\paragraph{\textbf{The star-forming GSMF}} An observable that provides us with better insight into the ability of simulations to properly model different galaxy populations is the GSMF for SFGs. As it has not been specifically calibrated against observations in any of the simulations analyzed here, it may also better discriminate between them. 
The GSMF for SFGs is shown in Fig. \ref{sfr_functions_simu.pdf} (2$^{\rm nd}$ from top). Simulations exhibit a different level of agreement with the observed GSMFs in a redshift-dependent way. 

\hagn has quite similar behavior at all redshifts. While it manages to reproduce the very massive end, it significantly overestimates the density of SFGs below $M_{\star}=10^{11}$ \MM. 
At $z=0$,  \simba, \eagle, and \tngone  reproduce the low-mass end of the star-forming GSMF reasonably well; however, they underpredict the density of massive SFGs.
At $z>0.5$, and as the redshift increases, the trend of each simulation accentuates.
\eagle and \tngone slightly overproduce the low- to intermediate-mass population and \eagle underestimates the massive end. \hagn continues to overestimate the GSMF in all but the most massive galaxies. Interestingly, \simba reproduces reasonably well all the mass regimes.

In summary, \hagn overpredicts the number of low- to intermediate-mass SFGs at all redshifts, similar to the global population. This is consistent with a possible interpretation that SN feedback quenching is too inefficient for intermediate- to low-mass galaxies.
\eagle and \tngone over-predict the low-mass end for the star-forming and the global population at $z \gtrsim 0.5$, and \eagle is the only one simulation to under-predict the number of SFGs at $z>1.5$.  Overall \simba appears to be the only simulation able to reproduce relatively well the GSMFs for the global and star-forming populations.
%
%%%%%
\paragraph{\textbf{The star-forming main sequence}}  Additional key observable for galaxy formation models is the relation between the SFR and stellar mass of galaxies. The cosmological simulations have been known to under-predict the amplitude of the SFR--\MS relation at high redshifts ($z\sim 2$) by a factor of a few, while the agreement improves at low redshift ($z\lesssim 0.5$). 
This can be seen in Fig. \ref{sfr_functions_simu.pdf} (bottom row), where the MS for all the simulations is compared to the observations based on the COSMOS2020 data set. The SFR is derived from the $NrK$ method and the width of the gray-shaded area corresponds to the observed scatter. 

 At $z=0$, all the simulations are close to the observations except for \hagn and \eagle, which slightly underestimate the SFR at low masses, \MS$\sim 10^{9.5}$ \MM. At increasing redshift, the three simulations, \hagn, \eagle, and \tng,  gradually deviate from the observed MS. The departure is more pronounced at low masses, where it reaches up to a factor of 10 underestimation of the SFR at $z=2$ for galaxies with \MS$\sim 10^9$ \MM and a factor of $\sim$3 for galaxies with \MS$\ge 10^{10}$ \MM. This trend is much less pronounced with \simba, which remains consistent with the observed MS at high masses up to $z=2$ and deviates from the MS mainly for the low-mass regime, \MS$\le 10^{10}$ \MM.

While a natural reason for this offset on the side of modeling could be a too strong stellar feedback in low-mass galaxies, and in the case of \hagn, \eagle and \tngone, possibly also AGN feedback at the high-mass end at high redshift ($z \gtrsim 1$), this seems unlikely considering the excess of simulated galaxies at intermediate- to low-mass in the GSMFs. This is probably not the whole story as simply boosting the SFR would for example bring \hagn in even bigger disagreement with the observed cosmic SFRD at all redshifts, and it would induce a disagreement for other simulations at $z \lesssim 0.5$, where they match the average SFRD derived in this work (see Fig. \ref{SFR_density_sfr_new6_sfr.pdf}, bottom panel). Figure~\ref{sfr_functions_simu.pdf} also suggests that systematic offset does not explain the discrepancy in the SFR--\MS relation as it would simply shift the SFR functions toward higher values, inducing a further disagreement with observed SFRF at all SFRs at $z=0$ for all simulations, at the bright end for \simba at all redshifts, and at the faint end and at intermediate SFR for all simulations at $z \gtrsim 0.5$. 
Alternatively, as suggested by \cite{Furlong2015}, a potential solution to low SFRs is an insufficient burstiness of star formation. Making star formation burstier could result in a higher SFR over shorter periods compared to the current models without significantly modifying the stellar mass of galaxies. It remains to be seen whether this solution could at the same time solve the offset of the SFR--\MS relation, and the discrepancies at the bright and low end of SFR functions while keeping GSMFs in relatively good agreement with observations.

Another interesting aspect is the scatter of the star-forming MS. This scatter is significantly broader for \simba ($\sigma \sim 0.4$) compared to the other simulations ($\sigma \sim 0.3$). This trend tends to broaden the SFR functions per stellar mass bin and could be in part responsible for the underestimation of the low-to-intermediate SFR end and the flattening of the faint end slope in \simba's SFRF.\\

%%%%%%%%%%%%%%%%%%%

In conclusion, this set of simulations, although not exhaustive, reflects today's state-of-the-art galaxy formation models and shows the complexity of simulating distributions of galaxies' SFR that match observations. Indeed, compared to stellar mass, SFR is a more instantaneous parameter and is subject to much more stochasticity. It is therefore not surprising that simulations do not yet fully agree with observations. 
Further SFR function predictions from simulations will provide an efficient testing ground, in addition to more standard used stellar mass functions and star-forming MS, %(SFR--\MS relation),
for converging toward more realistic feedback mechanisms implementation in simulation.

%
\iffalse
\textbf{Questions for Romeel or us}:
\begin{itemize}
\item  Why such a fast evolution of the slope in \simba ? To be discuss with the perspective of too many low mass galaxies turned passive. Vincent can you make the plots for the GSMF at z=0,0.5,1,1.5,2 for Passive+SF with Simba and observations. Other sims as well.   
\item Origin of this flattening of the slope (stellar wind, SN feedbacks) ... ? 
\item Why no difference between SFR 100Myr and 10Myr in TNG and HAGN ? True ?
\item Why \simba 100Myr SFR is not ok (to be checked! Vincent !), what is the difference with SFR real time
\item what about the SFR timescale in simulation and its impact on the SFR functions comparison. It is around 100 Myr consistent with NUV or FIR observations. 
\end{itemize}
\fi 
%
%
\section{Cosmic SFRD}\label{SFRD}
The cosmic SFRD, $\psi$, is derived by integrating the SFR Schechter functions as 
\begin{equation}\label{integ}
\psi(z) = \int_{0}^{\infty}   SFR\ \Phi(SFR,z)\ dSFR
.\end{equation}
The redshift evolution of $\psi(z)$ is shown in Fig. \ref{SFR_density_sfr_new6_sfr.pdf}. 
The light blue shaded region encloses the integration of the SFR functions based on the IRX calibration with and without stacking technique (lower and upper bounds, respectively). The open circles are the average SFRD resulting from the two calibrations, while the error bars reflect the impact of a changing slope  around $\alpha=-1.3$ ($\Delta\alpha=\pm 0.1$) and redshift residuals in the SFR calibration mentioned in Sect.~\ref{section4.4}.

Our measurements are compared with the SFRD from the literature obtained with different SFR estimators -- radio-based \citep{Malefahlo2020, Karim2011,Leslie2020}, UV-based \citep{Schiminovich2005, Dahlen2007, Cucciati2012}, and FIR-based \citep{ Sanders2003, Magnelli2011, Magnelli2013} -- and the compilation fit from \citet[][thick black line]{Madau2014}. There is an overall good agreement with a gradual decline of the cosmic SFRD since $z\sim 2$, the cosmic noon, up to the present day by a factor of 10. While \citet{Madau2014} predicts a decline scaling as $\psi(z)\propto (1+z)^{2.7}$ since $z=1.5-2$, our observations suggest a slightly steeper rate with $\psi(z)\propto (1+z)^{2.8}$ when using the same parametric form as their Eq. 15. 

The redshift evolution of $\psi(z)$ might be better captured by the parametric form  proposed by \citet{Katsianis2021}, which closely mimics the evolution of the gas reservoir model \citep[]{Bouche2010} described with two parameters \citep[see Eq. 16 in ][]{Katsianis2021}. The gray-filled area shows this alternative parametric form, with the upper bound corresponding to their parameters while the lower bound is an adaptation to fit our observations. This suggests a faster evolution in between $0\le z\le 1$ followed by a plateau between $1\le z\le 2$ rather than a peak at $z\sim 2$, in better agreement with our measurements and recent FIR estimates \citep{Gruppioni2013, Gruppioni2020, Katsianis2021}.

 We also show the contribution to the SFRD of different populations split into SFR (top panel) and stellar mass (bottom panel) bins.  The star-forming populations contributing the most to the SFRD evolve with redshift. At low-$z$, $z\le 0.5$, the main contributors are galaxies with low star formation activity ($0.3\le SFR\le 3 $\SFR) while moderate SFGs ($3\le SFR\le 30 $\SFR) take over at higher redshift. The contribution of the high star-forming population ($SFR\ge 30$\SFR) shows a sharp increase from $0.5\le z \le 2$ contributing equally at $z\sim 2$ than the moderate star-forming population, in qualitative agreement with previous FIR studies \citep{LeFloch2005}. 
 When split into stellar mass bins, the main contribution to the SFRD comes from intermediate-mass galaxies ($9.5\le Log(M_{\star}/M_{\odot})\le 10.5$),  namely, the population below the knee of the GSMF, at all redshift. All the SFRDs from the three stellar mass bins show a global decline with cosmic time but at different paces. By Adopting the SFRD parametrization from \citet{Madau2014}, we find an evolution rate, $\psi(z)\propto (1+z)^{\alpha}$,  with $\alpha=2.2$ for the lowest-mass population and $\alpha=3.2$ for the most massive one. These trends are consistent with the downsizing picture \citep[]{Cowie1996, Juneau2005} in which most massive DM halos have a rapid accretion rate at an early time followed by a quenching phase after their star formation onset. Low-mass systems accrete at a later time but are also sensitive to the global decline of the accretion rate at all mass at a later time, preventing them from becoming more dominant at low $z$ \citep[e.g., ][]{Bouche2010}. 
 This picture is consistent with \citet[]{Gruppioni2013}, based on the infrared luminosity functions, but slightly differ in the relative contribution to the total SFRD. When adopting their stellar mass bins, we both find that the high-mass bin $11\le Log(M_{\star}/M_{\odot})\le 12$ never exceeds 15\% of the total SFRD. While we find that the two low-mass bins ($Log(M_{\star}/M_{\odot})\le 10$ and $10\le  Log(M_{\star}/M_{\odot})\le 11$) contribute equally at $\sim$50-40\%, \citet[]{Gruppioni2013} find a higher contribution from the intermediate-mass bin. But we note that the parameters of the infrared luminosity function in their lowest-mass bin are poorly constrained at increasing redshift.

In the bottom panel of Fig.~\ref{SFR_density_sfr_new6_sfr.pdf}, we show the SFRD of the simulations by integrating their respective Schechter functions discussed in Sect. \ref{simu}. The evolution of $\psi(z)$ for \hagn is in excellent agreement with the observations above $z=0.5$, despite noticeable differences in the shape of the SFR functions with the observations. The lack of evolution in the $SFR^{\star}$ parameter is compensated for by a high normalization. 
All the other simulations are slightly below the observations. For the \tngone and \eagle simulations, this is essentially due to the scarcity of high SFGs at high redshift, while for the \simba simulation, which is the only one to reproduce the density of high SFGs, it is caused by the lack of intermediate star-forming systems. This illustrates that integrated quantities such as the cosmic SFRD (or stellar mass density) are a necessary test to be passed by the simulations but not sufficient to understand the origin of any discrepancies without the full characterization of the SFR (or stellar mass) functions.

\sfig{1}{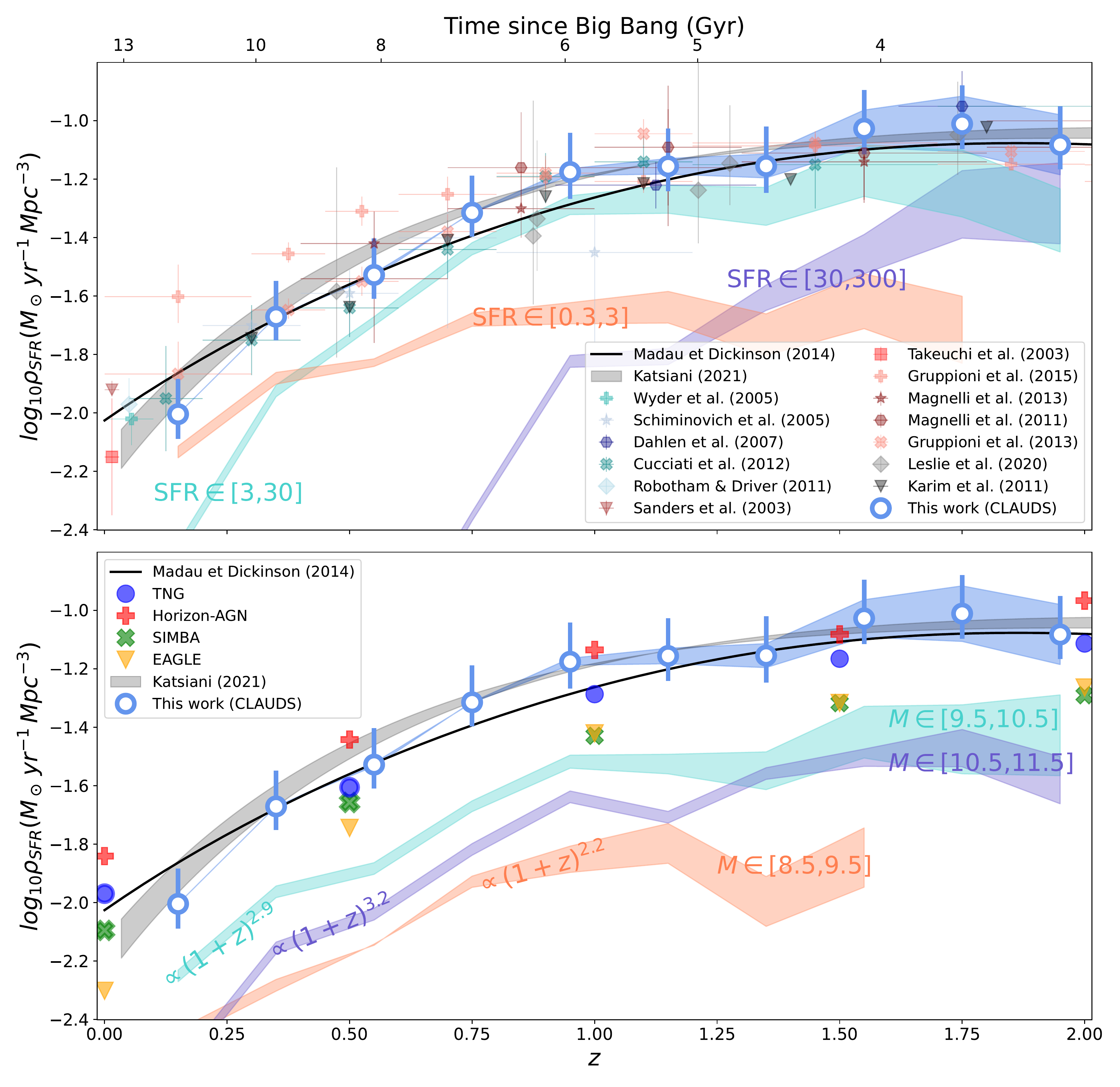}{
Cosmic SFRD. The blue dots correspond to the averaged SFRD resulting from the two IRX calibrations, with the upper and lower limits of the blue-filled area corresponding to the calibration with and without stacking, respectively. The error bars on the blue dots are computed using the faint-end slope uncertainty of the SFR functions, $\alpha=-1.3\pm0.1$. In the top panel, SFRDs from other observational studies are added with colors depending on their estimator: UV (blueish), infrared (reddish), and radio (black). In the bottom panel, SFRDs measured from the four hydrodynamical simulations are added. In both subplots, we also add the contributions to the cosmic SFRD of different stellar-mass and SFR regimes (color-shaded areas).}

\section{Conclusion}\label{discussion}

In this paper we investigate the evolution of the SFR functions up to $z\sim$2, based on stellar-mass-selected samples from deep HSC-CLAUDS observations. To derive the SFR of individual galaxies, we used an original method based on the UV$-$infrared energy budget that does not rely on SED fitting or uncertain assumptions regarding the dust attenuation laws.  This energy budget is calibrated by analyzing the  behavior of the IRX, IRX=$L_{\rm IR}/L_{\rm UV}$, in the rest-frame color-color diagram, $(NUV-r)$ versus $(r-K_s)$, for galaxies observed in the UV and FIR in the COSMOS2020 catalog. We show that the use of a single vector, $NrK$, is effective in predicting the IRX with a small scatter ($\sigma<0.2-0.3$\,dex) independently of the stellar mass. We further extended the IRX calibration to higher-redshift and lower-mass galaxies by stacking the FIR \textit{Spitzer} and \textit{Herschel} data. 

This approach allowed us to derive the individual SFRs of one million galaxies in the HSC-CLAUDS sample down to \MS $\sim 10^9$ \MM. The main results are as follows:

\begin{itemize}
\item Our estimated SFR nicely reproduces the evolution of the MS (SFR versus \MS) up to $z\sim$2 and the behavior of the attenuation (or $\langle IRX \rangle$) with stellar mass reported in the literature. 
\item  We reconstruct the SFR functions over a large range of SFRs ($10^{-2}-10^3$ \SFR) and redshifts (up to $z=2$), providing a constraint on the faint-end slope for the first time. The SFRFs are well fitted by a Schechter function after correcting for the Eddington bias. The high-SFR tails are in good agreement with previous FIR SFRFs and show a strong evolution of the Schechter parameter, $SFR_{\star}$, with redshift. 
 On the other hand, the slope of the SFRFs shows almost no evolution up to $z\sim 1.5-2$ with $\alpha=-1.3\pm 0.1$.
\item By integrating the SFRFs, we derive the cosmic SFRD from the current time up to $z=2$. Despite a relatively good agreement with \citet{Madau2014}, we find that our SFRD is a better fit with a plateau between $1\le z\le 2$ and a steeper drop below $z\sim$1 than a gradual decline since $z\sim2$. This has already been suggested in other studies \citep{Katsianis2021, Gruppioni2020}. 
\item The contributions to the total SFRD of  galaxy populations with different SFR regimes vary rapidly with redshift: Galaxies with moderate SFRs, $3 \le SFR\le 30$ \SFR, dominate the SFRD over most of cosmic time, $0.5\le z\le 2$, while low-SFR galaxies, $0.3 \le SFR\le 3$ \SFR, dominate at lower $z$. The contribution of high-SFR galaxies, $SFR\ge 30$\SFR, sharply increases since z$\sim$1 to contribute equally at $z\sim 2$ with the intermediate SFR population. 
% \ilbert{The list wintin the list is not easy to read. Better to remove the item $-$ here. }
%
\item  The contributions to the total SFRD of  galaxy populations in different stellar mass regimes vary in a similar way with redshift.  Galaxies with moderate stellar masses, \MS $=10^{9.5-10.5}$ \MM, dominate at all redshifts. The decline below $z\sim 1$ affects galaxies of all stellar masses, with a faster pace observed for the highest stellar masses. While feedback/outflows are well-observed phenomena and can be efficient at quenching the SFR in galaxies, there may be a more global origin for the decline at all masses, such as the decline of the cosmological accretion rate as a consequence of an expanding Universe that becomes dominant after cosmic noon.
 \item We compared the observed SFR functions  with four hydrodynamical simulations. Significant differences in the SFR functions are observed between the simulations, and none of them can reproduce the observations at all redshifts.
 They currently struggle to form high-SFR systems at high redshifts, with only one simulation able to reproduce the evolution of the density of high-SFR galaxies up to $z=2$. Large differences are also observed at intermediate and low SFRs. This reflects the fact that the SFR functions provide a powerful diagnostic in addition to the more commonly used integrated quantities, such as the stellar mass functions. The SFR functions that give an instantaneous view of the distribution of the in situ star formation at different epochs remain a challenge for the simulations despite the incorporation of diverse, physically motivated prescriptions for the star formation and feedback processes.
\end{itemize}
%
 \iffalse
 and the .    SFRFs are a very powerful tool, not only to improve our understanding of simulations by completing the Mass function / Main sequence view of galaxy evolution but to intent constraining the simulation underlying recipes. We think that SFR functions predictions from simulations should be more systematic as they will provide an efficient testing ground for converging towards more realistic feedback mechanisms implementation in simulation. This work already shows that they currently struggle to form high-SFR systems (which was a major discrepancy with observations) but that at least one simulation (\simba) finally manages to produce a bright-SFR end compatible with observations.

 There are more tensions between the SFR density predicted by simulations mostly resulting from normalization issues for \hagn, mid and low SFR under production for \simba and lack of high SFR objects for \eagle and \tngone.

  \item CLAUDS SFR functions provide an SFR density estimate that is consistent with the literature. Despite a relatively good agreement with \citet{Madau2014} we emphasize that our estimator is more compatible with a plateau between redshift 1 and 2 and a steeper drop below redshift 1 than a peak at redshift 2. This has been suggested in other studies \citep{Katsianis2021, Gruppioni2020}. There are more tensions between the SFR density predicted by simulations mostly resulting from normalization issues for \hagn, mid and low SFR under production for \simba and lack of high SFR objects for \eagle and \tngone. 
\fi

\begin{acknowledgements}
VP and SA wish to thank V\'eronique Buat for fruitful discussions. We are grateful to the referees for their constructive input. VP work was partly funded by NASA APRA grant 80NSSC20K0396. This work was supported by the Spin(e) ANR project (ANR-13-BS05-0005), the DEEPDIP ANR project (ANR-19-CE31-0023), and by the Programme National Cosmology et Galaxies (PNCG) of CNRS/INSU with INP and IN2P3, co-funded by CEA and CNES. \\
These data were obtained and processed as part of the CFHT Large Area U-band Deep Survey (CLAUDS), which is a collaboration between astronomers from Canada, France, and China described in \citet{Sawicki2019}.  CLAUDS is based on observations obtained with MegaPrime/ MegaCam, a joint project of CFHT and CEA/DAPNIA, at the CFHT which is operated by the National Research Council (NRC) of Canada, the Institut National des Science de l’Univers of the Centre National de la Recherche Scientifique (CNRS) of France, and the University of Hawaii. CLAUDS uses data obtained in part through the Telescope Access Program (TAP), which has been funded by the National Astronomical Observatories, Chinese Academy of Sciences, and the Special Fund for Astronomy from the Ministry of Finance of China. CLAUDS uses data products from CALET and the Canadian Astronomy Data Centre (CADC) and was processed using resources from Compute Canada and Canadian Advanced Network For Astrophysical Research (CANFAR) and the CANDIDE cluster at IAP maintained by Stephane Rouberol. The french members of the COSMOS team acknowledge the funding from the Centre National d'Etudes Spatiales (CNES).
The Hyper Suprime-Cam (HSC) collaboration includes the astronomical communities of Japan and Taiwan, and Princeton University. The HSC instrumentation and software were developed by the National Astronomical Observatory of Japan (NAOJ), the Kavli Institute for the Physics and Mathematics of the Universe (Kavli IPMU), the University of Tokyo, the High Energy Accelerator Research Organization (KEK), the Academia Sinica Institute for Astronomy and Astrophysics in Taiwan (ASIAA), and Princeton University. Funding was contributed by the FIRST program from Japanese Cabinet Office, the Ministry of Education, Culture, Sports, Science and Technology (MEXT), the Japan Society for the Promotion of Science (JSPS), Japan Science and Technology Agency (JST), the Toray Science Foundation, NAOJ, Kavli IPMU, KEK, ASIAA, and Princeton University. The Cosmic Dawn Center is funded by the Danish National Research Foundation under Grant No. 140. \\
We acknowledge the Virgo Consortium for making their simulation data available. The \eagle simulations were performed using the DiRAC-2 facility at Durham, managed by the ICC, and the PRACE facility Curie based in France at TGCC, CEA, Bruy\`eres-le-Ch\^atel.
We have benefited from the publicly available programming language Python, including the numpy \citep{harris2020},  matplotlib \citep{Hunter2007} packages, and the Topcat analysis tool \citep{Taylor2005}.
\end{acknowledgements}

\begin{table*}[H]
\scalebox{0.74}{\begin{tabular}{cccccccccc}
\label{tab:fit}
Redshift & $\alpha_{Schechter}$ & $\alpha_{Double}$ & $\Phi^\star_{Schechter}$ & $\Phi^\star_{Schechter,\alpha_{-1.4}}$ & $\Phi^\star_{Double}$ & $SFR^\star_{Schechter}$ & $SFR^\star_{Schechter,\alpha_{-1.4}}$ & $SFR^\star_{Double}$ & $\sigma_{-Double}$ \\
\hhline{==========}
$0.05 < z < 0.15$ & $-1.42 \pm 0.05$ & $-1.45 \pm 0.02$ & $1.93 \pm 0.60$ & $2.15 \pm 0.19$ & $3.24 \pm 1.32$ & $3.39 \pm 1.07$ & $3.10 \pm 0.49$ & $5.00 \pm 3.88$ & $0.20 \pm 0.10$ \\
$0.15 < z < 0.25$ & $-1.49 \pm 0.05$ & $-1.50 \pm 0.01$ & $1.58 \pm 0.58$ & $2.73 \pm 0.25$ & $2.93 \pm 0.83$ & $6.50 \pm 2.39$ & $4.11 \pm 0.66$ & $8.48 \pm 4.14$ & $0.21 \pm 0.07$ \\
$0.25 < z < 0.35$ & $-1.53 \pm 0.05$ & $-1.52 \pm 0.04$ & $1.11 \pm 0.51$ & $2.59 \pm 0.32$ & $5.22 \pm 2.93$ & $14.79 \pm 7.04$ & $7.40 \pm 1.66$ & $4.03 \pm 3.60$ & $0.49 \pm 0.17$ \\
$0.35 < z < 0.45$ & $-1.43 \pm 0.08$ & $-1.46 \pm 0.05$ & $1.50 \pm 0.77$ & $1.79 \pm 0.23$ & $3.29 \pm 1.55$ & $11.44 \pm 5.86$ & $9.84 \pm 2.26$ & $9.47 \pm 7.56$ & $0.29 \pm 0.12$ \\
$0.45 < z < 0.55$ & $-1.74 \pm 0.10$ & $-1.69 \pm 0.05$ & $0.39 \pm 0.44$ & $2.93 \pm 0.46$ & $1.68 \pm 1.41$ & $27.21 \pm 27.24$ & $6.15 \pm 1.40$ & $14.00 \pm 14.95$ & $0.33 \pm 0.18$ \\
$0.55 < z < 0.65$ & $-1.56 \pm 0.11$ & $-1.55 \pm 0.06$ & $1.02 \pm 0.66$ & $2.22 \pm 0.28$ & $3.57 \pm 2.16$ & $20.47 \pm 11.34$ & $11.36 \pm 2.20$ & $9.04 \pm 7.85$ & $0.42 \pm 0.15$ \\
$0.65 < z < 0.75$ & $-1.30 \pm 0.13$ & $-1.44 \pm 0.05$ & $3.12 \pm 1.38$ & $2.14 \pm 0.26$ & $3.05 \pm 1.54$ & $13.37 \pm 4.98$ & $17.90 \pm 3.37$ & $32.32 \pm 27.40$ & $0.18 \pm 0.10$ \\
$0.75 < z < 0.85$ & $-1.59 \pm 0.07$ & $-1.55 \pm 0.06$ & $1.15 \pm 0.38$ & $2.27 \pm 0.16$ & $4.15 \pm 1.63$ & $22.83 \pm 5.44$ & $14.47 \pm 1.32$ & $9.61 \pm 5.16$ & $0.44 \pm 0.09$ \\
$0.85 < z < 0.95$ & $-1.50 \pm 0.10$ & $-1.42 \pm 0.09$ & $1.17 \pm 0.42$ & $1.61 \pm 0.11$ & $4.53 \pm 1.50$ & $28.61 \pm 7.15$ & $23.12 \pm 1.97$ & $9.16 \pm 4.79$ & $0.47 \pm 0.07$ \\
$0.95 < z < 1.15$ & $-1.67 \pm 0.14$ & $-1.31 \pm 0.22$ & $0.96 \pm 0.59$ & $2.37 \pm 0.23$ & $8.89 \pm 3.78$ & $46.65 \pm 18.52$ & $26.28 \pm 2.98$ & $6.09 \pm 5.05$ & $0.53 \pm 0.08$ \\
$1.15 < z < 1.35$ & $-1.38 \pm 0.19$ & $-1.22 \pm 0.22$ & $1.45 \pm 0.63$ & $1.39 \pm 0.11$ & $5.03 \pm 1.51$ & $34.48 \pm 9.96$ & $35.46 \pm 3.00$ & $9.08 \pm 6.05$ & $0.48 \pm 0.06$ \\
$1.35 < z < 1.55$ & $-1.70 \pm 0.23$ & $0.20 \pm 17.46$ & $0.74 \pm 0.58$ & $1.67 \pm 0.23$ & $2.14 \pm 72.49$ & $81.37 \pm 38.62$ & $50.35 \pm 7.29$ & $1.09 \pm 27.11$ & $0.53 \pm 0.05$ \\
$1.55 < z < 1.75$ & $-2.09 \pm 0.19$ & $-0.65 \pm 5.36$ & $0.15 \pm 0.15$ & $1.26 \pm 0.19$ & $4.61 \pm 18.84$ & $184.36 \pm 96.81$ & $60.26 \pm 8.77$ & $3.36 \pm 32.40$ & $0.58 \pm 0.07$ \\
$1.75 < z < 2.00$ & $-1.79 \pm 0.20$ & $0.07 \pm 28.85$ & $0.26 \pm 0.21$ & $0.80 \pm 0.11$ & $1.24 \pm 69.49$ & $173.40 \pm 80.78$ & $91.74 \pm 13.78$ & $1.57 \pm 72.48$ & $0.57 \pm 0.13$ \\
\hline
\end{tabular}}
\end{table*}

\newpage


\bibliographystyle{aa}
\bibliography{./Masterlibrary}

%%%%%%%%%%%%%%%%% APPENDICES %%%%%%%%%%%%%%%%%%%%%

\appendix
\label{app_stacking}

\newpage

\section{IRX constraints from stacking analysis}
\label{App:stacking}

Surveys conducted in the thermal infrared regime are generally not sensitive enough for detecting low-mass sources individually. To explore the relationship between $NrK$ and IRX at low-to-intermediate stellar masses, we constrained galaxy total infrared properties using standard stacking procedures. The stacking technique consists in co-adding the signal arising from a  number of sources, and it is commonly used to characterize the average emission associated with a given galaxy sample. Constraints can be inferred down to flux levels much fainter than the typical 3$\sigma$ sensitivity limit of the initial data, depending on the number of co-added sources  \citep[e.g.,][]{Dole2006, Karim2011}.

\begin{figure*}[!hbp]
  \centering
  \includegraphics[width=\linewidth]{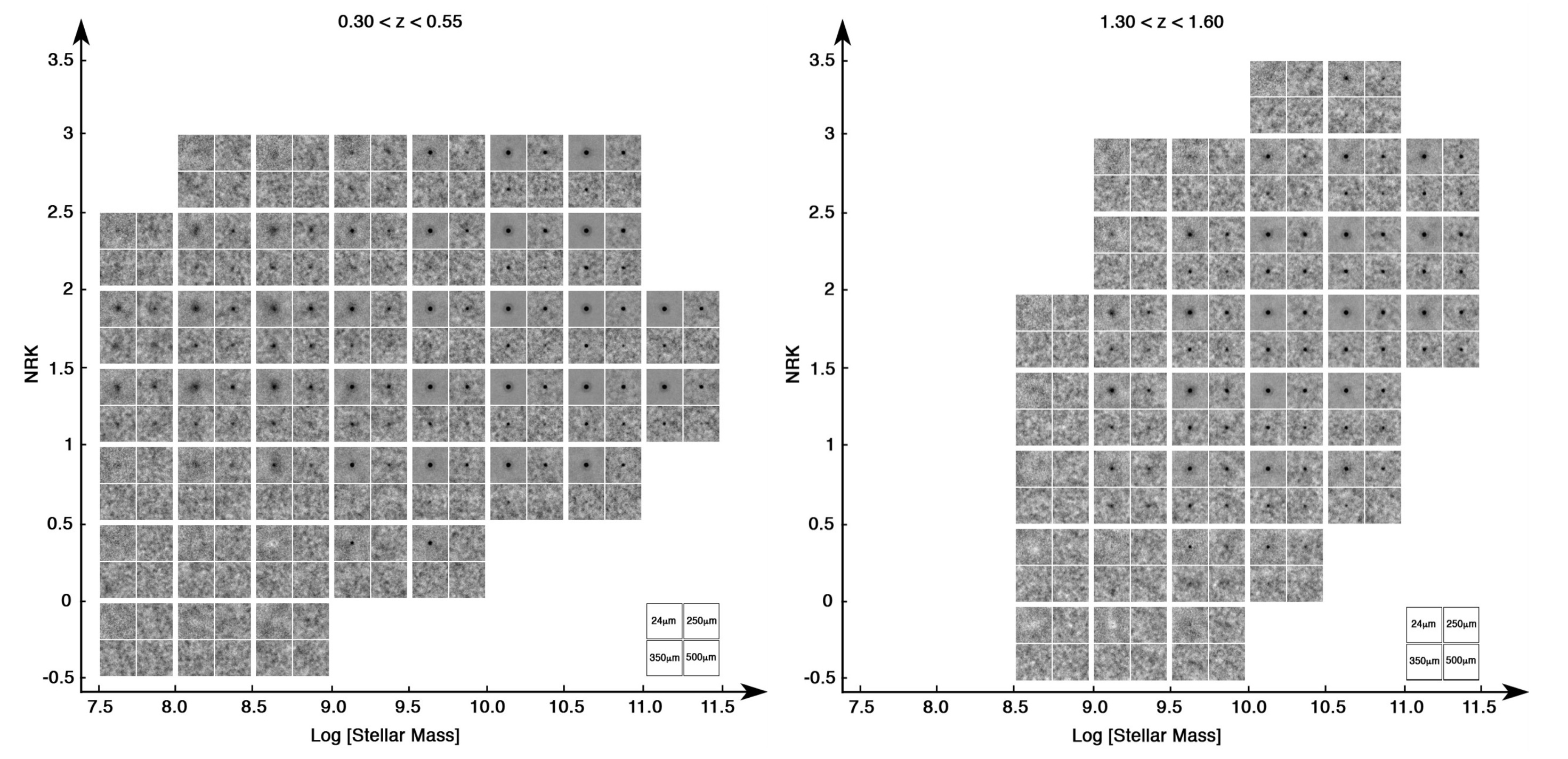}
  \caption{Results from MIR and FIR stacking in bins of $NrK$ and stellar mass in two redshift intervals. Each stamp includes the mean fluxes for the 24 $\mu$m data from \textit{Spitzer} and the 250, 350, and 500 $\mu$m data from \textit{Herschel}, as indicated in the inset}
  \label{stacking_both.jpg}
\end{figure*}

Our stacking was performed in the subsample lying in the COSMOS field, which benefits from exquisite MIR and FIR coverage from the {\it Spitzer\,} and {\it Herschel\,} satellites. Galaxies were stacked in various bins of redshift, mass, and $NrK$ colors, using the IDL routines of the library released by \cite{Bethermin2010}. The size of the redshift bin was fixed to $\Delta z = 0.25$ up to $z = 1.3$ and to $\Delta z = 0.3$ above. We adopted a color bin size of $\Delta NrK$\,=\,0.5 and a bin size of 0.5\,dex in logarithmic scale for the stellar mass. Typically,  the $NrK$ galaxy colors and the stellar masses span a range of 3 to 4\,magnitudes at the redshifts explored in this work.  The adopted bin sizes thus ensured a statistically large enough number of sources to be stacked in each bin,  while mitigating the effects of evolution within the bin. An illustration of the stacked signals that we obtained is shown in Fig. A.1 for two different redshift bins. The procedures we carried out are further described below.

\subsection{Stacking at 24 $\mu$m}
%\label{sect:stack_method1}

Given  the high space density of 24$\mu$m sources identified in the COSMOS field (6.5 arcmin$^{-2}$ down to 3$\sigma$), stacks were produced using the 24$\mu$m {\it residual\,} map
obtained by subtracting with PSF fitting each individual source found in the field \citep[see their Fig. 1]{Floch2009}.
For each bin of redshift, mass, and color, a stack was first created by mean-averaging 60\arcsec$\times$60\arcsec  sub-images centered at the sky position of each object in the bin (see Fig. A.1).
The averaged stacked signal was measured using aperture photometry, with the same aperture and aperture correction as used by \citet{LeFloch2009}  to perform the 24$\mu$m PSF fitting and photometry in COSMOS. To account for the emission arising from sources individually detected at 24$\mu$m and thus missed in the stack from the residual map, we then cross-correlated the galaxy subsamples associated with each bin with the list of sources initially subtracted from the 24 $\mu$m mosaic.

% \newpage

 This correlation was performed with a matching radius of 2\arcsec,  as also assumed by \citet{LeFloch2009}  in their identification of the COSMOS 24$\mu$m source optical counterparts. The fluxes of galaxies individually detected at 24$\mu$m were then weighted by the inverse of the total number of sources considered in the subsample, and they were finally co-added to the averaged stacked signal to obtain the average 24$\mu$m flux of the population selected in the bin.

The uncertainties associated with the stacked signal were estimated using bootstrap techniques as also described by \cite{Bethermin2010}. For each subsample, we stacked as many sub-images as the total number of sources to be stacked in the bin, but we randomly allowed some of these sub-images to appear several times in the stack. This process was repeated 200\,times for each bin of mass, redshift, and color, and the uncertainty was estimated from the dispersion of the averaged signal measured in each of the 200\,stacked images. We co-added in quadrature this uncertainty with the 1$\sigma$  flux uncertainties of the other sources  already detected at
 24$\mu$m, which gave us the final uncertainty associated with the mean 24$\mu$m flux characterizing the galaxy subsample of the bin.

To gain further confidence in the results obtained with this first approach, we inferred another estimate of the 24$\mu$m emission associated with each bin of redshift, mass, and color,  using a method based on {median} stacking and applied to the initial 24$\mu$m image of COSMOS. Here, each pixel in the final stack represents the median value of the distribution arising from the same pixel taken from the sub-images of the sample. The signal in the final stack thus corresponds to the median property of the stacked population. Systematic offsets may be expected with the mean estimate discussed above, depending on the underlying distribution of properties associated with the source population in the bin. This method has yet the advantage of making the contribution of neighboring contaminants almost negligible.  We measured the median stacked fluxes and their associated uncertainties with aperture photometry and bootstrapping, similar to the procedure already used for the mean stacking described earlier. Comparison between the two methods revealed a remarkable agreement down to faint fluxes (i.e.,  $\sim$\,10 $\mu$Jy), with a small systematic offset mostly noticeable at low redshift.

This offset between the median and the mean flux densities can be explained by the generally skewed distribution of galaxy luminosities in a given bin of redshift and stellar mass. The same effect can actually be seen by comparing in each bin the average and the median fluxes of 24$\mu$m sources individually reported in the COSMOS 24$\mu$m catalog. In the following analysis, we decided to use  the flux estimates obtained with the mean stacking,
mostly to remain  consistent with the mean values of the IRX considered for the population of sources individually detected in the infrared and discussed in Sect. \ref{IRX_calib_subsect}.

\bfig{1}{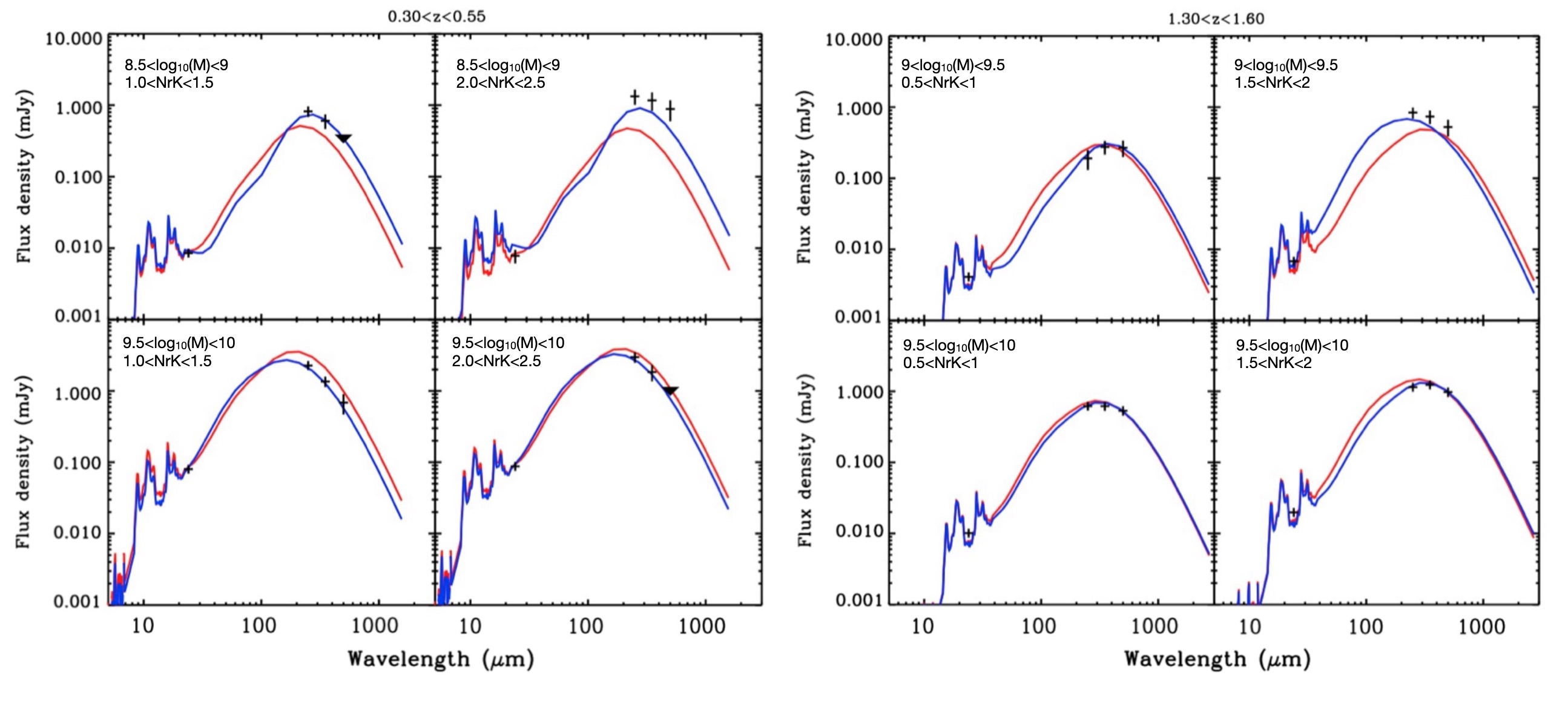}{Averaged MIR and FIR SED fitting of galaxy subsamples at $0.3<z<0.55$ (left panel) and $1.3<z<1.6$ (right panel), based on fluxes derived from the mean stacking technique described in Appendix~\ref{App:stacking}.  For each redshift range, SED fits are illustrated for two different bins of stellar mass and two bins of $NRK$ color. The best SED fits are shown with the solid blue curves, while the red curves represent SEDs extrapolated from the 24$\mu$m flux. Downward filled triangles depict upper limits for the stacked fluxes at 500$\mu$m.}

\subsection{Stacking at FIR wavelengths}

While the 24$\mu$m emission of galaxies allows us to probe their component of hot dust heated by young massive stars and/or radiations produced by BH accretion activity, the FIR wavelength regime is key for constraining the peak of their SED and estimating their total infrared luminosity with minimal uncertainty. We stacked the COSMOS imaging performed at 250, 350, and 500 $\mu$m with the SPIRE instrument on board the \textit{Herschel} satellite \citep{Oliver2012}, to determine the average FIR emission of sources for each bin of redshift, mass, and color as achieved at 24 $\mu$m. We followed the same approach except that the mean stacking was performed directly on the initial SPIRE images instead of the residual mosaics. This choice was motivated by the larger PSF characterizing the FIR \textit{Herschel} data, as well as the difficulty to associate individual FIR detection with their counterparts at shorter wavelengths, hence preventing in each bin of stacked sources a reliable control of the contribution of galaxies already subtracted from the residual map.
Photometry on the stacks was estimated in different ways (aperture photometry, PSF deconvolution, and Gaussian fitting), and we also compared  median and mean stacking with each other. All methods led to consistent results. Flux uncertainties were finally estimated with bootstrapping techniques, similar to the procedure employed at 24$\mu$m.

Given the size of the beam characterizing the \textit{Herschel} mosaics, the stacked emission at FIR wavelength is often affected by galaxy clustering, which results in a systematic overestimate of the associated signal due to contamination by neighboring sources. For each SPIRE band, the final flux was therefore corrected for this bias, which was inferred by decomposing the stacked signal into the contribution of the PSF  produced by the stacked sources, and that of a wider underlying component characterized by a Gaussian profile arising from the contaminating neighbors \citep{Bethermin2012}. As already noticed in the literature, we found that the bias can be neglected at the highest stellar masses, but typically reaches 30 to 40\% for galaxies at  \MS $\sim10^{9}$ \MM.

\newpage

\subsection{Total infrared luminosities and mean IRX estimates}
For each bin of redshift, stellar mass, and $NrK$ color, a fitting of SEDs at MIR and FIR wavelengths was performed with  LePhare so as to estimate the average Infrared luminosity of galaxies in the bin. This step was carried out following the same procedure as the one adopted earlier for the individual detections (see Sect. \ref{Far-InfraRed luminosities and FIR sample properties}). We fixed the  24 $\mu$m flux to the one obtained by combining the stacked signal from the residual 24 $\mu$m map and the contribution of each individual detection (see Sect. A.1). At FIR wavelengths, we used the peak flux of the mean stacked data, corrected from the clustering bias (see Sect. A.2). In several bins, especially at low stellar mass, we note that the stacked signal is only seen at 24 $\mu$m (see Fig. A.1). For these cases, the total infrared luminosity was inferred by extrapolating the average 24$\mu$m flux with the set of templates from \citet{Dale2002}, which adopts the locally observed dust temperature-luminosity relationship. Such extrapolations usually provide luminosity estimates consistent with the more accurate constraints obtained in the FIR where the peak of galaxy infrared SEDs is located, except for galaxies at very high luminosities in the starburst regime. We also verified the validity of this approach using our own galaxy sample, considering the bins with detections in both the MIR and FIR stacked data. For these cases, an overall agreement was found between the luminosity extrapolated from the 24$\mu$m flux and the one obtained with the combination of the  \textit{Spitzer} and \textit{Herschel} fluxes, except at high luminosity where the $L_{\rm IR}$ derived from the 24 $\mu$m tend to be overestimated (Sect.~\ref{Far-InfraRed luminosities and FIR sample properties}).

The average IRX for each bin of redshift, mass, and color was finally derived as the ratio between the average infrared luminosity estimated above and the mean of the UV luminosities measured individually for all the galaxies in the bin. The final uncertainty on the IRX was obtained by combining the uncertainties in quadrature.

% \newpage

\section{Description of the simulations}\label{simulation}

\subsection{\simba}

\paragraph*{\simba\footnote{\url{http://simba.roe.ac.uk/}} \citep{Dave2019}} was run with a modified version of the gravity and hydrodynamics solver \gizmo \citep{Hopkins2015}, relying on the \gad gravity solver \citep{Springel2005}.
The \simba run used in this work follows the evolution of 1024$^3$ DM and gas particles within a comoving volume of (100 $h^{-1}$ Mpc$)^3$. The simulation assumes a standard $\Lambda$CDM cosmology compatible with  \cite{PlanckCollaboration2016a} ($\Omega_{\rm m}=0.3$, $\Omega_{\Lambda}=0.7$, $\Omega_{\rm b}=0.048$, $H_{0}=68$ km s$^{-1}$ Mpc$^{-1}$, $\sigma_8=0.82$ and $n_{\rm s}=0.97$). The minimum gravitational softening length is 0.5 comoving $h^{-1}$ kpc, and the initial gas and DM particle mass is 1.82 $ \times 10^{7}$ \MM and 9.6 $\times 10^{7}$ \MM, respectively. 

% -------
Photoionization heating and radiative cooling models (\grac library; \citealt{Smith2017}) account for metal cooling and the nonequilibrium evolution of primordial elements. A spatially uniform UV ionizing background model  \citep{HaardtMadau2012} is modified to account for self-shielding based on the \cite{Rahmati2013} prescription.

% -------
Star formation is H$_2$-based and is only allowed to occur in gas with the hydrogen density $n_{\rm H} \geq$ 0.13 cm$^{-3}$.
The SFR is computed from the H$_2$ density and the dynamical time following the \cite{Schmidt1959} relation with 2 percent efficiency. The H$_2$ fraction follows the prescription of \citet{KrumholzGnedin2011}, based on the local column density and metallicity \citep{Dave2016}.

The stellar feedback is modeled using metal-enriched, two-phase galactic winds. 
In addition, the winds are decoupled, that is, hydrodynamics in the winds is turned off until they leave the ISM \citep{Springel2003}. Therefore, they do not deposit energy in the ISM on their way out.

The chemical enrichment model tracks 11 elements (H, He, C, N, O, Ne, Mg, Si, S, Ca, and Fe) from Type Ia and II SNe and asymptotic giant branch (AGB) stars. In addition, dust growth and destruction are tracked for each individual element on the fly. 

The growth of BH particles follows a two-mode accretion model. Hot gas ($T > 10^5$~K) is accreted in a spherically symmetric way following \citet{Bondi1952}, while cold gas accretion follows a torque-limited sub-grid prescription capturing the response of gas inflows near the BH to angular momentum loss due to dynamical instabilities \citep{HopkinsQuataert2011,Angles-Alcazar2017}. This combination of BH accretion modes determines the implementation of feedback from AGNs in the form of two-mode kinetic feedback, the so-called radiative mode and jet mode feedback.
Ten percent of the material accreted into the central region is assumed to fall onto the BH. These gas particles are immediately ejected in a purely kinetic and bipolar way (i.e., with zero opening angle w.r.t. the angular momentum of the inner disk) in two modes. At high accretion rates (above 0.2 times Eddington rate) and mass above 10$^{7.5}$ \MM, BHs eject material in $\sim 1000$ km/s winds without changing its temperature (radiative mode). As the BH accretion rate drops below 0.2 of the Eddington rate, jet feedback mode starts to turn on and is fully achieved below 0.02. Gas is ejected with a velocity increment proportional to the logarithm of the inverse of the accretion rate and capped at 7000 km/s. The temperature of the ejected particles is increased, consistently with observations \citep[][]{Fabian2012}.

In addition, \simba implements the X-ray radiation pressure feedback activated in galaxies with low cold gas content and when the jet mode is active. Gas with the hydrogen density $n_{\rm H} <$ 0.13 cm$^{-3}$ is heated by increasing its temperature, while for gas above this density, one-half of the X-ray energy is applied kinetically in the form of radial outward kick, and the second half is added as heat.
The main observational quantities used to calibrate the model include the observed $z=0$ GSMF \citep[][]{Dave2019}.

\subsection{\hagn}

\paragraph*{\hagn\footnote{\url{https://www.horizon-simulation.org/}} \citep{Dubois2014}} was run with the adaptive-mesh refinement code \ramses \citep{Teyssier2002} within a comoving volume of (100 $h^{-1}$ Mpc$)^3$ and assuming a standard $\Lambda$CDM cosmology compatible with WMAP  \citep{WMAP2011} ($\Omega_{\rm m}=0.272$, $\Omega_{\Lambda}=0.728$, $\Omega_{\rm b}=0.045$, $H_{0}=70.4$ km s$^{-1}$ Mpc$^{-1}$, $\sigma_8=0.81$ and $n_{\rm s}=0.967$). 
It contains $1024^3$ DM particles (i.e., a mass resolution of $M_\mathrm{DM,res}=8 \times 10^7$ \MM), and 
the initially coarse $1024^3$ grid (initial gas resolution is $M_\mathrm{gas,res}=1 \times 10^7$ \MM) is refined down to $1$~physical kpc. The refinement is triggered when the number of  particles becomes greater than 8 (or if the total baryonic mass reaches eight times the initial DM mass resolution in a cell). Heating of the gas from a uniform UV background is activated at $z_{\rm  reion} = 10$ following \cite{haardt&madau96} and gas is allowed to cool to $10^4\, \rm K$ via H, He, and metals \citep{sutherland&dopita93}.

Star formation only proceeds in regions with hydrogen number density $n_{\rm H} \geq$ 0.1 cm$^{-3}$ (the stellar mass resolution is $\simeq 2\times10^6$ \MM), following a Schmidt relation with 2 percent efficiency, that is to say, 2 percent of gas above the threshold density is converted into stars local free-fall time. 
\hagn implements sub-grid feedback from stellar winds and SN (both type Ia and  II) with mass, energy, and metal release. 

\hagn follows galactic BH formation and growth. BHs can grow by gas accretion at a Bondi-Hoyle-Lyttleton rate capped at the Eddington accretion rate when they form a tight enough binary.    
The AGN feedback is implemented as a combination of two different modes,
the so-called radio mode operating when the accretion rate is below 1\% of Eddington ratio and the quasar mode active otherwise. 
The quasar mode consists of an isotropic, spherically symmetric, injection of thermal energy. At low accretion rates, the radio
mode deposits AGN feedback energy into a bipolar outflow with
a jet velocity of 10$^4$ km/s. The efficiency of the radio mode is larger than the quasar mode with efficiencies tuned to match the BH-galaxy scaling relations at $z=0$ \citep[see][for details]{Dubois2012}.

\subsection{\eagle}
% Discuss \citet{Matthee2018,Furlong2015}\\

\paragraph*{\eagle\footnote{\url{https://icc.dur.ac.uk/Eagle/index.php}} \citep{Crain2015,Schaye2015,McAlpine2016}} was run using a modified version of the N-body Tree-Particle-Mesh smoothed particle hydrodynamics code \gad \citep{Springel2005}. 
The \eagle run used in this work follows the evolution of 1504$^3$ DM particles and an initially equal number of baryonic particles within a comoving volume of (100 Mpc$)^3$, yielding DM and baryonic mass of 9.7 $ \times 10^{6}$ \MM and 1.81 $\times 10^{6}$ \MM, respectively. 
The Plummer-equivalent gravitational softening length is 2.66 comoving kpc, limited to a maximum length of proper 0.7 kpc.
The simulation assumes a standard $\Lambda$CDM cosmology compatible with  \cite{PlanckCollaboration2014} ($\Omega_{\rm m}=0.307$, $\Omega_{\Lambda}=0.693$, $\Omega_{\rm b}=0.048$, $H_{0}=67.77$ km s$^{-1}$ Mpc$^{-1}$, $\sigma_8=0.8288$ and $n_{\rm s}=0.9611$).  Radiative cooling and photo-heating are implemented element-by-element for 11 species \citep{Wiersma2009} exposed to the cosmic microwave background and evolving UV/X-ray background radiation \citep{Haardt2001}. Star formation is implemented within the gas modeled as a single-phase fluid with a polytropic pressure floor \citep{Schaye2008} with a metallicity-dependent density threshold \citep{Schaye2004}, reproducing by construction the observed Kennicutt–Schmidt relation \citep{Kennicutt1998}. The seeding low-mass galaxies with BHs and their growth via gas accretion and merging are based on the method introduced by \cite{Springel2005b} and substantially modified by \cite{Booth2009} and \cite{Rosas-Guevara2015}.

%Stellar and AGN feedback:
Both stellar and AGN feedback are implemented as stochastic heating of gas particles \citep{DallaVecchia2012} with a temperature increase of 10$^{7.5}$ K and 10$^{8.5}$ K, respectively, chosen to minimize numerical radiative losses and to allow for self-regulation. 
Therefore, only a single mode of AGN feedback is implemented with a fixed efficiency and injected energy proportional to the gas accretion rate, following a scheme close to the so-called quasar-mode feedback.
 
Thermal energy is injected into the gas without turning off radiative cooling and without decoupling hydrodynamical forces. The main observational quantities used to calibrate the model include the observed $z=0$ GSMF, galaxy sizes, and stellar to BH mass \citep[see][for more details]{Schaye2015,Crain2015}.

\subsection{\tng}

\paragraph*{\tng\footnote{\url{https://www.tng-project.org/}} \citep{Pillepich2018,Nelson2019}} is a suite of cosmological magnetohydrodynamic simulations 
%at various resolutions and volumes, 
run with the moving-mesh code \arepo \citep{Springel2010}, assuming a $\Lambda$CDM
cosmology compatible with \cite{PlanckCollaboration2016a}. In this work, we use the \tngone simulation, with the box length of 75 $h^{-1}$Mpc ($\approx$ 110 comoving Mpc), with
1820$^3$ DM particles and 1820$^3$ initial gas cells corresponding to a mass resolution of $7.5 \times 10^6$ \MM for DM and $1.4 \times 10^6$ \MM for baryons. The minimum gravitational softening length is 0.75 kpc for DM and stars, and 190 comoving pc for gas.

% Radiative cooling and heating
\tng employs a galaxy formation model built upon the original \illustris simulation \citep{Genel2014,Vogelsberger2014}. 
It includes radiative gas cooling, both primordial and from metal lines, in the presence of a time-variable, spatially uniform, ionizing UV background instantaneously switched on at $z = 6$, with corrections for self-shielding in the dense ISM \citep{Katz1992,Faucher-Giguere2009}. In addition, cooling is further modulated by the radiation field of nearby AGNs.

% Star formation
Star formation occurs within the gas with the hydrogen density $n_{\rm H} \geq$ 0.1 cm$^{-3}$ following the empirically defined Kennicutt–Schmidt relation. Pressurization of the multiphase ISM from unresolved SNe is modeled for star-forming gas with a two-phase effective equation of state \citep{Springel2003}.

Stellar populations evolve and return mass and metals to their ambient ISM via Type Ia and II SNe and AGB
stars following tabulated mass and metal yields.
In practice, the model tracks the production and evolution of nine elements: H, He, C, N, O, Ne, Mg, Si, and Fe \citep{Pillepich2018}.

% Stellar feedback
Feedback associated with star formation drives galactic scale outflows implemented with a kinetic wind scheme. Wind particles are hydrodynamically decoupled until they leave the dense ISM. Once hydrodynamically recoupled outside the local ISM, they deposit their mass, momentum, metals, and thermal energy content.

% BH and AGN feedback
\tng follows the formation of massive BHs in sufficiently massive halos, which accrete gas from surrounding gas and inject feedback energy into their environment.
The two modes of AGN feedback are implemented: at low accretion rates \tng employs a kinetic AGN
feedback model producing BH-driven winds, while at high accretion rates thermal energy is injected into the gas surrounding the BH \citep{Weinberger2017}. \tng includes the magnetic fields that are followed with ideal magnetohydrodynamic and are dynamically coupled to the gas via the magnetic pressure \citep{Pakmor2013}. The main quantities used to calibrate the \tng model include the global SFRD as a function of cosmic time, the GSMF at $z = 0$, and the current stellar-to-halo mass relation \citep[see][for more details]{Pillepich2018}.

%

%%%%%%%%%%%%%%%%%%%%%%%%%%%%%%%%%%%%%%%%%%%%%%%%

%\bsp   % typesetting comment
%\label{lastpage}
%\end{document}


\end{document}